%% file: main.tex
\documentclass{article}

\usepackage[utf8]{inputenc} 
\usepackage[T1]{fontenc}    
\usepackage{hyperref}       
\usepackage{url}            
\usepackage{booktabs}       

\usepackage{array}
\usepackage[framemethod=TikZ]{mdframed}

\usepackage[normalem]{ulem}
\usepackage{amssymb, amsfonts, amsmath}
\usepackage[nointegrals]{wasysym}
\usepackage{array}
\usepackage{longtable}
\usepackage{xspace}
\usepackage{makecell}
\usepackage{boldline}
\usepackage{boxedminipage}
\usepackage{cite}
\usepackage{mathtools}
\usepackage{subcaption}

\usepackage{nicefrac}       
\usepackage{microtype}      

\usepackage{soul}
\usepackage{float}

\usepackage{adjustbox}
\usepackage{multirow}
\usepackage{pgfplots}
\usepackage{xcolor}

\usepackage{hyperref}

\usepackage[accepted]{mlsys2025}

\usepackage{verbatim}

\mlsystitlerunning{Intelligent Router for LLM Workloads: Improving Performance Through Workload-Aware Load Balancing}

\begin{document}

\twocolumn[
\mlsystitle{Intelligent Router for LLM Workloads: Improving Performance Through Workload-Aware Load Balancing}



\mlsyssetsymbol{equal}{*}

\begin{mlsysauthorlist}
\mlsysauthor{Kunal Jain}{msft}
\mlsysauthor{Anjaly Parayil}{msft}
\mlsysauthor{Ankur Mallick}{msft}
\mlsysauthor{Esha Choukse}{msft}
\mlsysauthor{Xiaoting Qin}{msft}
\mlsysauthor{Jue Zhang}{msft}
\mlsysauthor{Íñigo Goiri}{msft}
\mlsysauthor{Rujia Wang}{msft}
\mlsysauthor{Chetan Bansal}{msft}
\mlsysauthor{Victor Rühle}{msft}
\mlsysauthor{Anoop Kulkarni}{msft}
\mlsysauthor{Steve Kofsky}{msft}
\mlsysauthor{Saravan Rajmohan}{msft}
\end{mlsysauthorlist}

\mlsysaffiliation{msft}{Microsoft}

\mlsyscorrespondingauthor{Kunal Jain}{t-kunjain@microsoft.com}
\mlsyscorrespondingauthor{Anjaly Parayil}{aparayil@microsoft.com}


\vskip 0.3in

\begin{abstract}
Large Language Model (LLM) workloads have distinct prefill and decode phases with different compute and memory requirements which should ideally be accounted for when scheduling input queries across different LLM instances in a cluster. However existing scheduling algorithms treat LLM workloads as monolithic jobs without considering the distinct characteristics of the two phases in each workload. This leads to sub-optimal scheduling and increased response latency. In this work, we start by characterizing factors affecting the response latency during LLM inference serving. We establish that better load balancing of inference requests across the available LLM instances can improve the end-to-end latency to a larger extent than merely focusing on optimizing the instance-level scheduler.  Motivated by our findings, we propose a heuristic-guided reinforcement learning-based intelligent router for data-driven and workload-aware scheduling. Our router schedules queries across LLM instances by leveraging a trainable response-length predictor, and a novel formulation for estimating the impact of mixing different workloads and achieves over 11\% lower end-to-end latency than existing approaches on a mix of public datasets and 7.8\% lower end-to-end latency on real workload data with diverse input and output trends from Cloud Provider X. Additionally, the proposed framework can also serve as a standard for benchmarking different LLM inference schedulers since it provides the best latency for a given model, hardware, and instance-level scheduler combination. 
\end{abstract}
]




\input{sections/1_Intro}
\input{sections/2_Preliminaries}
\input{sections/3_RelatedWork}

\input{sections/4_EmpiricalStudy_v1}
\input{sections/5_WorkloadInfraawarerouter}
\input{sections/6_Experiments}

\input{sections/7_Conclusion}

\bibliography{references}
\bibliographystyle{mlsys2025}

\appendix
\input{sections/Appendix}


\end{document}

%% file: sections/1_Intro.tex
\section{Introduction}

The emergence of large language models (LLMs) and their generative ability has led to an increase in their usage in conversational engines, search engines, and code assistants \citet{chen2021evaluating, adiwardana2020towards,roller2020recipes}. The widespread usage of these large models, coupled with the significant GPU compute required for inference, has made LLM inference the dominant GPU workload. Optimizing LLM inference is thus critical for improving user experience, lowering the pressure on GPU resources, and reducing the environmental impact of running LLM workloads, and so there has been a flurry of recent work looking at various aspects of LLM inference optimization \citet{zhang2024h2o,spector2023accelerating,li2024blockllm, lin2024infinite}. 

LLM inference is usually performed on the cloud by model instances hosted by commercial cloud providers \citep{azureai,vertexai} or dedicated LLM inference platforms \citep{huggingface,openai} that serve inference requests from a variety of tenants. Owing to the widespread use of LLMs in chatbots, document summarization, and content creation, the requests vary in terms of their input and output characteristics. Each LLM instance that serves the inference request contains a scheduler, which is a batching system responsible for creating a batch of requests by retrieving requests from a queue and scheduling the execution engine. There exist multiple approaches in the literature that try to optimize the batching of these requests at a \emph{single} LLM instance \citep{orca, patel2023splitwise, agrawal2024taming, zhong2024distserve} with various goals like reducing queueing delay of requests, maximizing the utilization of the serving infrastructure, etc. For similar reasons, works like \citep{ding2024hybrid,ong2024routellm} have looked at routing requests across \emph{multiple LLMs} (route easy requests to a small model and hard requests to a big model). However neither of these lines of work have considered \emph{routing requests across multiple instances of a single LLM}.  

\begin{figure*}[t]
    \centering
     \begin{subfigure}{0.33\linewidth}
        \resizebox{\linewidth}{!}{%
        \input{plots/tikzplots/spiking_example}
        }
        \caption{Effects of mixing requests}\label{fig:mixing_requests}
    \end{subfigure}%
    \begin{subfigure}{0.33\linewidth}
        \resizebox{\linewidth}{!}{%
         \input{plots/tikzplots/completion_time_imporvements}
        }
        \caption{E2E latency }\label{fig:overall_improvement}
    \end{subfigure}%
    \begin{subfigure}{0.33\linewidth}
        \resizebox{\linewidth}{!}{%
   \input{plots/tikzplots/ttfb_time_series}
        }
        \caption{Average TTFT of requests served}\label{fig:ttft_time_series}
    \end{subfigure}
    \caption{Key Results: (a) The red curve indicates trend in the execution time when a LLM instance serves a single request, while the blue curve shows spikes in the total execution time of a request due to the addition of additional requests to that instance at fixed intervals.
    (b) Our RL-based approaches improve upon Round-robin (RR) routing in terms of overall latency with Workload Guided RL reducing average end-to-end latency by 19.18 seconds. (c) The average Time-To-First-Token is the lowest for the proposed approach.
    }
\end{figure*}

This is a significant gap since all cloud providers host multiple instances of each model and need to design policies for assigning requests to instances such that they can be served with low latency. The wide variety in the size of input queries and LLM responses across scenarios implies that sub-optimal request assignment can significantly increase inference latency. \autoref{fig:mixing_requests} shows spikes in execution time of a request when new requests are added while the LLM instance is still processing an initial request. The execution time of a request during each iteration is significantly impacted by the addition of new requests. 

\begin{figure*}[ht]
\centering
    \begin{subfigure}{0.66\linewidth}
    \resizebox{\linewidth}{!}{%
    \includegraphics[]{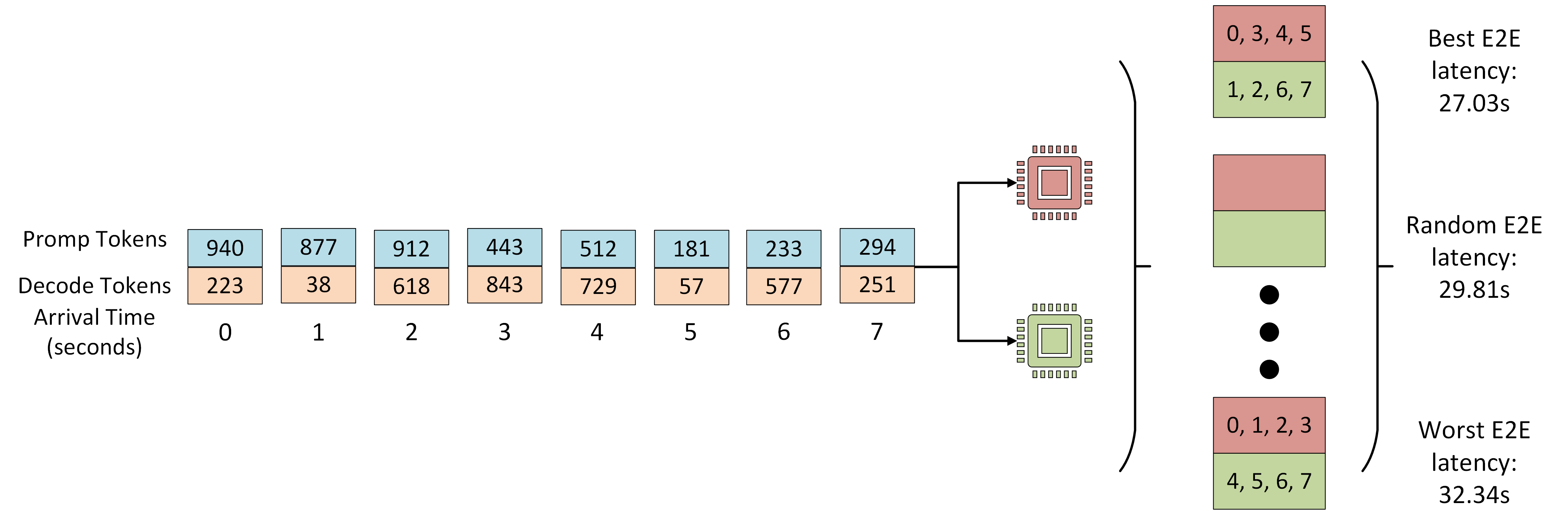}
    }\caption{
    Experimental Set Up
    }
    \end{subfigure}%
    \begin{subfigure}{0.33\linewidth}
        \centering
        \adjustbox{max width=\linewidth}{
        \begin{tabular}{|c|c|}
            \hline
             E2E Latency & \# of interference spikes \\
             \hline
             27.03 & 18 \\
             28.92 & 23 \\
             29.47 & 22 \\
             30.27 & 30 \\
             31.13 & 34 \\
             32.34 & 40 \\
             \hline
        \end{tabular}
        }
        \caption{E2E latency and spikes}
        \label{tab:my_label}
    \end{subfigure}%
    \caption{
      (a) We assigned 8 requests, arriving at a rate of 1 per second, with input and output lengths varying from 10 to 100, to two LLM instances using set partitioning. Through exhaustive search of all possible combinations, we found that the best achievable latency was 27.03 seconds, the worst was 32.34 seconds, and random assignment resulted in an expected latency of 29.81 seconds. The ideal case is approximately 10\% better than the average.
(b) The figure shows the number of spikes and their corresponding E2E latency during the simulation. The ideal scenario has a maximum of seven spikes, but we observe a significantly higher number of spikes, suggesting request preemption
    }\label{fig:mixing_small_scale_experiment}
\end{figure*}

We conducted an empirical study (see Figure \ref{fig:mixing_small_scale_experiment}) analyzing the disparity between optimal and random assignment of 8 requests arriving at a rate of 1 per second, with varying input and output lengths assigned to two LLM instances using set partitioning. Through exhaustive search of all possible combinations, we found that the best achievable latency was 27.03 seconds, the worst was 32.34 seconds, and a random assignment yielded a latency of 29.81 seconds. On average, a random assignment yields 10\% higher end-to-end latency than the optimal assignment.

In this work, we start by analyzing the prompt and decode phases of inference requests. We estimate the time to complete a request for a given prompt and decode length. 
Next, we analyze the factors that affect the end-to-end latency of requests running in an LLM instance. These factors include co-serving requests in the prompt and decode phase (determined by the instance-level scheduler), and the diversity in the prompt and decode distribution. By classifying requests based on prompt and decode characteristics, we model the latency impact of mixing incoming and existing requests and propose a latency impact estimator. To help with the latency impact estimation, we develop a lightweight decode length predictor backed by insights from the empirical study. 

Given that requests are served by one of many instances of an LLM in reality, we analyze the interplay between routing strategies and instance-level scheduling algorithms and how they affect the end-to-end latency of the requests. We assert that poor assignment of requests at the instance-level scheduler cannot improve the end-to-end latency, and hence we need better routing strategies that consider characteristics of incoming requests as well as those of requests at each model instance.  It is also necessary to treat routing as \emph{distinct} from instance-level scheduling so that we can consider the efficacy of routing strategies independent of instance level innovations such as prefill chunking and prompt caching \citep{agrawal2023sarathi,gim2024prompt}. We propose to optimize the routing strategies for \emph{any} optimizations that exist at each LLM instance. 

Finally, we use the above components to propose an intelligent router for a given instance level scheduler, LLM, and hardware combination. By strategically delaying routing and selecting the best model instance based on the current state of LLM instances and incoming requests, we reduce queuing at model instances which improves latency by 11.43\% on average for 2000 requests over 4 LLM instances. 
We present an extensive evaluation of the proposed framework's scalability with an increase in the available LLM instances, adaptability to different LLM-hardware combinations, and performance on a real production trace.



\textbf{Contributions: }1) We assert that
poor choices of requests at the instance-level scheduler cannot
improve the end-to-end latency beyond a point and established the impact of concurrently serving inference requests  with diverse characteristics (\autoref{Sec: Observation}). 2) We propose a novel formulation to model this impact (\autoref{Sec: Observation}). 3) We develop a lightweight model for predicting decode length, performing well across various prompt and decode characteristics
 (\autoref{sec:output_length_predictor}). 4) We propose a heuristic-guided, workload-aware reinforcement learning router that encodes prior knowledge of workload mixing effects and routes requests to the best-suited LLM instance, thereby achieving improved end-to-end performance (\autoref{subsec:RL_router}).
{Given that the intelligent router finds the best possible assignment for optimizing end-to-end latency for a given hardware, model, and instance-level scheduler, we present a standard for future benchmarking of inference schedulers.
 Additionally, the framework also offers flexibility to plug and play different optimizations, such as prefill chunking or prefix caching, to evaluate the best possible improvements in such scenarios.}

%% file: plots/tikzplots/spiking_example.tex

\begin{tikzpicture}

\definecolor{darkgray176}{RGB}{176,176,176}
\definecolor{darkorange25512714}{RGB}{255,127,14}
\definecolor{steelblue31119180}{RGB}{31,119,180}

\begin{axis}[
tick align=outside,
tick pos=left,
x grid style={darkgray176},
xlabel style={font=\LARGE},
xlabel={Iterations of first request},
ylabel style={font=\LARGE},
xmin=-50.45, xmax=1059.45,
xtick style={color=black},
ylabel={Execution Time (s)},
y grid style={darkgray176},
ymin=0.0055259108543396, ymax=0.392793571949005,
ytick style={color=black},
legend style={nodes={scale=1, transform shape}, at={(0.99, 0.985)}, font=\large},
xtick={0, 250, 500, 750, 1000},
xticklabel style={font=\Large},
yticklabel style={font=\Large},
]
\addplot [style={ultra thick}, steelblue31119180]
table {%
0 0.375190496444702
1 0.0262508392333984
2 0.0238125324249268
3 0.0236079692840576
4 0.0235526561737061
5 0.0235335826873779
6 0.0233528614044189
7 0.0233175754547119
8 0.0232584476470947
9 0.0232534408569336
10 0.0232579708099365
11 0.0232532024383545
12 0.0234289169311523
13 0.023564338684082
14 0.023406982421875
15 0.0232465267181396
16 0.023252010345459
17 0.0232152938842773
18 0.0231690406799316
19 0.0231809616088867
20 0.0231728553771973
21 0.0232706069946289
22 0.0232598781585693
23 0.0232090950012207
24 0.0232033729553223
25 0.0231525897979736
26 0.0231916904449463
27 0.0231289863586426
28 0.0231730937957764
29 0.0232181549072266
30 0.0232236385345459
31 0.0232131481170654
32 0.0231616497039795
33 0.0231750011444092
34 0.0232079029083252
35 0.0231688022613525
36 0.0231938362121582
37 0.0231392383575439
38 0.0233268737792969
39 0.0233237743377686
40 0.023237943649292
41 0.0232179164886475
42 0.023237943649292
43 0.0232114791870117
44 0.0232222080230713
45 0.0231723785400391
46 0.0232150554656982
47 0.0232012271881104
48 0.0232822895050049
49 0.172877788543701
50 0.027367115020752
51 0.026660680770874
52 0.026545524597168
53 0.0265793800354004
54 0.02651047706604
55 0.0264461040496826
56 0.0263955593109131
57 0.026226282119751
58 0.0262758731842041
59 0.0261774063110352
60 0.0261824131011963
61 0.0261063575744629
62 0.0261461734771729
63 0.0261127948760986
64 0.0260653495788574
65 0.0266542434692383
66 0.0263206958770752
67 0.0260264873504639
68 0.0260617733001709
69 0.0260941982269287
70 0.0260789394378662
71 0.0261454582214355
72 0.0260963439941406
73 0.0261509418487549
74 0.0260992050170898
75 0.0262672901153564
76 0.0263590812683105
77 0.0262601375579834
78 0.0261397361755371
79 0.0261528491973877
80 0.0261213779449463
81 0.0261216163635254
82 0.0261497497558594
83 0.026115894317627
84 0.0261967182159424
85 0.0260965824127197
86 0.0261263847351074
87 0.0261521339416504
88 0.026104211807251
89 0.0261185169219971
90 0.0261580944061279
91 0.026141881942749
92 0.0261151790618896
93 0.0262308120727539
94 0.0263462066650391
95 0.0262598991394043
96 0.0261790752410889
97 0.0262272357940674
98 0.0261478424072266
99 0.172266721725464
100 0.0278985500335693
101 0.0273447036743164
102 0.0271017551422119
103 0.027057409286499
104 0.027057409286499
105 0.0270476341247559
106 0.0270347595214844
107 0.0270867347717285
108 0.0269677639007568
109 0.0269944667816162
110 0.0270047187805176
111 0.0269229412078857
112 0.026949405670166
113 0.0269548892974854
114 0.0269584655761719
115 0.0269629955291748
116 0.0270133018493652
117 0.0270025730133057
118 0.0270256996154785
119 0.0270228385925293
120 0.0269753932952881
121 0.0269837379455566
122 0.0270359516143799
123 0.0270540714263916
124 0.0270628929138184
125 0.0270648002624512
126 0.027076244354248
127 0.0270953178405762
128 0.027061939239502
129 0.0270678997039795
130 0.0270180702209473
131 0.0270829200744629
132 0.0270476341247559
133 0.0270833969116211
134 0.0271050930023193
135 0.0271270275115967
136 0.0270414352416992
137 0.0270154476165771
138 0.027097225189209
139 0.0270781517028809
140 0.0270531177520752
141 0.027061939239502
142 0.0270318984985352
143 0.0270578861236572
144 0.0270810127258301
145 0.0271191596984863
146 0.0271265506744385
147 0.027108907699585
148 0.0270955562591553
149 0.171945810317993
150 0.0284545421600342
151 0.028339147567749
152 0.028052806854248
153 0.0280337333679199
154 0.0281107425689697
155 0.0280380249023438
156 0.0280194282531738
157 0.028212308883667
158 0.0280451774597168
159 0.0279474258422852
160 0.0279867649078369
161 0.0279159545898438
162 0.0278794765472412
163 0.0279483795166016
164 0.0279228687286377
165 0.0280392169952393
166 0.0282154083251953
167 0.0281946659088135
168 0.0282835960388184
169 0.0282995700836182
170 0.0281918048858643
171 0.0282154083251953
172 0.0281922817230225
173 0.0281727313995361
174 0.0281026363372803
175 0.0280599594116211
176 0.0280461311340332
177 0.0294990539550781
178 0.0280632972717285
179 0.0280249118804932
180 0.0280473232269287
181 0.0281291007995605
182 0.0280930995941162
183 0.0281364917755127
184 0.0281240940093994
185 0.0280718803405762
186 0.0280992984771729
187 0.0280971527099609
188 0.0281879901885986
189 0.0281796455383301
190 0.0281896591186523
191 0.0281269550323486
192 0.0281105041503906
193 0.0281803607940674
194 0.0282084941864014
195 0.0282044410705566
196 0.0281267166137695
197 0.0281152725219727
198 0.0281109809875488
199 0.171891450881958
200 0.030109167098999
201 0.0295615196228027
202 0.0294609069824219
203 0.0294444561004639
204 0.0294029712677002
205 0.0294334888458252
206 0.0292360782623291
207 0.0290584564208984
208 0.0290243625640869
209 0.029015064239502
210 0.0290665626525879
211 0.0290210247039795
212 0.0289919376373291
213 0.0290203094482422
214 0.0291345119476318
215 0.0291111469268799
216 0.0290839672088623
217 0.0290043354034424
218 0.0290536880493164
219 0.0289878845214844
220 0.0290606021881104
221 0.0291101932525635
222 0.0290496349334717
223 0.0290751457214355
224 0.0290560722351074
225 0.0291204452514648
226 0.0290446281433105
227 0.0290834903717041
228 0.0290849208831787
229 0.029094934463501
230 0.0291252136230469
231 0.0291037559509277
232 0.029111385345459
233 0.0290241241455078
234 0.0290367603302002
235 0.0290927886962891
236 0.0291264057159424
237 0.0291266441345215
238 0.0291404724121094
239 0.0291194915771484
240 0.0291481018066406
241 0.0292940139770508
242 0.0293722152709961
243 0.0293424129486084
244 0.029282808303833
245 0.0291743278503418
246 0.0291504859924316
247 0.0293645858764648
248 0.0291845798492432
249 0.172588348388672
250 0.0308165550231934
251 0.0306425094604492
252 0.0305585861206055
253 0.0302248001098633
254 0.0302202701568604
255 0.0301051139831543
256 0.0300276279449463
257 0.0300314426422119
258 0.0299303531646729
259 0.0298559665679932
260 0.0299050807952881
261 0.0299112796783447
262 0.0298502445220947
263 0.0298678874969482
264 0.029834508895874
265 0.0298388004302979
266 0.0298247337341309
267 0.0298137664794922
268 0.0298798084259033
269 0.0298738479614258
270 0.0298571586608887
271 0.0299134254455566
272 0.029876708984375
273 0.0299613475799561
274 0.0300736427307129
275 0.0301823616027832
276 0.0303664207458496
277 0.0305576324462891
278 0.0303833484649658
279 0.0303723812103271
280 0.030264139175415
281 0.0302019119262695
282 0.0301265716552734
283 0.0300471782684326
284 0.0299670696258545
285 0.0299711227416992
286 0.0300123691558838
287 0.0299334526062012
288 0.0299744606018066
289 0.0299854278564453
290 0.0300252437591553
291 0.0299298763275146
292 0.0300474166870117
293 0.0300025939941406
294 0.030001163482666
295 0.0299899578094482
296 0.0299844741821289
297 0.0299837589263916
298 0.0300171375274658
299 0.172605276107788
300 0.0315594673156738
301 0.0314624309539795
302 0.0313456058502197
303 0.0313065052032471
304 0.0308830738067627
305 0.0308835506439209
306 0.0308425426483154
307 0.0308806896209717
308 0.0308496952056885
309 0.0308642387390137
310 0.0307865142822266
311 0.0312483310699463
312 0.0308957099914551
313 0.0308845043182373
314 0.0307960510253906
315 0.0308332443237305
316 0.0308640003204346
317 0.0308146476745605
318 0.0309092998504639
319 0.0308842658996582
320 0.030827522277832
321 0.0309197902679443
322 0.0308492183685303
323 0.0308675765991211
324 0.030817985534668
325 0.0309455394744873
326 0.030949592590332
327 0.030198335647583
328 0.0305461883544922
329 0.030224084854126
330 0.0299789905548096
331 0.0300338268280029
332 0.0311527252197266
333 0.0300688743591309
334 0.0300767421722412
335 0.0302186012268066
336 0.0302534103393555
337 0.0301671028137207
338 0.0301535129547119
339 0.0301082134246826
340 0.0300905704498291
341 0.0300755500793457
342 0.0300843715667725
343 0.0300686359405518
344 0.0300881862640381
345 0.0300512313842773
346 0.0300800800323486
347 0.0301878452301025
348 0.0302224159240723
349 0.17256498336792
350 0.0318589210510254
351 0.0315637588500977
352 0.0313923358917236
353 0.0313024520874023
354 0.03102707862854
355 0.031031608581543
356 0.0310068130493164
357 0.0309944152832031
358 0.0309693813323975
359 0.0309023857116699
360 0.0309031009674072
361 0.0309000015258789
362 0.0308637619018555
363 0.0308454036712646
364 0.0308704376220703
365 0.0309000015258789
366 0.0308938026428223
367 0.0308635234832764
368 0.030853271484375
369 0.0309135913848877
370 0.0308816432952881
371 0.0309433937072754
372 0.030888557434082
373 0.0308966636657715
374 0.0309326648712158
375 0.0309574604034424
376 0.030996561050415
377 0.03104567527771
378 0.0311539173126221
379 0.0311131477355957
380 0.0310966968536377
381 0.0311098098754883
382 0.0310606956481934
383 0.0310599803924561
384 0.0310750007629395
385 0.0310044288635254
386 0.0309684276580811
387 0.031019926071167
388 0.03102707862854
389 0.0310051441192627
390 0.030998706817627
391 0.0310008525848389
392 0.0310554504394531
393 0.031146764755249
394 0.0311827659606934
395 0.0311391353607178
396 0.0311203002929688
397 0.0310323238372803
398 0.0310156345367432
399 0.172572135925293
400 0.0325930118560791
401 0.032423734664917
402 0.0323600769042969
403 0.0321023464202881
404 0.0321552753448486
405 0.0321354866027832
406 0.0320615768432617
407 0.0327513217926025
408 0.0323615074157715
409 0.0320520401000977
410 0.0320925712585449
411 0.0311729907989502
412 0.0308659076690674
413 0.0308570861816406
414 0.0308816432952881
415 0.0308554172515869
416 0.0308349132537842
417 0.0308573246002197
418 0.0309007167816162
419 0.03092360496521
420 0.0309112071990967
421 0.0309097766876221
422 0.0309083461761475
423 0.0309591293334961
424 0.0309147834777832
425 0.0309171676635742
426 0.0309510231018066
427 0.0309762954711914
428 0.030942440032959
429 0.030921459197998
430 0.0309450626373291
431 0.0309839248657227
432 0.0310139656066895
433 0.0309329032897949
434 0.0309567451477051
435 0.0309677124023438
436 0.0310001373291016
437 0.0309989452362061
438 0.0311558246612549
439 0.0309879779815674
440 0.0310585498809814
441 0.0310678482055664
442 0.0310492515563965
443 0.0310404300689697
444 0.0310330390930176
445 0.0310087203979492
446 0.0310096740722656
447 0.0310442447662354
448 0.0310509204864502
449 0.172927141189575
450 0.0326938629150391
451 0.0324611663818359
452 0.032059907913208
453 0.0319557189941406
454 0.0319638252258301
455 0.0320842266082764
456 0.0320034027099609
457 0.0320112705230713
458 0.0320291519165039
459 0.0319738388061523
460 0.0319557189941406
461 0.0318682193756104
462 0.0318853855133057
463 0.0318913459777832
464 0.0318853855133057
465 0.0319521427154541
466 0.0320072174072266
467 0.0320076942443848
468 0.0320308208465576
469 0.0312032699584961
470 0.0310149192810059
471 0.0309762954711914
472 0.0311110019683838
473 0.0313973426818848
474 0.0313313007354736
475 0.0312256813049316
476 0.0310826301574707
477 0.0311539173126221
478 0.0312263965606689
479 0.0325756072998047
480 0.0311644077301025
481 0.0311617851257324
482 0.0311810970306396
483 0.0310866832733154
484 0.0311441421508789
485 0.0310757160186768
486 0.031132698059082
487 0.0310330390930176
488 0.0310595035552979
489 0.031076192855835
490 0.031278133392334
491 0.0311827659606934
492 0.0310440063476562
493 0.0312302112579346
494 0.031200647354126
495 0.0312416553497314
496 0.0311019420623779
497 0.03104567527771
498 0.0310673713684082
499 0.172957897186279
500 0.0327515602111816
501 0.0322825908660889
502 0.032245397567749
503 0.0322906970977783
504 0.0322701930999756
505 0.0322165489196777
506 0.0321929454803467
507 0.0320587158203125
508 0.0319726467132568
509 0.0319533348083496
510 0.0319850444793701
511 0.0329089164733887
512 0.0319781303405762
513 0.0319845676422119
514 0.0319473743438721
515 0.0319936275482178
516 0.0319874286651611
517 0.0320215225219727
518 0.0320053100585938
519 0.0320405960083008
520 0.0320537090301514
521 0.0320687294006348
522 0.032041072845459
523 0.032045841217041
524 0.0320544242858887
525 0.0320229530334473
526 0.0320301055908203
527 0.0320992469787598
528 0.0320119857788086
529 0.0320298671722412
530 0.0320098400115967
531 0.0320332050323486
532 0.0320761203765869
533 0.0321052074432373
534 0.0321910381317139
535 0.032282829284668
536 0.0321123600006104
537 0.0321047306060791
538 0.0321075916290283
539 0.032109260559082
540 0.0322842597961426
541 0.0323059558868408
542 0.0322554111480713
543 0.0322465896606445
544 0.0321626663208008
545 0.0321304798126221
546 0.0321803092956543
547 0.0321626663208008
548 0.0322208404541016
549 0.0321624279022217
550 0.0322225093841553
551 0.032285213470459
552 0.0322480201721191
553 0.0321893692016602
554 0.0322186946868896
555 0.0323915481567383
556 0.0323746204376221
557 0.0323765277862549
558 0.0322582721710205
559 0.0322492122650146
560 0.0322105884552002
561 0.0322203636169434
562 0.0322613716125488
563 0.0323741436004639
564 0.0323455333709717
565 0.0322723388671875
566 0.0322349071502686
567 0.0321838855743408
568 0.0322692394256592
569 0.0322437286376953
570 0.0322577953338623
571 0.0327975749969482
572 0.0325217247009277
573 0.032289981842041
574 0.0323102474212646
575 0.0322854518890381
576 0.0322816371917725
577 0.0323278903961182
578 0.032264232635498
579 0.0324676036834717
580 0.0326054096221924
581 0.0324833393096924
582 0.0324263572692871
583 0.0324981212615967
584 0.0325264930725098
585 0.0324068069458008
586 0.0324087142944336
587 0.0323834419250488
588 0.0324573516845703
589 0.0324141979217529
590 0.0323641300201416
591 0.0323185920715332
592 0.0323348045349121
593 0.0323858261108398
594 0.0323779582977295
595 0.0323851108551025
596 0.0323643684387207
597 0.0325005054473877
598 0.0325636863708496
599 0.0325019359588623
600 0.0324749946594238
601 0.032423734664917
602 0.0325021743774414
603 0.0325686931610107
604 0.0324561595916748
605 0.032383918762207
606 0.0323796272277832
607 0.032397985458374
608 0.0323836803436279
609 0.032374382019043
610 0.0324020385742188
611 0.0324528217315674
612 0.0323781967163086
613 0.0324037075042725
614 0.0324292182922363
615 0.0324785709381104
616 0.0324337482452393
617 0.0324177742004395
618 0.0324387550354004
619 0.0324761867523193
620 0.0324907302856445
621 0.0325267314910889
622 0.0324945449829102
623 0.0324485301971436
624 0.0324792861938477
625 0.0324804782867432
626 0.032496452331543
627 0.032461404800415
628 0.0324177742004395
629 0.0325431823730469
630 0.0325794219970703
631 0.0345199108123779
632 0.0325512886047363
633 0.0327818393707275
634 0.0327670574188232
635 0.0326306819915771
636 0.0326151847839355
637 0.0325641632080078
638 0.0325100421905518
639 0.0325536727905273
640 0.0330913066864014
641 0.0326352119445801
642 0.0325415134429932
643 0.0326037406921387
644 0.0326390266418457
645 0.0326793193817139
646 0.0326976776123047
647 0.0325815677642822
648 0.0325489044189453
649 0.0326993465423584
650 0.0326800346374512
651 0.0326590538024902
652 0.0326979160308838
653 0.0326526165008545
654 0.0325727462768555
655 0.0328400135040283
656 0.0321123600006104
657 0.0315947532653809
658 0.0316205024719238
659 0.0316452980041504
660 0.0316207408905029
661 0.0316686630249023
662 0.03165602684021
663 0.0316686630249023
664 0.0316231250762939
665 0.031733512878418
666 0.031754732131958
667 0.0317928791046143
668 0.031787633895874
669 0.0317506790161133
670 0.0317392349243164
671 0.031693696975708
672 0.0317425727844238
673 0.031721830368042
674 0.0317237377166748
675 0.0316972732543945
676 0.03171706199646
677 0.0317268371582031
678 0.0317471027374268
679 0.0317859649658203
680 0.0317959785461426
681 0.031876802444458
682 0.0320122241973877
683 0.0318970680236816
684 0.0318386554718018
685 0.0320460796356201
686 0.0320992469787598
687 0.0319633483886719
688 0.0318925380706787
689 0.031898021697998
690 0.0318777561187744
691 0.0318965911865234
692 0.0318989753723145
693 0.0318818092346191
694 0.0318431854248047
695 0.0318434238433838
696 0.0318291187286377
697 0.0320043563842773
698 0.0320324897766113
699 0.0319435596466064
700 0.0319015979766846
701 0.0318703651428223
702 0.0318596363067627
703 0.0318963527679443
704 0.0318496227264404
705 0.0318312644958496
706 0.0318481922149658
707 0.0318400859832764
708 0.031853199005127
709 0.0319898128509521
710 0.0318319797515869
711 0.0320711135864258
712 0.0322813987731934
713 0.0323173999786377
714 0.0322611331939697
715 0.0321695804595947
716 0.0320863723754883
717 0.0320370197296143
718 0.0319504737854004
719 0.0319714546203613
720 0.0319387912750244
721 0.031947135925293
722 0.0319380760192871
723 0.0318999290466309
724 0.031944751739502
725 0.0319020748138428
726 0.0319445133209229
727 0.0319175720214844
728 0.0319373607635498
729 0.0319550037384033
730 0.0319480895996094
731 0.0319557189941406
732 0.0319840908050537
733 0.0319795608520508
734 0.0319886207580566
735 0.0320944786071777
736 0.031991720199585
737 0.032015323638916
738 0.0319652557373047
739 0.0319831371307373
740 0.031982421875
741 0.032395601272583
742 0.0322637557983398
743 0.0320513248443604
744 0.0320940017700195
745 0.0320093631744385
746 0.0320541858673096
747 0.0320122241973877
748 0.0320348739624023
749 0.0320563316345215
750 0.0320377349853516
751 0.032024621963501
752 0.0319907665252686
753 0.0321090221405029
754 0.0313146114349365
755 0.030925989151001
756 0.0308957099914551
757 0.0309157371520996
758 0.0309464931488037
759 0.0309276580810547
760 0.0309102535247803
761 0.0309240818023682
762 0.0309207439422607
763 0.0309348106384277
764 0.0309023857116699
765 0.0308797359466553
766 0.0309848785400391
767 0.0308973789215088
768 0.0309386253356934
769 0.0309162139892578
770 0.0309650897979736
771 0.0309286117553711
772 0.0309891700744629
773 0.0309383869171143
774 0.0309672355651855
775 0.0310003757476807
776 0.0310788154602051
777 0.0310215950012207
778 0.0310709476470947
779 0.0310111045837402
780 0.0310301780700684
781 0.0309925079345703
782 0.0311095714569092
783 0.0312027931213379
784 0.031285285949707
785 0.0313537120819092
786 0.0310332775115967
787 0.0309951305389404
788 0.0321393013000488
789 0.0309879779815674
790 0.0310220718383789
791 0.0310289859771729
792 0.0311355590820312
793 0.0311665534973145
794 0.0310800075531006
795 0.0311353206634521
796 0.031196117401123
797 0.0312063694000244
798 0.0311167240142822
799 0.031118631362915
800 0.0310454368591309
801 0.0310883522033691
802 0.0312261581420898
803 0.0304980278015137
804 0.0300483703613281
805 0.0300321578979492
806 0.0300309658050537
807 0.0300474166870117
808 0.0300705432891846
809 0.0300514698028564
810 0.0300478935241699
811 0.0300893783569336
812 0.0300793647766113
813 0.0300590991973877
814 0.0300698280334473
815 0.0300741195678711
816 0.0300841331481934
817 0.0300350189208984
818 0.030081033706665
819 0.0300626754760742
820 0.0300743579864502
821 0.0300114154815674
822 0.0300805568695068
823 0.030036449432373
824 0.0300896167755127
825 0.0301074981689453
826 0.0300679206848145
827 0.0301003456115723
828 0.0300986766815186
829 0.0313284397125244
830 0.0305800437927246
831 0.0301640033721924
832 0.0300877094268799
833 0.0301516056060791
834 0.0301496982574463
835 0.0301198959350586
836 0.0301826000213623
837 0.0295708179473877
838 0.0290045738220215
839 0.0290756225585938
840 0.0290019512176514
841 0.0290138721466064
842 0.0290322303771973
843 0.0290470123291016
844 0.0290017127990723
845 0.0289709568023682
846 0.0290036201477051
847 0.0290195941925049
848 0.0289998054504395
849 0.0290172100067139
850 0.0290606021881104
851 0.0291235446929932
852 0.0285825729370117
853 0.0281319618225098
854 0.0281162261962891
855 0.028085470199585
856 0.0280773639678955
857 0.028109073638916
858 0.0282042026519775
859 0.0281853675842285
860 0.0281767845153809
861 0.0283238887786865
862 0.0283920764923096
863 0.02829909324646
864 0.0283136367797852
865 0.0282444953918457
866 0.0282547473907471
867 0.0281879901885986
868 0.0282130241394043
869 0.0281844139099121
870 0.0281507968902588
871 0.0282018184661865
872 0.028221607208252
873 0.0282421112060547
874 0.0282104015350342
875 0.028256893157959
876 0.0282702445983887
877 0.0283524990081787
878 0.0283548831939697
879 0.0283753871917725
880 0.0283172130584717
881 0.02828049659729
882 0.0282695293426514
883 0.0282490253448486
884 0.0282237529754639
885 0.0282120704650879
886 0.0282077789306641
887 0.0282068252563477
888 0.0282192230224609
889 0.028228759765625
890 0.0282504558563232
891 0.0281882286071777
892 0.028226375579834
893 0.0282769203186035
894 0.0282633304595947
895 0.0282573699951172
896 0.0282418727874756
897 0.0282604694366455
898 0.0282254219055176
899 0.0282509326934814
900 0.0283036231994629
901 0.0273280143737793
902 0.0269207954406738
903 0.0268251895904541
904 0.0268411636352539
905 0.0268325805664062
906 0.0268828868865967
907 0.0268619060516357
908 0.0268921852111816
909 0.0269680023193359
910 0.0270161628723145
911 0.02691650390625
912 0.0269227027893066
913 0.0269231796264648
914 0.0269639492034912
915 0.0269744396209717
916 0.0269825458526611
917 0.02687668800354
918 0.0268762111663818
919 0.0268654823303223
920 0.0268573760986328
921 0.026939868927002
922 0.0269017219543457
923 0.0268876552581787
924 0.0268881320953369
925 0.0268805027008057
926 0.0271091461181641
927 0.0270347595214844
928 0.0268526077270508
929 0.0269150733947754
930 0.0268421173095703
931 0.0269136428833008
932 0.0268826484680176
933 0.0268890857696533
934 0.0268938541412354
935 0.0269021987915039
936 0.0268657207489014
937 0.0269055366516113
938 0.0270237922668457
939 0.0269110202789307
940 0.0268855094909668
941 0.0269374847412109
942 0.026869535446167
943 0.0269544124603271
944 0.0268790721893311
945 0.0269184112548828
946 0.0268688201904297
947 0.0268709659576416
948 0.026925802230835
949 0.027005672454834
950 0.0242893695831299
951 0.023810863494873
952 0.023775577545166
953 0.0237545967102051
954 0.0237390995025635
955 0.0237643718719482
956 0.0237588882446289
957 0.0237452983856201
958 0.0237627029418945
959 0.0237171649932861
960 0.0238058567047119
961 0.0237617492675781
962 0.0237171649932861
963 0.0237641334533691
964 0.0237393379211426
965 0.0237636566162109
966 0.0258855819702148
967 0.0237410068511963
968 0.0237479209899902
969 0.0237753391265869
970 0.0237991809844971
971 0.0237863063812256
972 0.0237796306610107
973 0.0237495899200439
974 0.0237622261047363
975 0.0238184928894043
976 0.0237765312194824
977 0.023768424987793
978 0.0237581729888916
979 0.0237672328948975
980 0.0237917900085449
981 0.0237736701965332
982 0.0237345695495605
983 0.023740291595459
984 0.0237407684326172
985 0.0237648487091064
986 0.0238478183746338
987 0.0237884521484375
988 0.0237374305725098
989 0.0237510204315186
990 0.0237467288970947
991 0.0237696170806885
992 0.0237491130828857
993 0.0237305164337158
994 0.0237791538238525
995 0.0237841606140137
996 0.0237390995025635
997 0.0237686634063721
998 0.0237784385681152
999 0.0237572193145752
1000 0.023754358291626
1001 0.0237717628479004
1002 0.0238156318664551
1003 0.0237905979156494
1004 0.0237605571746826
1005 0.0237538814544678
1006 0.0237805843353271
1007 0.0238244533538818
1008 0.0237667560577393
1009 0.0238871574401855
};
\addlegendentry{With mixing}

\addplot [style={ultra thick}, dashed, darkorange25512714]
table {%
0 0.373505115509033
1 0.0252466201782227
2 0.0237183570861816
3 0.0237312316894531
4 0.023751974105835
5 0.0237629413604736
6 0.0237216949462891
7 0.0236876010894775
8 0.0233843326568604
9 0.0233376026153564
10 0.0233473777770996
11 0.0232515335083008
12 0.0238223075866699
13 0.0234448909759521
14 0.0232400894165039
15 0.0232441425323486
16 0.0232067108154297
17 0.0232408046722412
18 0.0232264995574951
19 0.0232090950012207
20 0.023228645324707
21 0.0232326984405518
22 0.0232462882995605
23 0.0231614112854004
24 0.0231442451477051
25 0.0231993198394775
26 0.02315354347229
27 0.0232346057891846
28 0.0232079029083252
29 0.0232257843017578
30 0.0231590270996094
31 0.0231964588165283
32 0.0232117176055908
33 0.0232083797454834
34 0.0232245922088623
35 0.0231950283050537
36 0.0232012271881104
37 0.0232019424438477
38 0.0231838226318359
39 0.0231978893280029
40 0.0231795310974121
41 0.0231525897979736
42 0.0232062339782715
43 0.0232043266296387
44 0.0232489109039307
45 0.0232160091400146
46 0.0232381820678711
47 0.0233254432678223
48 0.0232486724853516
49 0.023235559463501
50 0.0232226848602295
51 0.0232102870941162
52 0.0232090950012207
53 0.0232870578765869
54 0.0232510566711426
55 0.0232505798339844
56 0.0232653617858887
57 0.0232582092285156
58 0.0232326984405518
59 0.0232124328613281
60 0.0231912136077881
61 0.0232243537902832
62 0.0232260227203369
63 0.0231971740722656
64 0.0231883525848389
65 0.0232491493225098
66 0.0232336521148682
67 0.0232162475585938
68 0.0232493877410889
69 0.0231654644012451
70 0.0231711864471436
71 0.0232393741607666
72 0.0232062339782715
73 0.023289680480957
74 0.023242712020874
75 0.0232346057891846
76 0.0232222080230713
77 0.023205041885376
78 0.0232536792755127
79 0.0231997966766357
80 0.0232200622558594
81 0.023200511932373
82 0.0232267379760742
83 0.0232410430908203
84 0.0232009887695312
85 0.0232095718383789
86 0.0232479572296143
87 0.0232398509979248
88 0.0232701301574707
89 0.023303747177124
90 0.0233421325683594
91 0.0235378742218018
92 0.0235288143157959
93 0.0235600471496582
94 0.0236730575561523
95 0.0237095355987549
96 0.0236389636993408
97 0.0234909057617188
98 0.0234072208404541
99 0.0233743190765381
100 0.0232851505279541
101 0.0233333110809326
102 0.0232813358306885
103 0.0232748985290527
104 0.0232570171356201
105 0.0232088565826416
106 0.0231986045837402
107 0.0232479572296143
108 0.0232532024383545
109 0.0232069492340088
110 0.0232114791870117
111 0.0232546329498291
112 0.0232319831848145
113 0.0232267379760742
114 0.0232260227203369
115 0.0232234001159668
116 0.0232775211334229
117 0.0233051776885986
118 0.0232346057891846
119 0.0232644081115723
120 0.0232369899749756
121 0.0232348442077637
122 0.0232095718383789
123 0.0232758522033691
124 0.0232088565826416
125 0.0232625007629395
126 0.0232908725738525
127 0.0232901573181152
128 0.02329421043396
129 0.0232834815979004
130 0.0234787464141846
131 0.0233986377716064
132 0.0232977867126465
133 0.0232484340667725
134 0.0232670307159424
135 0.0232281684875488
136 0.0232744216918945
137 0.0232374668121338
138 0.0232844352722168
139 0.0233209133148193
140 0.0232322216033936
141 0.0232491493225098
142 0.0232415199279785
143 0.0232884883880615
144 0.0232675075531006
145 0.0233848094940186
146 0.0233092308044434
147 0.0232887268066406
148 0.0232281684875488
149 0.0232553482055664
150 0.0232672691345215
151 0.0232372283935547
152 0.0232710838317871
153 0.0232508182525635
154 0.0232348442077637
155 0.0232322216033936
156 0.0232267379760742
157 0.023277759552002
158 0.0232541561126709
159 0.0232586860656738
160 0.0232853889465332
161 0.0232696533203125
162 0.0232939720153809
163 0.0233042240142822
164 0.023287296295166
165 0.023301362991333
166 0.0232779979705811
167 0.0233321189880371
168 0.0232617855072021
169 0.0232441425323486
170 0.0231962203979492
171 0.0232324600219727
172 0.0233156681060791
173 0.0233016014099121
174 0.0233054161071777
175 0.0233042240142822
176 0.0232884883880615
177 0.0233020782470703
178 0.0232882499694824
179 0.0233235359191895
180 0.0232596397399902
181 0.0232751369476318
182 0.0232303142547607
183 0.0232515335083008
184 0.0232529640197754
185 0.0232572555541992
186 0.0232532024383545
187 0.0232410430908203
188 0.0232856273651123
189 0.0232951641082764
190 0.0233108997344971
191 0.0232927799224854
192 0.0232374668121338
193 0.0232768058776855
194 0.0232605934143066
195 0.0233054161071777
196 0.0233008861541748
197 0.0232386589050293
198 0.0233111381530762
199 0.0232820510864258
200 0.023301362991333
201 0.0233373641967773
202 0.0233421325683594
203 0.0233490467071533
204 0.023252010345459
205 0.0232765674591064
206 0.0233397483825684
207 0.0232894420623779
208 0.0232582092285156
209 0.023289680480957
210 0.0232701301574707
211 0.0232927799224854
212 0.0232508182525635
213 0.0233232975006104
214 0.0232934951782227
215 0.0233097076416016
216 0.0247764587402344
217 0.0232656002044678
218 0.0232908725738525
219 0.0232775211334229
220 0.0232877731323242
221 0.0233023166656494
222 0.0232675075531006
223 0.023268461227417
224 0.0232882499694824
225 0.0232555866241455
226 0.023320198059082
227 0.023308277130127
228 0.0233132839202881
229 0.023296594619751
230 0.0233974456787109
231 0.0232880115509033
232 0.0233139991760254
233 0.0232710838317871
234 0.0232925415039062
235 0.0234503746032715
236 0.023446798324585
237 0.0233306884765625
238 0.023395299911499
239 0.023341178894043
240 0.0233695507049561
241 0.0233478546142578
242 0.0233504772186279
243 0.0233230590820312
244 0.0232813358306885
245 0.023303747177124
246 0.0233311653137207
247 0.0233292579650879
248 0.0236403942108154
249 0.0235195159912109
250 0.0233354568481445
251 0.0233604907989502
252 0.023345947265625
253 0.0233283042907715
254 0.0233297348022461
255 0.0233151912689209
256 0.0233249664306641
257 0.0233380794525146
258 0.0232856273651123
259 0.0233311653137207
260 0.0233564376831055
261 0.0233664512634277
262 0.023327112197876
263 0.0233311653137207
264 0.0233161449432373
265 0.0233213901519775
266 0.0232877731323242
267 0.0233590602874756
268 0.0233166217803955
269 0.0233449935913086
270 0.0233328342437744
271 0.0233263969421387
272 0.0233101844787598
273 0.0233154296875
274 0.0233118534088135
275 0.0233867168426514
276 0.0233385562896729
277 0.0233561992645264
278 0.0233275890350342
279 0.0233597755432129
280 0.0233001708984375
281 0.0233595371246338
282 0.0233254432678223
283 0.0233175754547119
284 0.023334264755249
285 0.0233967304229736
286 0.0234308242797852
287 0.0236063003540039
288 0.0235269069671631
289 0.0234599113464355
290 0.023369312286377
291 0.0233325958251953
292 0.0234076976776123
293 0.0233292579650879
294 0.0233426094055176
295 0.0233008861541748
296 0.0232856273651123
297 0.0233175754547119
298 0.0233352184295654
299 0.023406982421875
300 0.0234217643737793
301 0.0234501361846924
302 0.0234296321868896
303 0.0233707427978516
304 0.0233325958251953
305 0.0233821868896484
306 0.0233097076416016
307 0.023395299911499
308 0.0233197212219238
309 0.0233666896820068
310 0.0233602523803711
311 0.0233912467956543
312 0.0233466625213623
313 0.0233426094055176
314 0.0233154296875
315 0.0233423709869385
316 0.0233044624328613
317 0.0233626365661621
318 0.0234966278076172
319 0.0234532356262207
320 0.0234317779541016
321 0.0233564376831055
322 0.0233695507049561
323 0.0233235359191895
324 0.0233616828918457
325 0.0233290195465088
326 0.0233800411224365
327 0.023385763168335
328 0.0233113765716553
329 0.023381233215332
330 0.0233645439147949
331 0.023367166519165
332 0.0238800048828125
333 0.0234429836273193
334 0.0234129428863525
335 0.0232977867126465
336 0.0233800411224365
337 0.0233688354492188
338 0.0233268737792969
339 0.0233979225158691
340 0.0233311653137207
341 0.023334264755249
342 0.0233185291290283
343 0.0234122276306152
344 0.0233731269836426
345 0.0233161449432373
346 0.0233662128448486
347 0.0233612060546875
348 0.0233657360076904
349 0.0233643054962158
350 0.0233888626098633
351 0.023329496383667
352 0.0233445167541504
353 0.0233502388000488
354 0.0233480930328369
355 0.023425817489624
356 0.023409366607666
357 0.023343563079834
358 0.0233807563781738
359 0.02337646484375
360 0.023345947265625
361 0.0233368873596191
362 0.0233376026153564
363 0.02341628074646
364 0.023514986038208
365 0.0244143009185791
366 0.0239048004150391
367 0.0233783721923828
368 0.0233767032623291
369 0.0233860015869141
370 0.023409366607666
371 0.0234525203704834
372 0.0233964920043945
373 0.0233898162841797
374 0.0234811305999756
375 0.0234942436218262
376 0.0234072208404541
377 0.0233957767486572
378 0.0234534740447998
379 0.0236010551452637
380 0.0235590934753418
381 0.0235118865966797
382 0.0234887599945068
383 0.023456335067749
384 0.0234651565551758
385 0.0234367847442627
386 0.0234148502349854
387 0.0234134197235107
388 0.0233950614929199
389 0.0233795642852783
390 0.0233943462371826
391 0.0233633518218994
392 0.0233769416809082
393 0.0234205722808838
394 0.0233957767486572
395 0.0234289169311523
396 0.0233843326568604
397 0.0233662128448486
398 0.0233678817749023
399 0.0234112739562988
400 0.0233860015869141
401 0.023486852645874
402 0.0235958099365234
403 0.0237209796905518
404 0.0237245559692383
405 0.0235946178436279
406 0.0234670639038086
407 0.0234029293060303
408 0.0234200954437256
409 0.0234408378601074
410 0.0234200954437256
411 0.0234193801879883
412 0.023385763168335
413 0.0233798027038574
414 0.0233569145202637
415 0.0234167575836182
416 0.0233941078186035
417 0.0235130786895752
418 0.0234620571136475
419 0.0233650207519531
420 0.0233902931213379
421 0.0233721733093262
422 0.0234050750732422
423 0.0234150886535645
424 0.023411750793457
425 0.0233681201934814
426 0.0234134197235107
427 0.0234189033508301
428 0.0235879421234131
429 0.0237331390380859
430 0.0250847339630127
431 0.0234973430633545
432 0.023449182510376
433 0.0234355926513672
434 0.0233988761901855
435 0.0234432220458984
436 0.0234441757202148
437 0.0233957767486572
438 0.023444652557373
439 0.0234196186065674
440 0.0234625339508057
441 0.0234096050262451
442 0.0234510898590088
443 0.0234019756317139
444 0.0234038829803467
445 0.0234138965606689
446 0.0234940052032471
447 0.0234367847442627
448 0.0234503746032715
449 0.0235204696655273
450 0.023446798324585
451 0.0234367847442627
452 0.0234403610229492
453 0.0234262943267822
454 0.0234355926513672
455 0.023472785949707
456 0.0234367847442627
457 0.0234222412109375
458 0.0233838558197021
459 0.0239019393920898
460 0.0234575271606445
461 0.0233974456787109
462 0.0234549045562744
463 0.0234334468841553
464 0.0234472751617432
465 0.0234313011169434
466 0.0234856605529785
467 0.0235364437103271
468 0.0235092639923096
469 0.0235538482666016
470 0.0235240459442139
471 0.0234599113464355
472 0.0234513282775879
473 0.0234565734863281
474 0.0234477519989014
475 0.0234787464141846
476 0.0234265327453613
477 0.0235886573791504
478 0.0237078666687012
479 0.0236430168151855
480 0.0235552787780762
481 0.0236577987670898
482 0.0236005783081055
483 0.0234475135803223
484 0.0234525203704834
485 0.0234098434448242
486 0.02347731590271
487 0.0234558582305908
488 0.0234611034393311
489 0.0235834121704102
490 0.0234651565551758
491 0.0234620571136475
492 0.0234541893005371
493 0.0234696865081787
494 0.0234699249267578
495 0.0234329700469971
496 0.0234999656677246
497 0.0234360694885254
498 0.0234410762786865
499 0.0235061645507812
500 0.0234763622283936
501 0.0234851837158203
502 0.0235123634338379
503 0.0234513282775879
504 0.023536205291748
505 0.0236184597015381
506 0.0235140323638916
507 0.0234665870666504
508 0.023496150970459
509 0.0235049724578857
510 0.0235426425933838
511 0.0234348773956299
512 0.0234763622283936
513 0.0235292911529541
514 0.0235259532928467
515 0.0235185623168945
516 0.0234518051147461
517 0.0234653949737549
518 0.0234682559967041
519 0.0234470367431641
520 0.0234749317169189
521 0.0235879421234131
522 0.0236401557922363
523 0.0235161781311035
524 0.0234940052032471
525 0.0234625339508057
526 0.0234997272491455
527 0.0235145092010498
528 0.0235369205474854
529 0.0234906673431396
530 0.0234529972076416
531 0.0234990119934082
532 0.0234968662261963
533 0.023449182510376
534 0.0234968662261963
535 0.0234470367431641
536 0.0234835147857666
537 0.0234978199005127
538 0.0234658718109131
539 0.0235006809234619
540 0.023578405380249
541 0.0237467288970947
542 0.0237071514129639
543 0.0236978530883789
544 0.0236246585845947
545 0.0236382484436035
546 0.0235371589660645
547 0.0234334468841553
548 0.0234832763671875
549 0.0234775543212891
550 0.023442268371582
551 0.0235133171081543
552 0.0234477519989014
553 0.0235133171081543
554 0.0235166549682617
555 0.0235307216644287
556 0.0235030651092529
557 0.0234918594360352
558 0.0235202312469482
559 0.0234990119934082
560 0.0235633850097656
561 0.0235004425048828
562 0.0235183238983154
563 0.0235142707824707
564 0.0235359668731689
565 0.023489236831665
566 0.0235249996185303
567 0.0235433578491211
568 0.0235288143157959
569 0.0234956741333008
570 0.0234951972961426
571 0.0235002040863037
572 0.0235528945922852
573 0.0236656665802002
574 0.0238935947418213
575 0.024075984954834
576 0.0240213871002197
577 0.0238893032073975
578 0.0236067771911621
579 0.023543119430542
580 0.0235254764556885
581 0.02347731590271
582 0.0235400199890137
583 0.0235233306884766
584 0.0235164165496826
585 0.0234813690185547
586 0.023503303527832
587 0.0235583782196045
588 0.023503303527832
589 0.0235300064086914
590 0.0235509872436523
591 0.0235719680786133
592 0.0235333442687988
593 0.0235388278961182
594 0.0235459804534912
595 0.0235316753387451
596 0.0234827995300293
597 0.02431321144104
598 0.0238280296325684
599 0.0236318111419678
600 0.0235249996185303
601 0.0235664844512939
602 0.0235719680786133
603 0.0235300064086914
604 0.0235476493835449
605 0.023573637008667
606 0.0235939025878906
607 0.0235404968261719
608 0.0235183238983154
609 0.0235121250152588
610 0.0235288143157959
611 0.0235593318939209
612 0.0234982967376709
613 0.0236637592315674
614 0.0236895084381104
615 0.0236878395080566
616 0.0235867500305176
617 0.0235700607299805
618 0.0235590934753418
619 0.0235633850097656
620 0.0234842300415039
621 0.0235621929168701
622 0.023571252822876
623 0.0235445499420166
624 0.023547887802124
625 0.023554801940918
626 0.0235636234283447
627 0.0235390663146973
628 0.023547887802124
629 0.0241303443908691
630 0.0235366821289062
631 0.0235559940338135
632 0.0235309600830078
633 0.0235249996185303
634 0.0235419273376465
635 0.0235471725463867
636 0.0235679149627686
637 0.0235555171966553
638 0.0235905647277832
639 0.0235257148742676
640 0.0235655307769775
641 0.0235695838928223
642 0.0235388278961182
643 0.0258662700653076
644 0.0238208770751953
645 0.0239729881286621
646 0.0238049030303955
647 0.0236871242523193
648 0.023615837097168
649 0.023554801940918
650 0.023521900177002
651 0.0235230922698975
652 0.0235297679901123
653 0.0235996246337891
654 0.0235919952392578
655 0.0235981941223145
656 0.023773193359375
657 0.0238244533538818
658 0.0236976146697998
659 0.0236165523529053
660 0.023601770401001
661 0.0235929489135742
662 0.0235674381256104
663 0.0236363410949707
664 0.0236737728118896
665 0.0236814022064209
666 0.0236327648162842
667 0.023672342300415
668 0.0238308906555176
669 0.0238049030303955
670 0.0237114429473877
671 0.0236725807189941
672 0.0236532688140869
673 0.0235860347747803
674 0.0235967636108398
675 0.0235469341278076
676 0.0235257148742676
677 0.0235905647277832
678 0.0235328674316406
679 0.0235893726348877
680 0.0237793922424316
681 0.023712158203125
682 0.023622989654541
683 0.023613452911377
684 0.0235791206359863
685 0.0235624313354492
686 0.0235683917999268
687 0.0236325263977051
688 0.0235354900360107
689 0.0236251354217529
690 0.023578405380249
691 0.0235357284545898
692 0.023526668548584
693 0.0235695838928223
694 0.0235500335693359
695 0.0235507488250732
696 0.0235514640808105
697 0.0235781669616699
698 0.0235409736633301
699 0.0236053466796875
700 0.0236015319824219
701 0.0235862731933594
702 0.0235872268676758
703 0.0236160755157471
704 0.0235719680786133
705 0.0235669612884521
706 0.0236148834228516
707 0.0236480236053467
708 0.0236210823059082
709 0.0235946178436279
710 0.023557186126709
711 0.0235986709594727
712 0.0235824584960938
713 0.0237243175506592
714 0.0236265659332275
715 0.0236289501190186
716 0.0235865116119385
717 0.0236120223999023
718 0.0235493183135986
719 0.0235662460327148
720 0.0235931873321533
721 0.0235779285430908
722 0.0235934257507324
723 0.0236079692840576
724 0.0235664844512939
725 0.0236186981201172
726 0.0236191749572754
727 0.0236029624938965
728 0.0235805511474609
729 0.0235908031463623
730 0.0236084461212158
731 0.0235507488250732
732 0.0236234664916992
733 0.0235960483551025
734 0.0235862731933594
735 0.0235848426818848
736 0.0235843658447266
737 0.0236058235168457
738 0.0236203670501709
739 0.023587703704834
740 0.0235962867736816
741 0.0235834121704102
742 0.0235939025878906
743 0.02362060546875
744 0.0235793590545654
745 0.0236921310424805
746 0.0236399173736572
747 0.0236351490020752
748 0.0235910415649414
749 0.0235979557037354
750 0.0235974788665771
751 0.0235886573791504
752 0.0236821174621582
753 0.0236194133758545
754 0.0236344337463379
755 0.0236127376556396
756 0.0241703987121582
757 0.0235939025878906
758 0.0236191749572754
759 0.023604154586792
760 0.023592472076416
761 0.0236592292785645
762 0.0237166881561279
763 0.0236880779266357
764 0.0236248970031738
765 0.0236163139343262
766 0.0236008167266846
767 0.0236217975616455
768 0.0236372947692871
769 0.0236215591430664
770 0.0236902236938477
771 0.0237088203430176
772 0.0236868858337402
773 0.0236265659332275
774 0.0235998630523682
775 0.0236063003540039
776 0.023594856262207
777 0.0236084461212158
778 0.0236186981201172
779 0.0236139297485352
780 0.0236186981201172
781 0.0236339569091797
782 0.0236382484436035
783 0.0236737728118896
784 0.0236668586730957
785 0.0236220359802246
786 0.0236308574676514
787 0.0236446857452393
788 0.023637056350708
789 0.0236210823059082
790 0.023679256439209
791 0.0237045288085938
792 0.0237612724304199
793 0.023789644241333
794 0.0236601829528809
795 0.0236861705780029
796 0.0236759185791016
797 0.0236556529998779
798 0.023646354675293
799 0.0236668586730957
800 0.0236308574676514
801 0.0236613750457764
802 0.0236527919769287
803 0.0236525535583496
804 0.0236802101135254
805 0.0236735343933105
806 0.0238151550292969
807 0.02384352684021
808 0.0237228870391846
809 0.0236716270446777
810 0.0236902236938477
811 0.0237207412719727
812 0.0237009525299072
813 0.0236616134643555
814 0.02370285987854
815 0.023669958114624
816 0.0237035751342773
817 0.0236940383911133
818 0.0236690044403076
819 0.02372145652771
820 0.0236029624938965
821 0.0236730575561523
822 0.02362060546875
823 0.0235965251922607
824 0.0236384868621826
825 0.0236506462097168
826 0.0236508846282959
827 0.0236585140228271
828 0.0236570835113525
829 0.0240054130554199
830 0.0238077640533447
831 0.0236940383911133
832 0.0236515998840332
833 0.0236566066741943
834 0.0236971378326416
835 0.0236666202545166
836 0.0237152576446533
837 0.0237271785736084
838 0.0237019062042236
839 0.023643970489502
840 0.023634672164917
841 0.0236978530883789
842 0.0236399173736572
843 0.0236608982086182
844 0.0236496925354004
845 0.0236685276031494
846 0.0236959457397461
847 0.0236966609954834
848 0.0236711502075195
849 0.0236899852752686
850 0.0236258506774902
851 0.0237066745758057
852 0.023667573928833
853 0.0236499309539795
854 0.0236217975616455
855 0.0248920917510986
856 0.0236697196960449
857 0.0236945152282715
858 0.0236601829528809
859 0.0236735343933105
860 0.0236847400665283
861 0.0236554145812988
862 0.0237104892730713
863 0.0236761569976807
864 0.0236809253692627
865 0.0236923694610596
866 0.0236527919769287
867 0.0236494541168213
868 0.0236761569976807
869 0.023714542388916
870 0.0236649513244629
871 0.0236692428588867
872 0.0237052440643311
873 0.0236318111419678
874 0.0236940383911133
875 0.0236871242523193
876 0.0236806869506836
877 0.0236854553222656
878 0.0236263275146484
879 0.0237119197845459
880 0.0237419605255127
881 0.0236907005310059
882 0.0237157344818115
883 0.0237338542938232
884 0.0237016677856445
885 0.0237152576446533
886 0.0236666202545166
887 0.0236968994140625
888 0.0236778259277344
889 0.0237038135528564
890 0.0236649513244629
891 0.0236694812774658
892 0.023667573928833
893 0.023672342300415
894 0.0236697196960449
895 0.0237126350402832
896 0.0236983299255371
897 0.0237629413604736
898 0.0237295627593994
899 0.023726224899292
900 0.0237476825714111
901 0.02374267578125
902 0.023740291595459
903 0.0237414836883545
904 0.0236585140228271
905 0.0237507820129395
906 0.0237743854522705
907 0.0236914157867432
908 0.0237126350402832
909 0.023712158203125
910 0.023759126663208
911 0.0237128734588623
912 0.0237762928009033
913 0.0237293243408203
914 0.0237081050872803
915 0.0237407684326172
916 0.0237274169921875
917 0.0237035751342773
918 0.0236589908599854
919 0.0237336158752441
920 0.0237112045288086
921 0.0237195491790771
922 0.0237195491790771
923 0.0236732959747314
924 0.0237138271331787
925 0.0237264633178711
926 0.0237126350402832
927 0.0237853527069092
928 0.0237128734588623
929 0.0237658023834229
930 0.0237107276916504
931 0.0237946510314941
932 0.0237267017364502
933 0.0236814022064209
934 0.0237278938293457
935 0.0236976146697998
936 0.0236833095550537
937 0.023707389831543
938 0.0236935615539551
939 0.0237340927124023
940 0.0236806869506836
941 0.0237460136413574
942 0.023766040802002
943 0.0237479209899902
944 0.0247430801391602
945 0.0241670608520508
946 0.0237364768981934
947 0.0237209796905518
948 0.0239119529724121
949 0.0239419937133789
950 0.0238194465637207
951 0.0237560272216797
952 0.0237667560577393
953 0.0237643718719482
954 0.023707389831543
955 0.0237407684326172
956 0.0237667560577393
957 0.0237257480621338
958 0.0237178802490234
959 0.0237386226654053
960 0.0237383842468262
961 0.0237667560577393
962 0.0237648487091064
963 0.0237560272216797
964 0.0237298011779785
965 0.0237641334533691
966 0.0237069129943848
967 0.0237827301025391
968 0.0237734317779541
969 0.023806095123291
970 0.0237541198730469
971 0.0237405300140381
972 0.0237398147583008
973 0.0237112045288086
974 0.0236968994140625
975 0.0237584114074707
976 0.0237243175506592
977 0.0237452983856201
978 0.0237259864807129
979 0.0238277912139893
980 0.0238361358642578
981 0.0237689018249512
982 0.0237321853637695
983 0.0237841606140137
984 0.0237596035003662
985 0.0237703323364258
986 0.0237617492675781
987 0.0237889289855957
988 0.0239627361297607
989 0.0240330696105957
990 0.0238697528839111
991 0.0237956047058105
992 0.0237793922424316
993 0.0237700939178467
994 0.0237319469451904
995 0.023773193359375
996 0.0237216949462891
997 0.0237643718719482
998 0.0237908363342285
999 0.0239396095275879
};
\addlegendentry{Without mixing}

\end{axis}

\end{tikzpicture}

%% file: plots/tikzplots/completion_time_imporvements.tex
\begin{tikzpicture}

\definecolor{darkgray176}{RGB}{176,176,176}
\definecolor{steelblue31119180}{RGB}{31,119,180}

\begin{axis}[
tick align=outside,
tick pos=left,
x grid style={darkgray176},
xmin=-0.54, xmax=2.54,
xtick style={color=black},
xtick={0,1,2},
xticklabel style={align=center, font=\Large},
xticklabels={Baseline\\RL,Workload\\Aware RL,Workload\\Guided RL},
y grid style={darkgray176},
ylabel={Improvement over RR (s)},
ymin=0, ymax=20.1409797799587,
ytick style={color=black},
yticklabel style={font=\Large},
ylabel style={font=\LARGE},
]
\draw[draw=none,fill=steelblue31119180] (axis cs:-0.4,0) rectangle (axis cs:0.4,7.53867442607879);
\draw[draw=none,fill=steelblue31119180] (axis cs:0.6,0) rectangle (axis cs:1.4,13.5023439526558);
\draw[draw=none,fill=steelblue31119180] (axis cs:1.6,0) rectangle (axis cs:2.4,19.1818855047226);
\end{axis}

\end{tikzpicture}

%% file: sections/2_Preliminaries.tex
\section{Preliminaries}
\label{sec:preliminaries}

\subsection{LLM Inference} Large Language Models (LLMs) go through prompt/prefill and decode phases while serving a request. The prefill phase processes input tokens in parallel performing self attention computations at each layer and is thus typically compute bound. The decode phase generates subsequent output tokens sequentially based on the forward pass of the last output token and cached context ($\mathbf{K}$ and $\mathbf{V}$ matrices) from previous tokens and is thus memory bound \citep{agrawal2023sarathi}. The response generation ends either when the model produces an EOS token or if the request reaches its maximum token count. A single forward pass of the model is referred to as one iteration of the model \citep{orca}.

\subsection{Scheduler block at LLM instance} 
\begin{figure*}[t]
    \centering  \includegraphics[width=0.7\linewidth]{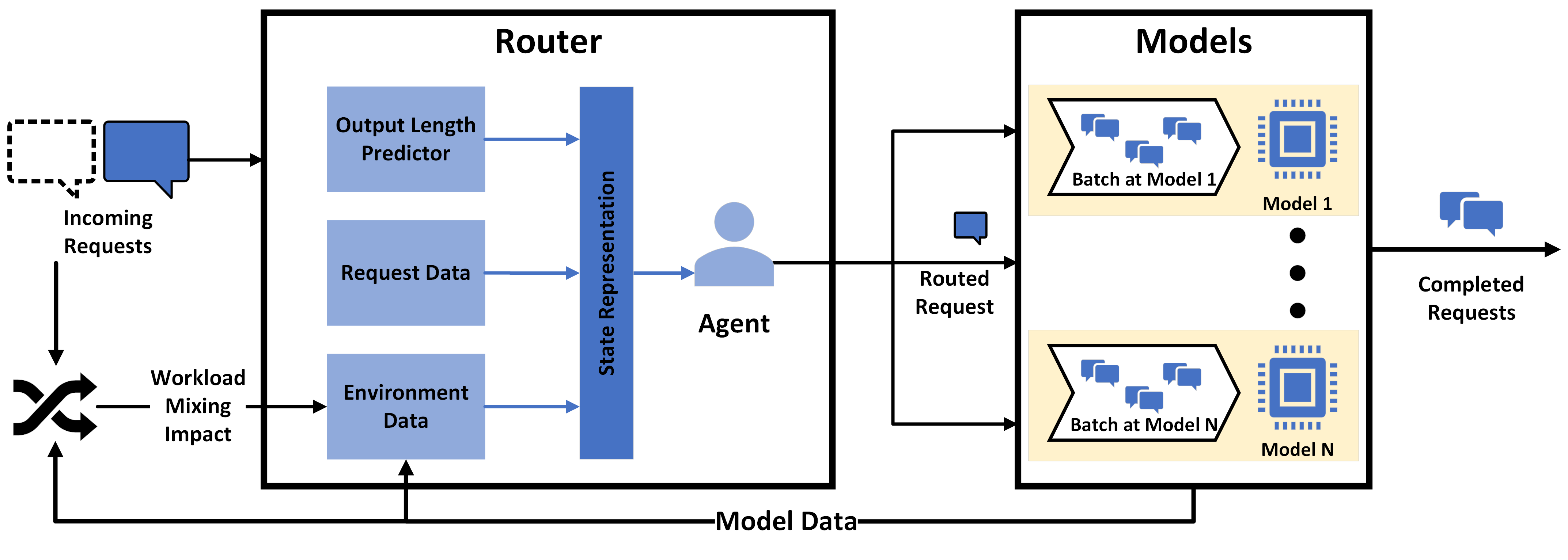}
    \caption{
    Our intelligent router optimizes request routing for end-to-end latency by using the output length predictor and workload impact estimator to route incoming requests to the appropriate model instance based on request characteristics and instance state. In contrast, current approaches focus on instance-level scheduling, as shown by highlighted regions around each model instance. Our router achieves optimal improvements regardless of the LLM instance's optimization strategy.}
    \label{fig:router}
\end{figure*}

Each LLM instance that serves the inference request contains a scheduler, which is a batching system responsible for creating a batch of requests by retrieving requests from a queue and scheduling the execution engine. The scheduler controls how many and which requests are processed in each iteration and may use techniques like iteration-level scheduling introduced in \citet{orca} to reduce queueing delay. The highlighted blocks in Figure \ref{fig:router} show the scheduler at each LLM instance. Often, the First-Come-First-Served policy is used for scheduling requests as online requests are latency-sensitive. 

\subsection{LLM Output Length Prediction} Prior works such as $S^3$ \citep{jin2023s3} also focus on optimizing the throughput of the LLM instance by predicting the output sequence length given an input prompt using a lightweight Distilbert model and batching inputs based on the predicted output length. The prediction is treated as a classification problem by dividing output length into $10$ uniform buckets and training the predictor to pick the correct bucket for an input. This lightweight approach predicts output length with 98.6\% accuracy on a QnA dataset and we build on it in this work. State-of-the-art approaches for predicting output length often rely on expensive models or on relative ranking and probability distributions, as noted by \citep{shahout2024skippredict,fu2024efficient,nie2024aladdin,zheng2024response,shahout2024don}. However, approaches such relative ranking may not effectively capture the performance degradation that occurs when serving requests with different characteristics together

\subsection{Problem Setup} We consider serving a stream of requests using multiple homogeneous LLM instances, each with a scheduler~\citet{orca} to iteratively batch requests using a First-Come-First-Served policy. Requests vary in tasks like summarization, QnA, and translation, each with different prompt and decode characteristics. Requests queue centrally and are routed one at a time to an LLM instance with available capacity. Due to memory constraints, a request may be preempted mid-process if its response exceeds expected size.
Our goal is to minimize end-to-end latency by assigning requests to LLM instances, assuming the request arrival rate maintains system equilibrium and given any optimization strategies present at the model-level scheduler.

%% file: sections/3_RelatedWork.tex
\section{Related Work}
\subsection{LLM Serving Systems}
Recent advancements in inference serving systems for LLMs have focused on optimizing throughput, latency, and resource management. ORCA \citet{orca}, Sarathi \citet{agrawal2023sarathi}, FlashAttention \citet{dao2022flashattention}, and vAttention \citet{prabhu2024vattention} are examples of systems that have achieved significant improvements in performance through techniques such as iteration-level scheduling, innovative batching, and IO-aware algorithms.

\subsection{LLM Serving Algorithms} This space has also seen several algorithmic innovations. QLM (\citet{jhaqlm}) utilizes Bayesian statistics and stochastic programming to manage non-deterministic execution times inherent in LLMs. Similarly, \citet{qiu2024efficient} advocates for speculative shortest-job-first scheduling, and \citet{wu2023fast} employs preemptive scheduling to improve performance.
DistServe and Splitwise (\citet{zhong2024distserve, patel2023splitwise}) optimize LLM serving performance by separating prefill and decoding computation for throughput enhancement while maintaining low latency.  In addressing system load and request patterns, \citet{jha2024learned} and \citet{mendoza2024model} utilize deep reinforcement learning to dynamically adjust service quality, increasing hardware utilization for cost-efficient serving. Additionally, \citet{liu2024andes} optimize Quality-of-Experience (QoE) for LLM serving systems, focusing on user-centric metrics to enhance individual experiences. \citet{jhaqlm, sun2024llumnix}  proposes a multi-model queue management framework for LLM serving and orchestrate the actions such as model swapping, request eviction, GPU-CPU state swapping, load balancing, and warm model start. 
While these works optimize request scheduling at the instance level, they ignore the diversity in the prompt and decode characteristics across requests.

\subsection{Hybrid LLM Inference} Recent works \citep{kag2022efficient,ding2022efficient,ding2024hybrid,ong2024routellm} have introduced a hybrid inference paradigm which uses two \emph{different models} instead of a single model for inference. The key idea is to save inference cost without compromising on response quality by routing easy queries to the smaller and less capable model (e.g. Mixtral \citep{jiang2024mixtral}) while the difficult queries are routed to the larger and more capable model (e.g. GPT-4 \citep{openai2023gpt4}). The routing is typically achieved by training a query-difficulty classifier and is thus different from our reinforcement learning based router which seeks to find the optimal assignment of requests across different instances of the \emph{same model.}

\subsection{Reinforcement Learning for routing jobs} Reinforcement Learning (RL) has been a natural choice for routing jobs in multi-server queues owing to the challenges in deriving exact policies. While previous works \citep{staffolani2023rlq,jali2024efficient} have looked at general jobs, in this work we leverage the specific characteristics of LLM requests and insights from our workload-study to design novel workload aware RL approaches for routing inference requests across LLM instances.

%% file: sections/4_EmpiricalStudy_v1.tex
\begin{figure*}
    \centering
        \begin{subfigure}{0.33\linewidth}
        \resizebox{\linewidth}{!}{%
         \input{plots/tikzplots/prompt_execution_time}
        }
        \caption{Effect of prompt length}\label{fig:prompt_length_execution_time}
    \end{subfigure}%
    \begin{subfigure}{0.33\linewidth}
        \resizebox{\linewidth}{!}{%
         \input{plots/tikzplots/decode_iteration_time}
        }
        \caption{Effect of  decode length}\label{fig:decode_length_execution_time}
    \end{subfigure}%
    \begin{subfigure}{0.33\linewidth}
        \resizebox{\linewidth}{!}{%
        \input{plots/tikzplots/spiking_example_cumsum}
        }
        \caption{Effect of request mixing}\label{fig:mixing_requestsv1}
    \end{subfigure}%
    \caption{\textbf{Effects of prompt and decode tokens on batch execution time.} (a) Execution time of batch with a request in prefill phase grows fast and linearly with the number of prefill tokens. (b) Execution time of batch with only decode tokens is affected to a much lesser degree with the number of tokens. (c) Increase in execution time on mixing requests with the arrival pattern of Figure \ref{fig:mixing_requests}  
    }
    \label{fig:mixing_overview}
\end{figure*}

\section{Observation}

\label{Sec: Observation}
\begin{table*}[t]
\adjustbox{max width=\textwidth}{%
    \centering
    \begin{tabular}{cccccccc}
        \hline
        \multirow{2}{*}{\textbf{Source}} & \multirow{2}{*}{\textbf{Task}} & \multirow{2}{*}{\textbf{Samples}} & \multicolumn{2}{c}{\textbf{Average Tokens}} & \multirow{2}{*}{\textbf{Heavy Decode}} & \multicolumn{2}{c}{\textbf{Accuracy}} \\
        & & & \textbf{Prompt} & \textbf{Decode} & & \textbf{SOTA} & \textbf{Ours} \\
        \hline
        Books~\citet{tiedemann-2012-parallel} & Translation & 7351 & 29.09 & 61.76 & 9.18\% & 4.47\% & 93.10\% \\
        
        Eli5 (Reddit subset)~\citet{eli5_lfqa} & QnA & 6988 & 29.83 & 334.40 & 58.18\% & 5.91\% & 70.36\% \\
        IMDb~\citet{maas-EtAl:2011:ACL-HLT2011} & Sentiment Analysis & 6564 & 211.54 & 142.53 & 41.01\% & 6.81\% & 79.92\% \\
        SQuAD~\citet{rajpurkar-etal-2016-squad} & In-context QnA & 7122 & 125.16 & 220.02 & 47.95\% & 6.22\% & 65.27\% \\
        WNUT~\citet{derczynski-etal-2017-results} & Entity Recognition & 3304 & 26.41 & 64.10 & 8.71\% & 2.76\% & 95.06\% \\
        \hline
        \textbf{Total} & - & 31329 & 89.03 & 175.71 & 35.54\% & 5.5\% & 79.15\% \\
        \hline
    \end{tabular}}
    \caption{\textbf{Dataset and Performance of Output Length Predictor.} Average prompt and decode tokens varies across data sources. The second last column indicates the percentage of requests with heavy decodes ($\geq 5$ seconds estimated completion time) and the last column indicates the accuracy of our decode length predictor model described in \autoref{sec:output_length_predictor} for each source.
    }
    \label{tab:dataset_distrib}
\end{table*}
\subsection{Dataset}
In general, LLM queries come from different tasks and  differ in terms of their prompt and decode distribution. We simulate the prompt and decode distribution using data from five different tasks: sentiment analysis, entity recognition, in-context QnA, general QnA, and translation (prompt details in Appendix~\ref{app:dataset_details}). Table~\ref{tab:dataset_distrib} summarizes the distribution of input (prompt) and generated output (decode) tokens for the requests from these tasks. We see a clear distinction in the average length of prompt and decode tokens, and in the percentage of requests with heavy decode, across tasks. 
\subsection{Prompt-Decode characteristics of requests and their execution time} 
{In this section, we start by  characterizing  the processing time of a request in terms of their prompt and decode token count. } Figures~\ref{fig:prompt_length_execution_time} and ~\ref{fig:decode_length_execution_time} show that batch execution time increases linearly with the number of prompt tokens due to it's compute bound nature, and the growth is much slower during the decode phase. 
Thus, we estimate the processing time for a request with $p$ prompt and $d$ decode tokens as {\color{black}$ p \times (\texttt{time per prompt token}) + d \times (\texttt{average decode batch time})$}. Similarly, the earliest time any model instance will become available is {\color{black}$(\texttt{iterations left})\times(\texttt{average batch time})$}. 
{ It is to be noted that Figures~\ref{fig:prompt_length_execution_time} and ~\ref{fig:decode_length_execution_time} correspond to the profiles for Llama-2-7b models on V100. A similar profiling approach can be followed for the processing time of a different LLM and hardware combination. }

\textbf{Consistency Across Hardware and Model Combinations}: 
For consistency across different LLM and hardware combinations, we classify prompt and decode phases as either heavy or light based on their processing time for that LLM and hardware. Requests that take 0.5 seconds or more to complete their prompt phases are defined as heavy prompts, while requests that take 5 seconds or more in the decode phase are defined as heavy decodes. These values are hyperparameters that can be tuned to the provider's needs. It should be noted that input prompts with 1024 tokens may be heavy for a V100, but they may not be classified as heavy for an H100 due to the latter's better processing capability. We then divide all incoming requests into four categories: light prompt-light decode (LL), light prompt-heavy decode (LH), heavy prompt-light decode (HL), and heavy prompt-heavy decode (HH). In the following section, we quantify the factors affecting the latency and end-to-end performance of LLM inference.

\subsection{Effects of Mixing Requests} 
\textbf{Effect of co-serving requests in the prompt and decode phase on a single LLM instance.} To analyze the effect of processing requests in their prompt and decode phase on a single machine, we served a single request on a LLM instance and added requests at fixed intervals while the first request is still in its decode phase. As shown in Figure \ref{fig:mixing_requests}, the execution time of the first request experienced spikes when new requests were added while the LLM instance was still processing an initial request.
The orange curve in Figure \ref{fig:mixing_requestsv1} shows the ideal latency for the first request, which is 17 seconds for a request with $p=1000$ and $d=1000$. However, the end-to-end latency increases to 31 seconds due to the arrival of new requests of $p=500$ and $d=500$ at every $50^{\mbox{th}}$ iteration. The decode phase of the first request experienced increase in execution time due to the latency spikes caused by the prompt phase of each of the incoming requests \citep{hu2024inference}. 

\textbf{Effect of serving requests with distinct prompt and decode characteristics on a single LLM instance.} Recall that: (i) the latency of the incoming request during the prompt phase increases rapidly and linearly with an increase in the number of prompt tokens, and (ii) the decode phase has a much lower impact, and the mean iteration time varies slowly with an increase in total tokens. As we can see from Figure~\ref{fig:mixing_overview} (Llama-2-7b models profiled on V100 GPUs), the gradient for our configuration can be calculated as $3.2 \times 10^{-4}$ and $3.3 \times 10^{-5}$.  Thus mixing requests with different prompt and decode characteristics in a batch at a model instance can impact overall latency due to mismatch and interference between the prompt and decode phase of the requests. We see a roughly linear increase in batch execution with an increase in token count. The same approach can be followed for other model and hardware combinations.


\begin{table*}
\adjustbox{max width=\textwidth}{%
    \centering
    \begin{tabular}{cccccc}
        \hline
        \multirow{2}{*}{\textbf{Batching Algorithm}} & \multirow{2}{*}{\textbf{Routing Algorithm}} & \multicolumn{4}{c}{\textbf{Total End to End Latency (seconds)}} \\
        \cmidrule(lr){3-6}
        & & \textbf{(LH, HL random)} & \textbf{(Random)} & \textbf{(LH, then HL)} & \textbf{(HL, then LH)} \\
        \hline
        \multirow{3}{*}{Bin Packing~\citet{jin2023s3}} & Dedicated Small-Large & 704.5 & 644.75 & 566.25 & 588.15 \\
        & Round Robin & \textbf{581.5} & 559.3 & 424.8 & \textbf{440.68} \\
        & Decode Balancer & 595.82 & 555.4 & 424.82 & 440.81 \\
        \hline
        \multirow{3}{*}{Least Work Left} & Dedicated Small-Large & 704.5 & 641.81 & 566.25 & 588.15 \\
        & Round Robin & 585.14 & \textbf{554.00} & \textbf{424.64} & 440.82 \\
        & Decode Balancer & 596.95 & 559.97 & 424.66 & 440.81 \\
        \hline
        \multirow{3}{*}{FCFS~\citet{orca}} & Dedicated Small-Large & 704.5 & 648.66 & 566.25 & 588.15 \\
        & Round Robin & 607.45 & 572.16 & 424.80 & 440.82 \\
        & Decode Balancer & 605.65 & 573.17 & 424.82 & 440.81 \\
        \hline
    \end{tabular}}
    \caption{\textbf{Performance of batching and routing algorithm combinations.} We simulate arrival of requests with distinct characteristics  using the request classification discussed in ~\autoref{Sec: Observation} and test the combined affect of routing and batching strategies. 
    Good routing algorithm on an average shows greater end-to-end latency improvement compared to the batching algorithm on all the scenarios with distinct characteristics and arrival sequence. 
    }
    \label{tab:batching_routing_combo}
\end{table*}

\subsection{Factors affecting E2E Latency}
Given that requests are served by multiple LLM instances in reality, we analyze the interplay between routing strategies and model-level scheduling algorithms and how they affect the end-to-end latency of the requests.
We take two LLM instances and consider the following arrival patterns of 3000 requests: 1) LH and HL requests arriving in a random fashion, 2) Requests from all four classes arriving in a random fashion, 3) LH requests arriving first followed by HL requests, and 4) HL requests arriving first followed by LH requests. We study the end-to-end latency in each scenario with different combinations of batching and routing algorithms. Details of these algorithms are added to Appendix \ref{app:heuristic}. We can see from Table \ref{tab:batching_routing_combo} that it is the combination of routing and batching algorithms that affects the overall end-to-end latency of processing the requests. 

For example, the bin-packing scheduler finishes Scenario I six seconds faster when using round-robin routing compared to decode balancer routing but the same combination is four seconds slower when benchmarked on Scenario II. We also observe that sub-optimal routing strategies, such as having dedicated servers for small and large requests, can have a severely adverse effect on the overall performance of our system. This highlights the importance of optimizing the routing strategy and shows that \emph{scheduling algorithms can only provide results as good as the choices provided to it by the router}. Another interesting insight is that, for Scenario III and IV, all the batching algorithms show the exact same end-to-end latency, and it is the routing algorithm that improves the end-to-end latency which hints at the need to adapt scheduling policies based on the characteristics of LLM requests and their arrival sequence.


\subsection{Insights} Quantitative analysis 
highlights the importance of considering request characteristics and avoiding serving requests with diverse characteristics concurrently at a single LLM instance, especially when multiple LLM instances are available. Further, the scheduling algorithms and optimizations at the LLM instance can only provide results as good as the choice provided to it by the router. 
Anchoring on our thesis that a batching algorithm can only provide results as good as the choices provided to it, and that different optimizations can exist in the instance-level scheduler, in the next section, we propose to find optimal routing strategies for a given model-level scheduler, hardware, and model combination. The next section will discuss the overall design of the router and the individual building blocks required.

%% file: plots/tikzplots/prompt_execution_time.tex
\begin{tikzpicture}

\definecolor{darkgray176}{RGB}{176,176,176}
\definecolor{steelblue31119180}{RGB}{31,119,180}

\begin{axis}[
tick align=outside,
tick pos=left,
x grid style={darkgray176},
xlabel={Prompt Length},
xmin=-136, xmax=2856,
xtick style={color=black},
y grid style={darkgray176},
ylabel={Execution Time (s)},
ymin=-0.0129016399383545, ymax=0.841398096084595,
ytick style={color=black},
xlabel style={font=\LARGE},
ylabel style={font=\LARGE, align=center},
xtick={0, 500, 1500, 2500},
xticklabel style={font=\Large},
yticklabel style={font=\Large},
]
\addplot [style={ultra thick}, steelblue31119180]
table {%
0 0.0259301662445068
16 0.0289239883422852
32 0.0416414737701416
48 0.0565421581268311
64 0.046886682510376
80 0.0474047660827637
96 0.0571253299713135
112 0.0588667392730713
128 0.059441089630127
144 0.0596709251403809
160 0.0598535537719727
176 0.0702650547027588
192 0.0714938640594482
208 0.072718620300293
224 0.0744292736053467
240 0.0756652355194092
256 0.0771186351776123
272 0.0941257476806641
288 0.095346212387085
304 0.0963103771209717
320 0.0986580848693848
336 0.0988497734069824
352 0.105891942977905
368 0.107746124267578
384 0.109459638595581
400 0.110300064086914
416 0.111846685409546
432 0.137118816375732
448 0.137981176376343
464 0.137346744537354
480 0.14100170135498
496 0.143091917037964
512 0.158323764801025
528 0.158967256546021
544 0.16115403175354
560 0.162318229675293
576 0.163281679153442
592 0.164844751358032
608 0.175516843795776
624 0.177906513214111
640 0.178899049758911
656 0.180448532104492
672 0.182017803192139
688 0.202052116394043
704 0.20245099067688
720 0.203096389770508
736 0.206321477890015
752 0.207566738128662
768 0.212887525558472
784 0.213969945907593
800 0.21585750579834
816 0.217315196990967
832 0.219423055648804
848 0.220771312713623
864 0.248662948608398
880 0.2490394115448
896 0.251743078231812
912 0.252279996871948
928 0.254705190658569
944 0.268722534179688
960 0.27169942855835
976 0.272392511367798
992 0.274324178695679
1008 0.275379657745361
1024 0.288097620010376
1040 0.290544986724854
1056 0.290775775909424
1072 0.29224967956543
1088 0.295177698135376
1104 0.295514106750488
1120 0.319012403488159
1136 0.319695711135864
1152 0.321484327316284
1168 0.322445869445801
1184 0.323533535003662
1200 0.332821846008301
1216 0.334495544433594
1232 0.335256814956665
1248 0.338485479354858
1264 0.340180635452271
1280 0.366076946258545
1296 0.367207050323486
1312 0.371614933013916
1328 0.374088525772095
1344 0.376471996307373
1360 0.388513326644897
1376 0.393062829971313
1392 0.393398284912109
1408 0.394390821456909
1424 0.396713972091675
1440 0.399199247360229
1456 0.411638975143433
1472 0.41431188583374
1488 0.415409088134766
1504 0.417757272720337
1520 0.419057846069336
1536 0.438719749450684
1552 0.439294338226318
1568 0.441414833068848
1584 0.444800138473511
1600 0.445283651351929
1616 0.447707891464233
1632 0.454000949859619
1648 0.454030752182007
1664 0.458595514297485
1680 0.462724447250366
1696 0.46373438835144
1712 0.490664005279541
1728 0.493332624435425
1744 0.495943069458008
1760 0.497720718383789
1776 0.499418258666992
1792 0.514575719833374
1808 0.51724100112915
1824 0.518572568893433
1840 0.522361755371094
1856 0.525505065917969
1872 0.538833141326904
1888 0.540992259979248
1904 0.541810512542725
1920 0.545202493667603
1936 0.547177314758301
1952 0.549756526947021
1968 0.565767049789429
1984 0.567928552627563
2000 0.56971001625061
2016 0.572404384613037
2032 0.57493257522583
2048 0.582447528839111
2064 0.586772441864014
2080 0.588979959487915
2096 0.591349840164185
2112 0.592999458312988
2128 0.624629735946655
2144 0.62796425819397
2160 0.626991033554077
2176 0.630064249038696
2192 0.636774301528931
2208 0.634358644485474
2224 0.6483473777771
2240 0.6484215259552
2256 0.653899908065796
2272 0.652481555938721
2288 0.658568859100342
2304 0.678625106811523
2320 0.683353662490845
2336 0.684181690216064
2352 0.6873459815979
2368 0.685320377349854
2384 0.701105356216431
2400 0.705122947692871
2416 0.706488847732544
2432 0.709446430206299
2448 0.712699890136719
2464 0.712011098861694
2480 0.722255229949951
2496 0.728476047515869
2512 0.730871438980103
2528 0.731582403182983
2544 0.73551869392395
2560 0.768270969390869
2576 0.772233486175537
2592 0.773378133773804
2608 0.779909610748291
2624 0.782583475112915
2640 0.789651393890381
2656 0.791654348373413
2672 0.795733213424683
2688 0.79619288444519
2704 0.801101684570312
2720 0.802566289901733
};
\end{axis}

\end{tikzpicture}

%% file: plots/tikzplots/decode_iteration_time.tex
\begin{tikzpicture}

\definecolor{darkgray176}{RGB}{176,176,176}
\definecolor{steelblue31119180}{RGB}{31,119,180}

\begin{axis}[
tick align=outside,
tick pos=left,
x grid style={darkgray176},
xlabel={Total Tokens},
xmin=1360.4, xmax=6631.6,
xtick style={color=black},
y grid style={darkgray176},
ylabel={Execution Time (s)},
ymin=0.0332032918930054, ymax=0.0375164270401001,
ytick style={color=black},
xlabel style={font=\LARGE},
ylabel style={font=\LARGE, align=center},
xtick={2000, 4000, 6000},
xticklabel style={font=\Large},
yticklabel style={font=\Large},
]
\addplot [style={ultra thick}, steelblue31119180]
table {%
1600 0.0345757007598877
1608 0.0346066951751709
1616 0.0345878601074219
1624 0.0345983505249023
1632 0.0345587730407715
1640 0.0346050262451172
1648 0.0345609188079834
1656 0.0345604419708252
1664 0.0345211029052734
1672 0.0345680713653564
1680 0.0345842838287354
1688 0.0346939563751221
1696 0.0358595848083496
1704 0.0351295471191406
1712 0.0346081256866455
1720 0.0346443653106689
1728 0.0346453189849854
1736 0.0346505641937256
1744 0.0346279144287109
1752 0.0346074104309082
1760 0.0346143245697021
1768 0.0346827507019043
1776 0.0347011089324951
1784 0.0346372127532959
1792 0.034632682800293
1800 0.0346438884735107
1808 0.0348646640777588
1816 0.0348098278045654
1824 0.0347206592559814
1832 0.0347106456756592
1840 0.0372304916381836
1848 0.0347309112548828
1856 0.034684419631958
1864 0.0347127914428711
1872 0.0346951484680176
1880 0.0347192287445068
1888 0.0348827838897705
1896 0.0348608493804932
1904 0.0348563194274902
1912 0.0348923206329346
1920 0.0348935127258301
1928 0.0348231792449951
1936 0.0347602367401123
1944 0.0353631973266602
1952 0.0349969863891602
1960 0.0347890853881836
1968 0.0348160266876221
1976 0.0350334644317627
1984 0.0351386070251465
1992 0.0348720550537109
2000 0.0348031520843506
2008 0.0348353385925293
2016 0.0347616672515869
2024 0.0348091125488281
2032 0.0347685813903809
2040 0.0347392559051514
2048 0.0347962379455566
2056 0.0347647666931152
2064 0.0348398685455322
2072 0.0349678993225098
2080 0.0348856449127197
2088 0.0348775386810303
2096 0.034825325012207
2104 0.0348577499389648
2112 0.0349526405334473
2120 0.0349502563476562
2128 0.0348665714263916
2136 0.034930944442749
2144 0.0348377227783203
2152 0.0348753929138184
2160 0.034881591796875
2168 0.034848690032959
2176 0.0348689556121826
2184 0.0348362922668457
2192 0.0350081920623779
2200 0.0338644981384277
2208 0.0338723659515381
2216 0.0337021350860596
2224 0.0334420204162598
2232 0.033466100692749
2240 0.0334949493408203
2248 0.0334880352020264
2256 0.0335624217987061
2264 0.0334978103637695
2272 0.0334815979003906
2280 0.0333993434906006
2288 0.0334579944610596
2296 0.0335030555725098
2304 0.0334274768829346
2312 0.0334088802337646
2320 0.0334820747375488
2328 0.0334765911102295
2336 0.0334694385528564
2344 0.0346887111663818
2352 0.0340180397033691
2360 0.0335044860839844
2368 0.0335421562194824
2376 0.0336153507232666
2384 0.0336980819702148
2392 0.0337600708007812
2400 0.0336685180664062
2408 0.033560037612915
2416 0.0335047245025635
2424 0.0335602760314941
2432 0.0335667133331299
2440 0.0334982872009277
2448 0.0335400104522705
2456 0.0335812568664551
2464 0.0337181091308594
2472 0.0336174964904785
2480 0.0335731506347656
2488 0.0337150096893311
2496 0.0336995124816895
2504 0.033564567565918
2512 0.0335705280303955
2520 0.0335371494293213
2528 0.0335590839385986
2536 0.0336024761199951
2544 0.0335345268249512
2552 0.0335893630981445
2560 0.0335655212402344
2568 0.0335586071014404
2576 0.0335876941680908
2584 0.0336267948150635
2592 0.0336449146270752
2600 0.0336263179779053
2608 0.0335867404937744
2616 0.0336370468139648
2624 0.0336120128631592
2632 0.0336332321166992
2640 0.0335807800292969
2648 0.034146785736084
2656 0.0338058471679688
2664 0.0337049961090088
2672 0.033808708190918
2680 0.0337593555450439
2688 0.0337824821472168
2696 0.033726692199707
2704 0.0336935520172119
2712 0.0337569713592529
2720 0.0338320732116699
2728 0.0336902141571045
2736 0.0336878299713135
2744 0.0336430072784424
2752 0.0336465835571289
2760 0.0336999893188477
2768 0.0337154865264893
2776 0.0336506366729736
2784 0.0336766242980957
2792 0.0336532592773438
2800 0.0336744785308838
2808 0.0336778163909912
2816 0.0339376926422119
2824 0.0338466167449951
2832 0.0337975025177002
2840 0.0337686538696289
2848 0.0337071418762207
2856 0.0337502956390381
2864 0.0337808132171631
2872 0.0337827205657959
2880 0.0337271690368652
2888 0.0337982177734375
2896 0.0337395668029785
2904 0.0337316989898682
2912 0.0337076187133789
2920 0.0337450504302979
2928 0.0337553024291992
2936 0.0336589813232422
2944 0.0337276458740234
2952 0.0337278842926025
2960 0.0338010787963867
2968 0.0338747501373291
2976 0.0338699817657471
2984 0.0343060493469238
2992 0.0340938568115234
3000 0.0338644981384277
3008 0.0338499546051025
3016 0.0362815856933594
3024 0.0338563919067383
3032 0.0338289737701416
3040 0.0338199138641357
3048 0.0338218212127686
3056 0.0338168144226074
3064 0.0337996482849121
3072 0.0338053703308105
3080 0.0338201522827148
3088 0.0338563919067383
3096 0.0338828563690186
3104 0.0339105129241943
3112 0.0340301990509033
3120 0.0341014862060547
3128 0.0339486598968506
3136 0.0339040756225586
3144 0.0351336002349854
3152 0.0344130992889404
3160 0.0338644981384277
3168 0.0339615345001221
3176 0.0339081287384033
3184 0.0339040756225586
3192 0.0339105129241943
3200 0.033890962600708
3208 0.0339386463165283
3216 0.0339515209197998
3224 0.0339422225952148
3232 0.0340485572814941
3240 0.0339138507843018
3248 0.0340256690979004
3256 0.0339367389678955
3264 0.0339727401733398
3272 0.0339303016662598
3280 0.0338895320892334
3288 0.0339686870574951
3296 0.0339698791503906
3304 0.0339279174804688
3312 0.034127950668335
3320 0.0340571403503418
3328 0.033951997756958
3336 0.0339255332946777
3344 0.0339689254760742
3352 0.0339810848236084
3360 0.0340700149536133
3368 0.0340063571929932
3376 0.0340218544006348
3384 0.034005880355835
3392 0.0339779853820801
3400 0.034013032913208
3408 0.0339796543121338
3416 0.0340001583099365
3424 0.0339791774749756
3432 0.0339846611022949
3440 0.0340051651000977
3448 0.0344457626342773
3456 0.0345220565795898
3464 0.0342473983764648
3472 0.0342936515808105
3480 0.0343160629272461
3488 0.0352084636688232
3496 0.0341804027557373
3504 0.0347084999084473
3512 0.0343930721282959
3520 0.0341441631317139
3528 0.0342695713043213
3536 0.0342929363250732
3544 0.0341880321502686
3552 0.0341434478759766
3560 0.0341506004333496
3568 0.0341122150421143
3576 0.0341970920562744
3584 0.0341899394989014
3592 0.0341763496398926
3600 0.0342445373535156
3608 0.0343194007873535
3616 0.0341854095458984
3624 0.0342261791229248
3632 0.0341877937316895
3640 0.0342135429382324
3648 0.0341711044311523
3656 0.0341558456420898
3664 0.0341897010803223
3672 0.0341739654541016
3680 0.0341885089874268
3688 0.0341720581054688
3696 0.0343358516693115
3704 0.0342259407043457
3712 0.0342123508453369
3720 0.0341906547546387
3728 0.0341920852661133
3736 0.034266471862793
3744 0.034243106842041
3752 0.0342392921447754
3760 0.0342113971710205
3768 0.0342752933502197
3776 0.0342371463775635
3784 0.0342230796813965
3792 0.0342555046081543
3800 0.0342051982879639
3808 0.0342216491699219
3816 0.0342845916748047
3824 0.0342562198638916
3832 0.0343136787414551
3840 0.0342645645141602
3848 0.0343084335327148
3856 0.0342564582824707
3864 0.034296989440918
3872 0.0342779159545898
3880 0.0343430042266846
3888 0.0342893600463867
3896 0.0348153114318848
3904 0.0345206260681152
3912 0.0343129634857178
3920 0.0343351364135742
3928 0.0342948436737061
3936 0.0343058109283447
3944 0.0343015193939209
3952 0.0342915058135986
3960 0.0343220233917236
3968 0.0342986583709717
3976 0.0343365669250488
3984 0.0343344211578369
3992 0.0343809127807617
4000 0.0344116687774658
4008 0.0343544483184814
4016 0.0344507694244385
4024 0.0343825817108154
4032 0.0343654155731201
4040 0.0343735218048096
4048 0.0343873500823975
4056 0.034433126449585
4064 0.0344388484954834
4072 0.0343797206878662
4080 0.0343689918518066
4088 0.0356500148773193
4096 0.0348613262176514
4104 0.0343565940856934
4112 0.0343785285949707
4120 0.0344369411468506
4128 0.0344550609588623
4136 0.0344221591949463
4144 0.0344829559326172
4152 0.0344157218933105
4160 0.0343942642211914
4168 0.0344016551971436
4176 0.0344464778900146
4184 0.0344524383544922
4192 0.0369253158569336
4200 0.0344662666320801
4208 0.0343887805938721
4216 0.0345830917358398
4224 0.0346279144287109
4232 0.0345239639282227
4240 0.0345258712768555
4248 0.0346128940582275
4256 0.0347068309783936
4264 0.0345785617828369
4272 0.0345346927642822
4280 0.0346915721893311
4288 0.0345621109008789
4296 0.0345432758331299
4304 0.0345149040222168
4312 0.0345044136047363
4320 0.0345025062561035
4328 0.0344963073730469
4336 0.0344350337982178
4344 0.0344910621643066
4352 0.0345485210418701
4360 0.0345399379730225
4368 0.0345611572265625
4376 0.0345938205718994
4384 0.0345602035522461
4392 0.0345029830932617
4400 0.034520149230957
4408 0.0345253944396973
4416 0.0345892906188965
4424 0.0345668792724609
4432 0.0345213413238525
4440 0.0345289707183838
4448 0.0345420837402344
4456 0.0345413684844971
4464 0.0345933437347412
4472 0.0345251560211182
4480 0.0345277786254883
4488 0.0345547199249268
4496 0.0345561504364014
4504 0.0345947742462158
4512 0.0351917743682861
4520 0.0348711013793945
4528 0.0346150398254395
4536 0.0346500873565674
4544 0.0346066951751709
4552 0.0345950126647949
4560 0.0346047878265381
4568 0.0346198081970215
4576 0.0345838069915771
4584 0.0346364974975586
4592 0.0346119403839111
4600 0.0345790386199951
4608 0.034588098526001
4616 0.0347287654876709
4624 0.0347344875335693
4632 0.03475022315979
4640 0.0347375869750977
4648 0.0347030162811279
4656 0.0346860885620117
4664 0.0346801280975342
4672 0.0346333980560303
4680 0.0346992015838623
4688 0.034675121307373
4696 0.0346517562866211
4704 0.0346770286560059
4712 0.0346527099609375
4720 0.0346953868865967
4728 0.0346496105194092
4736 0.0346636772155762
4744 0.0347974300384521
4752 0.0348868370056152
4760 0.03489089012146
4768 0.034876823425293
4776 0.0347955226898193
4784 0.0347225666046143
4792 0.0347335338592529
4800 0.0346858501434326
4808 0.0347318649291992
4816 0.0347139835357666
4824 0.0346724987030029
4832 0.0346870422363281
4840 0.0347073078155518
4848 0.0347015857696533
4856 0.0346734523773193
4864 0.0347466468811035
4872 0.0347158908843994
4880 0.0347232818603516
4888 0.0347578525543213
4896 0.0347363948822021
4904 0.0347585678100586
4912 0.0347557067871094
4920 0.0347363948822021
4928 0.034759521484375
4936 0.0347294807434082
4944 0.0347650051116943
4952 0.0348634719848633
4960 0.0350165367126465
4968 0.0350847244262695
4976 0.0355191230773926
4984 0.0354454517364502
4992 0.0351769924163818
5000 0.0349626541137695
5008 0.035128116607666
5016 0.0351181030273438
5024 0.0350046157836914
5032 0.0348763465881348
5040 0.0348448753356934
5048 0.0347812175750732
5056 0.034794807434082
5064 0.0347921848297119
5072 0.0347480773925781
5080 0.0348906517028809
5088 0.0348184108734131
5096 0.0348565578460693
5104 0.0348339080810547
5112 0.0348069667816162
5120 0.0348057746887207
5128 0.034813404083252
5136 0.0348532199859619
5144 0.0348348617553711
5152 0.0348279476165771
5160 0.0349416732788086
5168 0.0350310802459717
5176 0.0349395275115967
5184 0.034926176071167
5192 0.0347967147827148
5200 0.0362064838409424
5208 0.0353713035583496
5216 0.0348494052886963
5224 0.0348803997039795
5232 0.0348196029663086
5240 0.0348212718963623
5248 0.0348465442657471
5256 0.0348708629608154
5264 0.0348918437957764
5272 0.0349855422973633
5280 0.0350468158721924
5288 0.0349094867706299
5296 0.0349094867706299
5304 0.034883975982666
5312 0.0348527431488037
5320 0.0349009037017822
5328 0.0348529815673828
5336 0.0348923206329346
5344 0.0373203754425049
5352 0.0349082946777344
5360 0.0348660945892334
5368 0.0349504947662354
5376 0.0349109172821045
5384 0.034923791885376
5392 0.0349369049072266
5400 0.0350136756896973
5408 0.0349383354187012
5416 0.0349311828613281
5424 0.0349175930023193
5432 0.0351285934448242
5440 0.0350470542907715
5448 0.0350246429443359
5456 0.0349972248077393
5464 0.0349690914154053
5472 0.0349416732788086
5480 0.0349509716033936
5488 0.0349407196044922
5496 0.0349512100219727
5504 0.0349249839782715
5512 0.034956693649292
5520 0.0349786281585693
5528 0.0350110530853271
5536 0.0350871086120605
5544 0.0350360870361328
5552 0.0350162982940674
5560 0.0349936485290527
5568 0.035027027130127
5576 0.0349991321563721
5584 0.0349805355072021
5592 0.0350341796875
5600 0.0350608825683594
5608 0.0350508689880371
5616 0.0350356101989746
5624 0.0350346565246582
5632 0.0350062847137451
5640 0.0350055694580078
5648 0.0350871086120605
5656 0.0351047515869141
5664 0.0350210666656494
5672 0.0351297855377197
5680 0.0351016521453857
5688 0.0351252555847168
5696 0.0350492000579834
5704 0.0356075763702393
5712 0.0352940559387207
5720 0.0351235866546631
5728 0.0351576805114746
5736 0.0353183746337891
5744 0.0352911949157715
5752 0.035222053527832
5760 0.0352404117584229
5768 0.0352606773376465
5776 0.0353403091430664
5784 0.0354225635528564
5792 0.0353691577911377
5800 0.0352103710174561
5808 0.0351741313934326
5816 0.0351834297180176
5824 0.0351674556732178
5832 0.035111665725708
5840 0.0351567268371582
5848 0.0351240634918213
5856 0.0351154804229736
5864 0.0351133346557617
5872 0.0351243019104004
5880 0.0351512432098389
5888 0.0351366996765137
5896 0.0351352691650391
5904 0.0353074073791504
5912 0.035322904586792
5920 0.0352303981781006
5928 0.0352222919464111
5936 0.0352416038513184
5944 0.0351526737213135
5952 0.0351731777191162
5960 0.0351836681365967
5968 0.0351588726043701
5976 0.0353624820709229
5984 0.0352170467376709
5992 0.035236120223999
6000 0.0351986885070801
6008 0.0351438522338867
6016 0.0351948738098145
6024 0.0351917743682861
6032 0.0351617336273193
6040 0.035297155380249
6048 0.0352969169616699
6056 0.0352704524993896
6064 0.035306453704834
6072 0.035240650177002
6080 0.0352437496185303
6088 0.0352749824523926
6096 0.0352480411529541
6104 0.0352821350097656
6112 0.0352935791015625
6120 0.0353200435638428
6128 0.0352818965911865
6136 0.0352408885955811
6144 0.0352704524993896
6152 0.0352823734283447
6160 0.0353083610534668
6168 0.0353608131408691
6176 0.0353174209594727
6184 0.0353288650512695
6192 0.0353531837463379
6200 0.035388708114624
6208 0.0353491306304932
6216 0.0352849960327148
6224 0.0353274345397949
6232 0.0353105068206787
6240 0.0353274345397949
6248 0.0357789993286133
6256 0.0355706214904785
6264 0.0353238582611084
6272 0.0353617668151855
6280 0.0353410243988037
6288 0.0353105068206787
6296 0.0354149341583252
6304 0.0354604721069336
6312 0.0353837013244629
6320 0.0353662967681885
6328 0.0353844165802002
6336 0.0353989601135254
6344 0.0353953838348389
6352 0.0353751182556152
6360 0.0353832244873047
6368 0.0353488922119141
6376 0.0353853702545166
6384 0.0353755950927734
6392 0.0353870391845703
};
\end{axis}

\end{tikzpicture}

%% file: plots/tikzplots/spiking_example_cumsum.tex
\begin{tikzpicture}

\definecolor{darkgray176}{RGB}{176,176,176}
\definecolor{darkorange25512714}{RGB}{255,127,14}
\definecolor{steelblue31119180}{RGB}{31,119,180}

\begin{axis}[
tick align=outside,
tick pos=left,
x grid style={darkgray176},
xlabel={Iteration of first request},
xlabel style={font=\LARGE},
xmin=-50.45, xmax=1059.45,
xtick style={color=black},
y grid style={darkgray176},
ylabel={Total Time Elapsed (s)},
ylabel style={font=\LARGE},
ymin=-1.19743618965149, ymax=33.0250929355621,
ytick style={color=black},
legend style={nodes={scale=1, transform shape}, at={(0.55, 0.985)}, font=\large},
xtick={0, 250, 500, 750, 1000},
xticklabel style={font=\Large},
yticklabel style={font=\Large},
]
\addplot [style={ultra thick}, steelblue31119180]
table {%
0 0.36432147026062
1 0.38903546333313
2 0.412749290466309
3 0.436093807220459
4 0.459347009658813
5 0.482573270797729
6 0.505788564682007
7 0.528976440429688
8 0.552211761474609
9 0.575435400009155
10 0.598613977432251
11 0.621795177459717
12 0.644926309585571
13 0.668387651443481
14 0.691851854324341
15 0.715174913406372
16 0.738372325897217
17 0.761771678924561
18 0.785238265991211
19 0.808583736419678
20 0.831787586212158
21 0.854949235916138
22 0.878297328948975
23 0.901693344116211
24 0.924891471862793
25 0.948061227798462
26 0.971243858337402
27 0.994405031204224
28 1.01756286621094
29 1.04069995880127
30 1.06381678581238
31 1.08700394630432
32 1.11051797866821
33 1.13370561599731
34 1.15687489509583
35 1.18001055717468
36 1.20318531990051
37 1.22632789611816
38 1.24950051307678
39 1.27267813682556
40 1.29581880569458
41 1.31898331642151
42 1.34218382835388
43 1.36535859107971
44 1.3885338306427
45 1.41169762611389
46 1.43482041358948
47 1.45796513557434
48 1.48114585876465
49 1.65291810035706
50 1.6797034740448
51 1.7060444355011
52 1.73239398002625
53 1.75868988037109
54 1.78486299514771
55 1.81172394752502
56 1.83778500556946
57 1.8643217086792
58 1.89052939414978
59 1.91656804084778
60 1.94256949424744
61 1.96857142448425
62 1.9946072101593
63 2.02061152458191
64 2.04664182662964
65 2.07271456718445
66 2.09887742996216
67 2.12493300437927
68 2.15098977088928
69 2.17745637893677
70 2.20403385162354
71 2.23023247718811
72 2.25633835792542
73 2.28280711174011
74 2.30924820899963
75 2.33550190925598
76 2.36185479164124
77 2.38818740844727
78 2.41438412666321
79 2.44045209884644
80 2.46662282943726
81 2.49279761314392
82 2.51893615722656
83 2.54503202438354
84 2.57113099098206
85 2.59717321395874
86 2.62327814102173
87 2.64931869506836
88 2.67539715766907
89 2.70152950286865
90 2.72763419151306
91 2.75371146202087
92 2.77980327606201
93 2.80593466758728
94 2.83203291893005
95 2.85813117027283
96 2.88426375389099
97 2.91035437583923
98 2.93644213676453
99 3.10865426063538
100 3.1363410949707
101 3.16365194320679
102 3.190913438797
103 3.21821975708008
104 3.24549412727356
105 3.27280497550964
106 3.29978251457214
107 3.32673931121826
108 3.35373115539551
109 3.38070106506348
110 3.40764284133911
111 3.4345874786377
112 3.46149730682373
113 3.48843336105347
114 3.51536226272583
115 3.54232048988342
116 3.56981658935547
117 3.59726238250732
118 3.62442779541016
119 3.65147423744202
120 3.678786277771
121 3.70618939399719
122 3.73332762718201
123 3.76056027412415
124 3.78799223899841
125 3.81531739234924
126 3.84247636795044
127 3.86968350410461
128 3.89685320854187
129 3.92397379875183
130 3.95100784301758
131 3.97810506820679
132 4.00513792037964
133 4.03218579292297
134 4.05922389030457
135 4.08624053001404
136 4.11326265335083
137 4.140291929245
138 4.16731810569763
139 4.19436645507812
140 4.22135663032532
141 4.24841594696045
142 4.27544832229614
143 4.30248141288757
144 4.32949280738831
145 4.35653424263
146 4.38355445861816
147 4.41059994697571
148 4.43770503997803
149 4.61034727096558
150 4.63879561424255
151 4.66701793670654
152 4.69524240493774
153 4.72344207763672
154 4.75164151191711
155 4.77986454963684
156 4.80782318115234
157 4.83571362495422
158 4.86364984512329
159 4.89156866073608
160 4.9194643497467
161 4.9473512172699
162 4.97768759727478
163 5.00560855865479
164 5.03350877761841
165 5.06137752532959
166 5.08931851387024
167 5.11722350120544
168 5.14512276649475
169 5.17311263084412
170 5.20112872123718
171 5.22911834716797
172 5.25712609291077
173 5.2856719493866
174 5.31413149833679
175 5.34240937232971
176 5.37095332145691
177 5.39938998222351
178 5.42752003669739
179 5.45551443099976
180 5.48380064964294
181 5.51212930679321
182 5.54026436805725
183 5.56840801239014
184 5.59655332565308
185 5.62463879585266
186 5.65267181396484
187 5.68075680732727
188 5.70889759063721
189 5.73710680007935
190 5.76524066925049
191 5.79337239265442
192 5.82148504257202
193 5.84959030151367
194 5.87772464752197
195 5.9058518409729
196 5.93397378921509
197 5.96225094795227
198 5.99053192138672
199 6.16259956359863
200 6.19230532646179
201 6.22167444229126
202 6.25100755691528
203 6.28029823303223
204 6.3093695640564
205 6.33841443061829
206 6.36738753318787
207 6.39635491371155
208 6.42531132698059
209 6.4543309211731
210 6.48334431648254
211 6.51224303245544
212 6.54118394851685
213 6.57016587257385
214 6.59909296035767
215 6.62806868553162
216 6.65700793266296
217 6.68602132797241
218 6.71498537063599
219 6.74394583702087
220 6.77299189567566
221 6.80209565162659
222 6.8311402797699
223 6.86034345626831
224 6.88985800743103
225 6.91918540000916
226 6.94832181930542
227 6.97793173789978
228 7.00736284255981
229 7.03651738166809
230 7.06590485572815
231 7.09542608261108
232 7.12467312812805
233 7.15388631820679
234 7.18292808532715
235 7.21196031570435
236 7.24101233482361
237 7.27014780044556
238 7.29928064346313
239 7.32843232154846
240 7.35766887664795
241 7.3867518901825
242 7.41587328910828
243 7.44496512413025
244 7.47413468360901
245 7.50327754020691
246 7.53237581253052
247 7.56156945228577
248 7.59067606925964
249 7.76290535926819
250 7.79316353797913
251 7.8232045173645
252 7.85314154624939
253 7.88309741020203
254 7.91301608085632
255 7.94299674034119
256 7.97301983833313
257 8.00300407409668
258 8.03292679786682
259 8.06282305717468
260 8.09270262718201
261 8.12265634536743
262 8.15260195732117
263 8.18248391151428
264 8.21231842041016
265 8.24211859703064
266 8.2719099521637
267 8.30172681808472
268 8.3316011428833
269 8.36142992973328
270 8.39127707481384
271 8.42110204696655
272 8.45095324516296
273 8.480792760849
274 8.5106565952301
275 8.54050636291504
276 8.57053256034851
277 8.60053992271423
278 8.63050818443298
279 8.6604688167572
280 8.69042253494263
281 8.72033858299255
282 8.75022649765015
283 8.78027534484863
284 8.8103654384613
285 8.84034276008606
286 8.8703134059906
287 8.90029668807983
288 8.9303286075592
289 8.9603328704834
290 8.99050784111023
291 9.02048230171204
292 9.05047082901001
293 9.08041954040527
294 9.11041688919067
295 9.1403284072876
296 9.1703405380249
297 9.20026230812073
298 9.2302143573761
299 9.40262961387634
300 9.43399214744568
301 9.46522498130798
302 9.49608850479126
303 9.52684497833252
304 9.55765438079834
305 9.58844971656799
306 9.61922192573547
307 9.6500358581543
308 9.68079733848572
309 9.71150326728821
310 9.74224901199341
311 9.77296304702759
312 9.80369997024536
313 9.83442401885986
314 9.86514210700989
315 9.89592218399048
316 9.92667031288147
317 9.95740485191345
318 9.98928785324097
319 10.0200748443604
320 10.0508966445923
321 10.082004070282
322 10.1128733158112
323 10.1436564922333
324 10.1744756698608
325 10.2053446769714
326 10.2362184524536
327 10.2663745880127
328 10.296324968338
329 10.3262550830841
330 10.3561894893646
331 10.3861360549927
332 10.416069984436
333 10.4459872245789
334 10.4759314060211
335 10.5058846473694
336 10.5358273983002
337 10.5658226013184
338 10.5961966514587
339 10.626583814621
340 10.6568081378937
341 10.6873710155487
342 10.7177703380585
343 10.7479033470154
344 10.7783391475677
345 10.8086774349213
346 10.8388204574585
347 10.8692057132721
348 10.8995358943939
349 11.0721800327301
350 11.1037168502808
351 11.134991645813
352 11.1659796237946
353 11.1967942714691
354 11.2276298999786
355 11.2585105895996
356 11.2893130779266
357 11.3201625347137
358 11.3509714603424
359 11.381795167923
360 11.4125661849976
361 11.4433269500732
362 11.4741635322571
363 11.5049498081207
364 11.5357885360718
365 11.5666015148163
366 11.5974018573761
367 11.6282188892365
368 11.6590666770935
369 11.6899125576019
370 11.7207255363464
371 11.7516012191772
372 11.7824563980103
373 11.813309431076
374 11.8441643714905
375 11.8750371932983
376 11.9065079689026
377 11.9373087882996
378 11.9681663513184
379 11.9990141391754
380 12.0299088954926
381 12.0607719421387
382 12.0916781425476
383 12.1225597858429
384 12.1534299850464
385 12.1843016147614
386 12.2152206897736
387 12.2461574077606
388 12.2773010730743
389 12.3084745407104
390 12.3394861221313
391 12.3710312843323
392 12.4026322364807
393 12.4339497089386
394 12.4655025005341
395 12.4970779418945
396 12.5283584594727
397 12.5595655441284
398 12.5908441543579
399 12.7640018463135
400 12.7966320514679
401 12.8298063278198
402 12.8621468544006
403 12.8940696716309
404 12.9259300231934
405 12.9579029083252
406 12.989844083786
407 13.0217115879059
408 13.0535659790039
409 13.0853996276855
410 13.1173400878906
411 13.1485252380371
412 13.1793110370636
413 13.210081577301
414 13.2408394813538
415 13.2716462612152
416 13.3024218082428
417 13.3332016468048
418 13.3640165328979
419 13.3948059082031
420 13.4256064891815
421 13.456435918808
422 13.4872698783875
423 13.5181438922882
424 13.5490329265594
425 13.5799255371094
426 13.6107811927795
427 13.6416845321655
428 13.6725673675537
429 13.7034854888916
430 13.7343904972076
431 13.7652566432953
432 13.7961633205414
433 13.8270547389984
434 13.8579580783844
435 13.8888621330261
436 13.9197859764099
437 13.9506525993347
438 13.9817578792572
439 14.0127124786377
440 14.0436570644379
441 14.0745565891266
442 14.1055066585541
443 14.136488199234
444 14.1675179004669
445 14.1987569332123
446 14.2299280166626
447 14.2611639499664
448 14.2925329208374
449 14.4662916660309
450 14.4990241527557
451 14.5312070846558
452 14.563268661499
453 14.5953333377838
454 14.6274838447571
455 14.6597092151642
456 14.6915555000305
457 14.7234065532684
458 14.7552602291107
459 14.7870948314667
460 14.8189251422882
461 14.8507182598114
462 14.882550239563
463 14.914370059967
464 14.9462141990662
465 14.9780967235565
466 15.011137008667
467 15.0430023670197
468 15.0749154090881
469 15.1060020923615
470 15.1369013786316
471 15.1678328514099
472 15.1987578868866
473 15.2296752929688
474 15.2605946063995
475 15.2915031909943
476 15.3224332332611
477 15.353372335434
478 15.3843014240265
479 15.4151844978333
480 15.4461200237274
481 15.4770374298096
482 15.5079891681671
483 15.5396826267242
484 15.5707817077637
485 15.6017220020294
486 15.6326634883881
487 15.6636905670166
488 15.6947121620178
489 15.7257702350616
490 15.756826877594
491 15.7878568172455
492 15.8188631534576
493 15.8498868942261
494 15.8808941841125
495 15.9118845462799
496 15.9430091381073
497 15.9740092754364
498 16.0050387382507
499 16.1781921386719
500 16.2105271816254
501 16.2427871227264
502 16.2750284671783
503 16.3072848320007
504 16.3395068645477
505 16.3714451789856
506 16.4033968448639
507 16.4353494644165
508 16.467310667038
509 16.499294757843
510 16.5311863422394
511 16.5632240772247
512 16.5958547592163
513 16.6282367706299
514 16.6605296134949
515 16.6931128501892
516 16.7254819869995
517 16.7577164173126
518 16.7901639938354
519 16.8225200176239
520 16.8546597957611
521 16.8867111206055
522 16.9187290668488
523 16.9507105350494
524 16.9826903343201
525 17.0146682262421
526 17.0466694831848
527 17.0786242485046
528 17.1105952262878
529 17.1425771713257
530 17.1745672225952
531 17.2065505981445
532 17.2385442256927
533 17.2705075740814
534 17.302547454834
535 17.3346416950226
536 17.3666803836823
537 17.398754119873
538 17.4308207035065
539 17.4628875255585
540 17.4949071407318
541 17.5269575119019
542 17.5590279102325
543 17.591057062149
544 17.6231043338776
545 17.6551678180695
546 17.6872336864471
547 17.7193210124969
548 17.7513749599457
549 17.7834393978119
550 17.8155488967896
551 17.8476490974426
552 17.8797717094421
553 17.9118633270264
554 17.9440257549286
555 17.9761536121368
556 18.0082478523254
557 18.0403747558594
558 18.0727233886719
559 18.1051073074341
560 18.13738322258
561 18.170209646225
562 18.2030856609344
563 18.2356126308441
564 18.2684645652771
565 18.3018913269043
566 18.3346228599548
567 18.3672647476196
568 18.3998355865479
569 18.432186126709
570 18.4644827842712
571 18.4966983795166
572 18.5289587974548
573 18.5611965656281
574 18.5934104919434
575 18.6256198883057
576 18.657808303833
577 18.6899831295013
578 18.7222397327423
579 18.7544596195221
580 18.7866697311401
581 18.8189823627472
582 18.8512217998505
583 18.8834652900696
584 18.9156839847565
585 18.9479205608368
586 18.9801614284515
587 19.012401342392
588 19.0446465015411
589 19.0768988132477
590 19.1091754436493
591 19.1414270401001
592 19.1737186908722
593 19.205991268158
594 19.2382373809814
595 19.2705433368683
596 19.3027670383453
597 19.334972858429
598 19.3672659397125
599 19.3995013237
600 19.4317755699158
601 19.4640653133392
602 19.49636054039
603 19.5286512374878
604 19.5611612796783
605 19.5940475463867
606 19.626838684082
607 19.6595001220703
608 19.6922841072083
609 19.724925994873
610 19.757533788681
611 19.7905390262604
612 19.8233969211578
613 19.8559119701385
614 19.8883943557739
615 19.9208018779755
616 19.9532194137573
617 19.985631942749
618 20.0201344490051
619 20.0525209903717
620 20.0848972797394
621 20.1172761917114
622 20.1496534347534
623 20.1820068359375
624 20.2144079208374
625 20.2467758655548
626 20.2791316509247
627 20.3115804195404
628 20.343948841095
629 20.3763353824615
630 20.4087059497833
631 20.4411704540253
632 20.4735908508301
633 20.5060577392578
634 20.538535118103
635 20.5710263252258
636 20.6034741401672
637 20.6359243392944
638 20.6683914661407
639 20.7009048461914
640 20.7334518432617
641 20.7659776210785
642 20.798449754715
643 20.8309442996979
644 20.8634207248688
645 20.8958652019501
646 20.9283359050751
647 20.9608278274536
648 20.9933531284332
649 21.0260167121887
650 21.058798789978
651 21.0919411182404
652 21.124950170517
653 21.1577651500702
654 21.1909165382385
655 21.2239763736725
656 21.2561037540436
657 21.288313627243
658 21.3203845024109
659 21.3521411418915
660 21.3837926387787
661 21.4154078960419
662 21.4470295906067
663 21.4787266254425
664 21.5104515552521
665 21.5421226024628
666 21.5737903118134
667 21.6054265499115
668 21.6370832920074
669 21.6687240600586
670 21.7003238201141
671 21.7319481372833
672 21.7635514736176
673 21.7951712608337
674 21.8267467021942
675 21.8583822250366
676 21.8900036811829
677 21.9216663837433
678 21.9533002376556
679 21.9849829673767
680 22.0166537761688
681 22.0483245849609
682 22.0808987617493
683 22.1126198768616
684 22.1442847251892
685 22.1759769916534
686 22.2077035903931
687 22.2393972873688
688 22.2711052894592
689 22.3028185367584
690 22.3345274925232
691 22.3662638664246
692 22.3979828357697
693 22.429673910141
694 22.4613695144653
695 22.4931118488312
696 22.5248885154724
697 22.5566840171814
698 22.5884342193604
699 22.6201674938202
700 22.6519408226013
701 22.6843135356903
702 22.7166290283203
703 22.7486214637756
704 22.7810204029083
705 22.8133599758148
706 22.845312833786
707 22.8777949810028
708 22.9101452827454
709 22.9422273635864
710 22.9740929603577
711 23.0059766769409
712 23.0378086566925
713 23.0696589946747
714 23.1015212535858
715 23.133421421051
716 23.1652574539185
717 23.1971325874329
718 23.2290279865265
719 23.2608814239502
720 23.2927670478821
721 23.3246548175812
722 23.3565447330475
723 23.3883562088013
724 23.4202325344086
725 23.4520833492279
726 23.4839158058167
727 23.5157618522644
728 23.5476512908936
729 23.5795962810516
730 23.6114971637726
731 23.6434082984924
732 23.6753215789795
733 23.7072083950043
734 23.7391250133514
735 23.7716891765594
736 23.8038713932037
737 23.835813999176
738 23.8677091598511
739 23.8996198177338
740 23.9315059185028
741 23.9634191989899
742 23.9953677654266
743 24.0272996425629
744 24.0592257976532
745 24.0911946296692
746 24.1231379508972
747 24.1551024913788
748 24.1875743865967
749 24.2201144695282
750 24.2523982524872
751 24.2850623130798
752 24.3176300525665
753 24.3499517440796
754 24.3818831443787
755 24.4132912158966
756 24.4443383216858
757 24.4752526283264
758 24.5061817169189
759 24.5370717048645
760 24.5680148601532
761 24.5989141464233
762 24.6298260688782
763 24.6607382297516
764 24.6916892528534
765 24.7225739955902
766 24.7534563541412
767 24.7843589782715
768 24.8152511119843
769 24.8461360931396
770 24.877021074295
771 24.9079627990723
772 24.9388563632965
773 24.9697771072388
774 25.0006680488586
775 25.0327460765839
776 25.0636975765228
777 25.0946183204651
778 25.1256279945374
779 25.1565525531769
780 25.1874949932098
781 25.2184262275696
782 25.2493503093719
783 25.2803630828857
784 25.3112878799438
785 25.3424015045166
786 25.373379945755
787 25.4043335914612
788 25.4352793693542
789 25.4662494659424
790 25.4972109794617
791 25.5281887054443
792 25.5591449737549
793 25.5902538299561
794 25.6213386058807
795 25.6524815559387
796 25.6841740608215
797 25.7157802581787
798 25.7471485137939
799 25.7789680957794
800 25.810558795929
801 25.8417792320251
802 25.8732891082764
803 25.9040281772614
804 25.9341781139374
805 25.9641950130463
806 25.9942116737366
807 26.0241661071777
808 26.0541939735413
809 26.0842802524567
810 26.1143846511841
811 26.1444449424744
812 26.1744561195374
813 26.2044563293457
814 26.2344851493835
815 26.2644999027252
816 26.2945222854614
817 26.3245539665222
818 26.3545794487
819 26.3845987319946
820 26.4146063327789
821 26.4446296691895
822 26.4746649265289
823 26.5059762001038
824 26.5364856719971
825 26.5665721893311
826 26.5966365337372
827 26.6267154216766
828 26.6567900180817
829 26.6869053840637
830 26.7169828414917
831 26.7470605373383
832 26.7771573066711
833 26.8072364330292
834 26.8372859954834
835 26.8673627376556
836 26.8974936008453
837 26.9269652366638
838 26.955911397934
839 26.9848630428314
840 27.0138506889343
841 27.0428647994995
842 27.071857213974
843 27.1008453369141
844 27.1298642158508
845 27.1588203907013
846 27.1877562999725
847 27.2166967391968
848 27.2457127571106
849 27.2752342224121
850 27.3047647476196
851 27.3340766429901
852 27.3628745079041
853 27.3915550708771
854 27.4199948310852
855 27.44815325737
856 27.4767205715179
857 27.5052547454834
858 27.533566236496
859 27.561785697937
860 27.5899496078491
861 27.6181800365448
862 27.6463398933411
863 27.6744785308838
864 27.7026314735413
865 27.7308101654053
866 27.7589521408081
867 27.7871055603027
868 27.8152420520782
869 27.8433811664581
870 27.8715300559998
871 27.8996131420135
872 27.9277484416962
873 27.9559323787689
874 27.9841253757477
875 28.0122973918915
876 28.0404734611511
877 28.0687055587769
878 28.0968806743622
879 28.1251051425934
880 28.1533002853394
881 28.1814842224121
882 28.2096889019012
883 28.2379024028778
884 28.2661547660828
885 28.2943639755249
886 28.3225994110107
887 28.3507919311523
888 28.3789355754852
889 28.407124042511
890 28.4353039264679
891 28.4634954929352
892 28.4916994571686
893 28.5199201107025
894 28.5481345653534
895 28.5763838291168
896 28.6045398712158
897 28.6327154636383
898 28.6608996391296
899 28.6894052028656
900 28.7180082798004
901 28.7452638149261
902 28.7726631164551
903 28.800107717514
904 28.8272767066956
905 28.8544125556946
906 28.8818774223328
907 28.9091453552246
908 28.9362444877625
909 28.9633810520172
910 28.9905557632446
911 29.0175483226776
912 29.0444695949554
913 29.0713369846344
914 29.0981643199921
915 29.1249704360962
916 29.1518168449402
917 29.1786599159241
918 29.2056276798248
919 29.2329819202423
920 29.2600333690643
921 29.2869231700897
922 29.3137974739075
923 29.3406772613525
924 29.3676187992096
925 29.394508600235
926 29.4213848114014
927 29.4482035636902
928 29.4751238822937
929 29.5020866394043
930 29.5289046764374
931 29.5557589530945
932 29.5825917720795
933 29.6094574928284
934 29.6362822055817
935 29.6631166934967
936 29.6899456977844
937 29.7168130874634
938 29.7436699867249
939 29.7705199718475
940 29.7973532676697
941 29.8242537975311
942 29.8511126041412
943 29.8779499530792
944 29.9047908782959
945 29.9315752983093
946 29.9583780765533
947 29.9852492809296
948 30.0121145248413
949 30.0413541793823
950 30.0655052661896
951 30.0892295837402
952 30.1129777431488
953 30.136647939682
954 30.1603565216064
955 30.1840531826019
956 30.2077887058258
957 30.2315022945404
958 30.2552425861359
959 30.278947353363
960 30.3026509284973
961 30.3263897895813
962 30.350109577179
963 30.3738675117493
964 30.3975319862366
965 30.4212274551392
966 30.4449415206909
967 30.4686737060547
968 30.4923963546753
969 30.5160839557648
970 30.5398042201996
971 30.5635857582092
972 30.5872583389282
973 30.6109776496887
974 30.6347200870514
975 30.6585109233856
976 30.6822946071625
977 30.7060258388519
978 30.7297449111938
979 30.7534301280975
980 30.7777671813965
981 30.8020648956299
982 30.8260674476624
983 30.8499104976654
984 30.8741312026978
985 30.8983111381531
986 30.9221432209015
987 30.9459543228149
988 30.9701597690582
989 30.9943578243256
990 31.0182664394379
991 31.042044878006
992 31.0657887458801
993 31.0895512104034
994 31.1133284568787
995 31.1370720863342
996 31.1608107089996
997 31.1845393180847
998 31.2082695960999
999 31.2319829463959
1000 31.255738735199
1001 31.2794930934906
1002 31.3032171726227
1003 31.3269517421722
1004 31.3506953716278
1005 31.3744506835938
1006 31.398209810257
1007 31.4219560623169
1008 31.4456691741943
1009 31.4695234298706
};
\addlegendentry{With mixing}

\addplot [style={ultra thick}, darkorange25512714]
table {%
0 0.358133316040039
1 0.382001876831055
2 0.40561318397522
3 0.42914867401123
4 0.452669620513916
5 0.476268768310547
6 0.499554395675659
7 0.522764205932617
8 0.545953989028931
9 0.569137811660767
10 0.592327356338501
11 0.615492343902588
12 0.638647079467773
13 0.661825656890869
14 0.684976100921631
15 0.708139181137085
16 0.731337308883667
17 0.75457239151001
18 0.777805089950562
19 0.801465034484863
20 0.825169086456299
21 0.84852147102356
22 0.87172269821167
23 0.895298480987549
24 0.918877363204956
25 0.942162752151489
26 0.965330600738525
27 0.98890495300293
28 1.012535572052
29 1.03598046302795
30 1.05929517745972
31 1.08250093460083
32 1.10566258430481
33 1.1288013458252
34 1.15193963050842
35 1.1750705242157
36 1.19825053215027
37 1.22140383720398
38 1.24454808235168
39 1.26768231391907
40 1.29087853431702
41 1.31404829025269
42 1.33718299865723
43 1.36037445068359
44 1.38350558280945
45 1.40672755241394
46 1.42991518974304
47 1.45306038856506
48 1.47623825073242
49 1.4994432926178
50 1.522620677948
51 1.54583978652954
52 1.5689902305603
53 1.59218668937683
54 1.61537837982178
55 1.63861703872681
56 1.66177678108215
57 1.68495845794678
58 1.70814847946167
59 1.73128819465637
60 1.75448107719421
61 1.77766394615173
62 1.80084228515625
63 1.82402205467224
64 1.84719252586365
65 1.87041473388672
66 1.8935694694519
67 1.91675758361816
68 1.93996071815491
69 1.96312093734741
70 1.98626708984375
71 2.0094530582428
72 2.03264260292053
73 2.05583834648132
74 2.07900929450989
75 2.10218048095703
76 2.12538075447083
77 2.14859247207642
78 2.17177963256836
79 2.19502520561218
80 2.21825265884399
81 2.24143242835999
82 2.26456999778748
83 2.28814792633057
84 2.31182765960693
85 2.33524799346924
86 2.35854387283325
87 2.38193821907043
88 2.40579438209534
89 2.42954540252686
90 2.45296716690063
91 2.47628307342529
92 2.49975538253784
93 2.52327585220337
94 2.54657196998596
95 2.56976747512817
96 2.59296178817749
97 2.61620545387268
98 2.63940358161926
99 2.66260480880737
100 2.68580937385559
101 2.70902323722839
102 2.73224782943726
103 2.75545382499695
104 2.77863883972168
105 2.80185461044312
106 2.82506656646729
107 2.8482654094696
108 2.87145042419434
109 2.89468908309937
110 2.91786766052246
111 2.9410662651062
112 2.96429705619812
113 2.98749613761902
114 3.01076984405518
115 3.03400206565857
116 3.05721592903137
117 3.08046412467957
118 3.1036856174469
119 3.12691068649292
120 3.15013861656189
121 3.17333841323853
122 3.19655299186707
123 3.22010612487793
124 3.24350547790527
125 3.26673698425293
126 3.28994083404541
127 3.3131411075592
128 3.33634781837463
129 3.35961270332336
130 3.38286781311035
131 3.40610432624817
132 3.42931485176086
133 3.4525671005249
134 3.47581648826599
135 3.49900412559509
136 3.52225995063782
137 3.54551315307617
138 3.56995749473572
139 3.59321928024292
140 3.61643218994141
141 3.63971018791199
142 3.66293239593506
143 3.68622493743896
144 3.70965909957886
145 3.73301458358765
146 3.7563362121582
147 3.77962064743042
148 3.80338478088379
149 3.82710957527161
150 3.85056853294373
151 3.87385201454163
152 3.89749455451965
153 3.92111539840698
154 3.94448614120483
155 3.96779084205627
156 3.99108266830444
157 4.01438283920288
158 4.0376455783844
159 4.06090068817139
160 4.08424496650696
161 4.10752081871033
162 4.13081479072571
163 4.15403032302856
164 4.17725729942322
165 4.20053172111511
166 4.22378468513489
167 4.24702000617981
168 4.27026796340942
169 4.29350733757019
170 4.31671357154846
171 4.33998250961304
172 4.36325287818909
173 4.38655424118042
174 4.40978932380676
175 4.43301367759705
176 4.45627808570862
177 4.47953772544861
178 4.5028293132782
179 4.52612352371216
180 4.54941892623901
181 4.57270407676697
182 4.59594774246216
183 4.61920046806335
184 4.64244174957275
185 4.66568231582642
186 4.68890595436096
187 4.7121913433075
188 4.73543548583984
189 4.75871515274048
190 4.7819504737854
191 4.80522513389587
192 4.82846570014954
193 4.85175895690918
194 4.87504148483276
195 4.89833331108093
196 4.92159414291382
197 4.94486308097839
198 4.96814680099487
199 4.99138641357422
200 5.01462173461914
201 5.03783464431763
202 5.0610716342926
203 5.08435416221619
204 5.10760855674744
205 5.13086366653442
206 5.15408825874329
207 5.17731714248657
208 5.20103001594543
209 5.22471141815186
210 5.24810695648193
211 5.27138638496399
212 5.29510998725891
213 5.31884217262268
214 5.34226751327515
215 5.3655846118927
216 5.38926959037781
217 5.41307091712952
218 5.43658423423767
219 5.45996141433716
220 5.48329734802246
221 5.50666284561157
222 5.52993559837341
223 5.55322670936584
224 5.57655024528503
225 5.59985899925232
226 5.62312269210815
227 5.64637398719788
228 5.66963601112366
229 5.69294214248657
230 5.71621656417847
231 5.73952388763428
232 5.76282429695129
233 5.78609609603882
234 5.80941152572632
235 5.83265542984009
236 5.85599064826965
237 5.87928342819214
238 5.90258264541626
239 5.92584276199341
240 5.95009446144104
241 5.97377753257751
242 5.99703121185303
243 6.02031111717224
244 6.04357695579529
245 6.06687617301941
246 6.09012460708618
247 6.11342024803162
248 6.13665723800659
249 6.15998959541321
250 6.18324708938599
251 6.20659160614014
252 6.22986030578613
253 6.25310873985291
254 6.27634501457214
255 6.29962611198425
256 6.32290554046631
257 6.34615564346313
258 6.36946988105774
259 6.39271974563599
260 6.41601800918579
261 6.43928408622742
262 6.46259498596191
263 6.4858705997467
264 6.5091814994812
265 6.53246307373047
266 6.55574893951416
267 6.57905292510986
268 6.60244274139404
269 6.62583065032959
270 6.64911389350891
271 6.67240071296692
272 6.6961886882782
273 6.7199718952179
274 6.74345397949219
275 6.76683235168457
276 6.79049110412598
277 6.81424283981323
278 6.83777070045471
279 6.86128997802734
280 6.88510251045227
281 6.90885829925537
282 6.93230962753296
283 6.95563554763794
284 6.97901582717896
285 7.0023193359375
286 7.02560234069824
287 7.04891729354858
288 7.07228064537048
289 7.09566211700439
290 7.1191611289978
291 7.14261794090271
292 7.16604733467102
293 7.1894223690033
294 7.2127103805542
295 7.23603677749634
296 7.25933003425598
297 7.28269004821777
298 7.30596876144409
299 7.32929158210754
300 7.35260510444641
301 7.37596368789673
302 7.39925146102905
303 7.42259621620178
304 7.44589900970459
305 7.46925473213196
306 7.49255609512329
307 7.51591062545776
308 7.53924798965454
309 7.56259536743164
310 7.5858736038208
311 7.60921311378479
312 7.63252139091492
313 7.65584087371826
314 7.67920804023743
315 7.70253396034241
316 7.72584629058838
317 7.7491774559021
318 7.77247023582458
319 7.79581022262573
320 7.81911373138428
321 7.8424608707428
322 7.86574625968933
323 7.88908076286316
324 7.912428855896
325 7.93578815460205
326 7.95910382270813
327 7.98239254951477
328 8.00570058822632
329 8.02899074554443
330 8.05230784416199
331 8.07561755180359
332 8.09937953948975
333 8.12313580513
334 8.14664077758789
335 8.17004609107971
336 8.19380259513855
337 8.21754932403564
338 8.24099898338318
339 8.26435947418213
340 8.2881076335907
341 8.31204128265381
342 8.3356945514679
343 8.35917544364929
344 8.38257718086243
345 8.40594339370728
346 8.42931628227234
347 8.45263457298279
348 8.47600626945496
349 8.49933576583862
350 8.52266621589661
351 8.5459566116333
352 8.57075381278992
353 8.59409761428833
354 8.617431640625
355 8.64076399803162
356 8.66413044929504
357 8.68762421607971
358 8.71106910705566
359 8.73438739776611
360 8.7577006816864
361 8.78104186058044
362 8.80437636375427
363 8.82774877548218
364 8.85109782218933
365 8.87443828582764
366 8.89773678779602
367 8.92110371589661
368 8.9444408416748
369 8.96781539916992
370 8.99118494987488
371 9.01457762718201
372 9.03795623779297
373 9.061283826828
374 9.08460640907288
375 9.10797333717346
376 9.13130283355713
377 9.15468692779541
378 9.17801547050476
379 9.20132493972778
380 9.22465467453003
381 9.24801278114319
382 9.27135300636292
383 9.2947211265564
384 9.31808805465698
385 9.34146213531494
386 9.36482405662537
387 9.3882462978363
388 9.41164588928223
389 9.4350368976593
390 9.45837473869324
391 9.48168420791626
392 9.50500583648682
393 9.52835416793823
394 9.55170702934265
395 9.57508707046509
396 9.59891986846924
397 9.62278890609741
398 9.64631676673889
399 9.6696891784668
400 9.69354200363159
401 9.71740865707397
402 9.74104142189026
403 9.76449918746948
404 9.78826761245728
405 9.81210994720459
406 9.83582234382629
407 9.85935091972351
408 9.88282346725464
409 9.90623545646667
410 9.92964553833008
411 9.95302248001099
412 9.97639441490173
413 9.99982380867004
414 10.0231950283051
415 10.0465571880341
416 10.0699834823608
417 10.0933649539948
418 10.1167435646057
419 10.1401889324188
420 10.1635627746582
421 10.1869473457336
422 10.2103173732758
423 10.2336800098419
424 10.2570431232452
425 10.2804083824158
426 10.3037900924683
427 10.3271682262421
428 10.350537776947
429 10.373911857605
430 10.3972897529602
431 10.4206268787384
432 10.444043636322
433 10.4674305915833
434 10.4908494949341
435 10.5141754150391
436 10.5375211238861
437 10.5609452724457
438 10.5843544006348
439 10.6077432632446
440 10.6311542987823
441 10.6545264720917
442 10.6778492927551
443 10.7012560367584
444 10.7246141433716
445 10.7479939460754
446 10.7713484764099
447 10.7947733402252
448 10.8181915283203
449 10.841625213623
450 10.8650212287903
451 10.8883953094482
452 10.9118161201477
453 10.9351742267609
454 10.9585587978363
455 10.9819345474243
456 11.0052878856659
457 11.0286419391632
458 11.0520675182343
459 11.0754957199097
460 11.0994050502777
461 11.1232914924622
462 11.1468803882599
463 11.17032289505
464 11.1941714286804
465 11.2179939746857
466 11.2415437698364
467 11.2650127410889
468 11.2887649536133
469 11.3126759529114
470 11.3363351821899
471 11.3598551750183
472 11.383282661438
473 11.4067056179047
474 11.4305818080902
475 11.4542074203491
476 11.4775638580322
477 11.5009412765503
478 11.5243675708771
479 11.5477890968323
480 11.5712175369263
481 11.5946772098541
482 11.6180818080902
483 11.6415028572083
484 11.6649467945099
485 11.6883504390717
486 11.7117357254028
487 11.7351627349854
488 11.7585971355438
489 11.7820751667023
490 11.8055222034454
491 11.8289623260498
492 11.8524055480957
493 11.8758409023285
494 11.8992726802826
495 11.9227106571198
496 11.9460923671722
497 11.9695355892181
498 11.992924451828
499 12.0163660049438
500 12.039844751358
501 12.0632445812225
502 12.0866365432739
503 12.1100578308105
504 12.1334595680237
505 12.1568620204926
506 12.1802959442139
507 12.2037267684937
508 12.2271621227264
509 12.2506172657013
510 12.2739996910095
511 12.2974047660828
512 12.3208553791046
513 12.3443033695221
514 12.3677341938019
515 12.3911831378937
516 12.4146256446838
517 12.4380552768707
518 12.4614381790161
519 12.484885931015
520 12.5086333751678
521 12.5323922634125
522 12.5559368133545
523 12.5795571804047
524 12.603488445282
525 12.627334356308
526 12.6509115695953
527 12.6744022369385
528 12.6983246803284
529 12.7221982479095
530 12.7457466125488
531 12.76921916008
532 12.7929337024689
533 12.8166222572327
534 12.8401968479156
535 12.8638002872467
536 12.8872997760773
537 12.9107892513275
538 12.9342744350433
539 12.9577403068542
540 12.9812064170837
541 13.0046494007111
542 13.0281248092651
543 13.0515594482422
544 13.0750327110291
545 13.0984652042389
546 13.1219005584717
547 13.1453132629395
548 13.1688187122345
549 13.1922972202301
550 13.2157263755798
551 13.2391860485077
552 13.2626786231995
553 13.2861142158508
554 13.3095095157623
555 13.3329708576202
556 13.3564217090607
557 13.379890203476
558 13.4033510684967
559 13.426922082901
560 13.4504244327545
561 13.4739236831665
562 13.4973952770233
563 13.5208914279938
564 13.544383764267
565 13.5678732395172
566 13.5932309627533
567 13.6167132854462
568 13.6401793956757
569 13.6636228561401
570 13.6871018409729
571 13.7105927467346
572 13.7340593338013
573 13.7575354576111
574 13.7810022830963
575 13.8045392036438
576 13.8280763626099
577 13.8515605926514
578 13.8750705718994
579 13.898503780365
580 13.9219739437103
581 13.9454748630524
582 13.9689388275146
583 13.9928507804871
584 14.0167479515076
585 14.040408372879
586 14.0639979839325
587 14.0878968238831
588 14.1119010448456
589 14.1356692314148
590 14.1594431400299
591 14.1831345558167
592 14.2071259021759
593 14.2309908866882
594 14.254593372345
595 14.278110742569
596 14.3016269207001
597 14.325156211853
598 14.348655462265
599 14.3722200393677
600 14.3957488536835
601 14.4192552566528
602 14.4427890777588
603 14.4663026332855
604 14.4898281097412
605 14.5133721828461
606 14.5368680953979
607 14.5603985786438
608 14.5839207172394
609 14.6074321269989
610 14.6309888362885
611 14.654492855072
612 14.6780052185059
613 14.7015233039856
614 14.7250611782074
615 14.7485837936401
616 14.7721018791199
617 14.7956330776215
618 14.8191576004028
619 14.8427066802979
620 14.8661794662476
621 14.8897576332092
622 14.9132828712463
623 14.9367952346802
624 14.9603276252747
625 14.9839103221893
626 15.007422208786
627 15.030992269516
628 15.0544798374176
629 15.0779840946198
630 15.1015200614929
631 15.124995470047
632 15.1485409736633
633 15.1720907688141
634 15.1955859661102
635 15.2191565036774
636 15.2426652908325
637 15.2661817073822
638 15.289737701416
639 15.3132479190826
640 15.3367650508881
641 15.3603487014771
642 15.3838829994202
643 15.4073765277863
644 15.4309117794037
645 15.454451084137
646 15.4779760837555
647 15.5015199184418
648 15.5250618457794
649 15.5486409664154
650 15.5721747875214
651 15.5959401130676
652 15.6196844577789
653 15.6433436870575
654 15.6669421195984
655 15.6909375190735
656 15.7150385379791
657 15.7388455867767
658 15.7625393867493
659 15.7863783836365
660 15.8104228973389
661 15.8343513011932
662 15.8580532073975
663 15.8818259239197
664 15.9055871963501
665 15.9292290210724
666 15.9528033733368
667 15.9763462543488
668 15.9999303817749
669 16.0235049724579
670 16.0470283031464
671 16.0706117153168
672 16.0941529273987
673 16.1176688671112
674 16.14124751091
675 16.1647698879242
676 16.1883113384247
677 16.2118837833405
678 16.2354326248169
679 16.2589955329895
680 16.2825508117676
681 16.3060710430145
682 16.3296275138855
683 16.3532211780548
684 16.3767266273499
685 16.4003026485443
686 16.4239187240601
687 16.4475810527802
688 16.4712550640106
689 16.4949135780334
690 16.5184934139252
691 16.5420551300049
692 16.5656032562256
693 16.5891661643982
694 16.6126978397369
695 16.6362748146057
696 16.659835100174
697 16.6833600997925
698 16.7069008350372
699 16.7304368019104
700 16.7539494037628
701 16.7774930000305
702 16.8010547161102
703 16.8245782852173
704 16.848171710968
705 16.8717455863953
706 16.8956434726715
707 16.9195101261139
708 16.9431781768799
709 16.9668970108032
710 16.9905068874359
711 17.0141270160675
712 17.0377948284149
713 17.0614783763885
714 17.085143327713
715 17.1088597774506
716 17.1325166225433
717 17.1561796665192
718 17.1798501014709
719 17.2039637565613
720 17.2280511856079
721 17.2518272399902
722 17.2754762172699
723 17.2995195388794
724 17.3234879970551
725 17.3472249507904
726 17.3708574771881
727 17.394838809967
728 17.4187707901001
729 17.4424428939819
730 17.4660515785217
731 17.4897062778473
732 17.513307094574
733 17.5369913578033
734 17.5606455802917
735 17.5843176841736
736 17.6080179214478
737 17.6316313743591
738 17.6552529335022
739 17.678863286972
740 17.7024385929108
741 17.7260599136353
742 17.7496259212494
743 17.7732393741608
744 17.7968118190765
745 17.820374250412
746 17.8439748287201
747 17.8675079345703
748 17.8910932540894
749 17.9146616458893
750 17.9381744861603
751 17.9617555141449
752 17.985404253006
753 18.0089757442474
754 18.0325622558594
755 18.0561408996582
756 18.079735994339
757 18.1033699512482
758 18.1270248889923
759 18.1506404876709
760 18.1742310523987
761 18.197868347168
762 18.2214572429657
763 18.2450487613678
764 18.2686414718628
765 18.2922439575195
766 18.315851688385
767 18.3394682407379
768 18.363050699234
769 18.386647939682
770 18.4102053642273
771 18.4337840080261
772 18.4573791027069
773 18.4810011386871
774 18.5046713352203
775 18.5283019542694
776 18.5518853664398
777 18.5754652023315
778 18.6002180576324
779 18.6238281726837
780 18.6474351882935
781 18.6711022853851
782 18.6951887607574
783 18.7192847728729
784 18.7430691719055
785 18.7666976451874
786 18.7907931804657
787 18.8149971961975
788 18.8389399051666
789 18.862731218338
790 18.8867180347443
791 18.9108040332794
792 18.9346876144409
793 18.9584112167358
794 18.9821112155914
795 19.005824804306
796 19.029492855072
797 19.0531120300293
798 19.0767071247101
799 19.1004159450531
800 19.1241176128387
801 19.1477909088135
802 19.1713931560516
803 19.1950290203094
804 19.2186000347137
805 19.2422239780426
806 19.2658312320709
807 19.28946185112
808 19.3131070137024
809 19.3367345333099
810 19.3603701591492
811 19.3840386867523
812 19.4077050685883
813 19.431391954422
814 19.4550261497498
815 19.4787263870239
816 19.5023972988129
817 19.5260813236237
818 19.5497877597809
819 19.5734469890594
820 19.5970461368561
821 19.6206719875336
822 19.645280122757
823 19.6693568229675
824 19.6930193901062
825 19.7166662216187
826 19.7403402328491
827 19.7639815807343
828 19.7876508235931
829 19.8113417625427
830 19.8349883556366
831 19.8586394786835
832 19.8822906017303
833 19.9059307575226
834 19.9295673370361
835 19.9531805515289
836 19.9767985343933
837 20.000426530838
838 20.0240683555603
839 20.0476953983307
840 20.0713419914246
841 20.0950005054474
842 20.1186382770538
843 20.1422262191772
844 20.1658246517181
845 20.1894314289093
846 20.2131145000458
847 20.2368569374084
848 20.2605516910553
849 20.2845251560211
850 20.3085782527924
851 20.3324775695801
852 20.3562321662903
853 20.380026102066
854 20.4042251110077
855 20.4283766746521
856 20.4522061347961
857 20.4759836196899
858 20.5000698566437
859 20.5240864753723
860 20.5478532314301
861 20.571578502655
862 20.5953087806702
863 20.619074344635
864 20.6427583694458
865 20.6664683818817
866 20.6901142597198
867 20.7137634754181
868 20.7374041080475
869 20.7610569000244
870 20.7847156524658
871 20.8083915710449
872 20.8320803642273
873 20.8557243347168
874 20.8794276714325
875 20.9031291007996
876 20.9267978668213
877 20.9505126476288
878 20.9741673469543
879 20.9978563785553
880 21.0215306282043
881 21.0451722145081
882 21.0688629150391
883 21.0925447940826
884 21.1162710189819
885 21.1399624347687
886 21.1636657714844
887 21.1873180866241
888 21.2109913825989
889 21.2346744537354
890 21.258317232132
891 21.2819862365723
892 21.3056645393372
893 21.3293483257294
894 21.3529930114746
895 21.376665353775
896 21.4003396034241
897 21.4240298271179
898 21.4477243423462
899 21.471449136734
900 21.4951255321503
901 21.5188052654266
902 21.542454957962
903 21.5661444664001
904 21.5898399353027
905 21.6135284900665
906 21.6372246742249
907 21.6609108448029
908 21.6845471858978
909 21.7085130214691
910 21.7324769496918
911 21.7563061714172
912 21.7801473140717
913 21.8043487071991
914 21.828446149826
915 21.8523857593536
916 21.8761761188507
917 21.9004564285278
918 21.9245774745941
919 21.9484338760376
920 21.9721748828888
921 21.9959995746613
922 22.019788980484
923 22.0435345172882
924 22.0672724246979
925 22.0910046100616
926 22.1146984100342
927 22.1384179592133
928 22.1622338294983
929 22.1859588623047
930 22.2096741199493
931 22.2333626747131
932 22.2570104598999
933 22.2806816101074
934 22.3044018745422
935 22.3281259536743
936 22.3518257141113
937 22.3760781288147
938 22.4000465869904
939 22.4237534999847
940 22.4474911689758
941 22.4712469577789
942 22.4949758052826
943 22.5186831951141
944 22.5424082279205
945 22.5661194324493
946 22.5898199081421
947 22.6135907173157
948 22.6373102664948
949 22.6610126495361
950 22.6847116947174
951 22.7083895206451
952 22.732090473175
953 22.7558200359344
954 22.7794919013977
955 22.8032598495483
956 22.8269860744476
957 22.8506550788879
958 22.8743674755096
959 22.8980648517609
960 22.9217791557312
961 22.9455463886261
962 22.9692587852478
963 22.9929578304291
964 23.0166800022125
965 23.040376663208
966 23.0641269683838
967 23.0878648757935
968 23.1116840839386
969 23.1354887485504
970 23.1592402458191
971 23.1831867694855
972 23.2073585987091
973 23.2314476966858
974 23.2553372383118
975 23.2792294025421
976 23.3034946918488
977 23.3276720046997
978 23.3515539169312
979 23.3753733634949
980 23.3994498252869
981 23.4235966205597
982 23.4474222660065
983 23.471198797226
984 23.4949896335602
985 23.5187604427338
986 23.5424990653992
987 23.5662317276001
988 23.5899744033813
989 23.6160516738892
990 23.6397578716278
991 23.663519859314
992 23.6872761249542
993 23.7110567092896
994 23.7347738742828
995 23.7585477828979
996 23.7822637557983
997 23.8059797286987
998 23.8297173976898
999 23.853600025177
};
\addlegendentry{Without mixing}

\end{axis}

\end{tikzpicture}

%% file: sections/5_WorkloadInfraawarerouter.tex
\section{Intelligent Router: Design}
\label{sec:router}
\autoref{fig:router} shows the overall design of the intelligent router. Based on the insights from \autoref{Sec: Observation}, an intelligent router should: a) classify requests based on prompt-decode characteristics and be able to estimate decode length, b) estimate the adverse effect of mixing diverse requests at a single model instance on end-to-end latency, c) leverage prior knowledge of these adverse effects for decision making, and d) possess lightweight modules for efficient processing. To achieve this, we develop an output length predictor and workload impact estimator for intelligent routing. Additionally, we propose a reinforcement learning framework to utilize accumulated context and prior knowledge, improving end-to-end latency.

\subsection{Output length predictor}\label{sec:output_length_predictor}
Similar to \citep{jin2023s3}, we generate responses for each request in the dataset discussed in \autoref{Sec: Observation} and categorize each request into a bucket based on the number of output tokens in its  response. We use these buckets as labels and input prompts as inputs to fine-tune a DistillBERT model for predicting the range of output tokens for new requests. However, instead of using buckets of equal size, we define the bucket ranges based on the estimated completion time for the request. Following the heavy-light decode logic described in~\autoref{Sec: Observation}, we predict buckets with ranges $0-0.5$ seconds, $0.5-0.5\times 4$ seconds, $0.5\times 4 - 0.5\times 8$ seconds and so forth. Our experimental set up produces roughly 500 decode tokens per second on average, to bucket bucket sizes of $0-250$, $250-1000$, $1000-4000$ and so on. Our choice of (unequal) bucket sizes helps us better distribute the requests among the buckets. This approach provides our routing strategy with actionable information by aligning the bucket ranges with the expected completion times.

We observe that the approach by~\citet{jin2023s3} does not directly generalize well to our dataset, which comprises requests from distinct prompt/decode distributions. The model achieves an accuracy of only 5.5\% in predicting unequal-sized buckets and 9.3\% in predicting equal-sized buckets of 250 tokens.
Leveraging our insight from \autoref{Sec: Observation} that input and output characteristics depend on the task type, we enhance the model's performance by appending the task type as a hint to the prompt provided to the DistillBERT model. Employing this technique, we achieve an accuracy of 79.15\% in predicting unequally sized buckets and 68.23\% in predicting equal-sized buckets. We can predict task type from the input prompt with an accuracy of 93.79\% (c.f. \autoref{subsec:task_predictability}).
\subsection{Workload impact estimator}
\label{subsec:impact_estimator}
Next, we use the profiling approach from Section 4 to obtain the analytic expression for the processing times of the prompt and decode phase. Let there be $n$ requests within model instance $m$, and $p_j^m$ and $d_j^m$ indicate the number of prompt and decode tokens processed by the $j$-th existing request at model $m$. As the impact on the prompt phase is directly proportional to the number of prompt tokens in the request and the total number of tokens already running in the decode phase, we can model the impact on the prompt phase (time to process $p_i$, $T_{p_i}^m)$ of an incoming request with $p_i$ tokens when added to the model instance with n requests, and corresponding penalty as:
\begin{eqnarray}
    T_{{p_i}}^m &=  \texttt{grad}_1*\left((p_i^2 + \sum_{j=1}^{n} \left(p_j^m+d_j^m\right) \right)\nonumber 
    \end{eqnarray}
    \begin{eqnarray}
\label{eqn:1}
    r_{p_i}^m  &=  
    \begin{cases}
        1 & \text{if } T_{p_i}^m \leq \epsilon\\
        1 - \frac{T_{p_i}^m}{{\epsilon}} & \text{otherwise}
    \end{cases}
\end{eqnarray}
Here, we introduce a penalty if the latency impact exceeds $\epsilon$. Similarly, the impact on existing requests beyond the prompt phase is directly proportional to the total number of requests in the model. We can model the penalty due to the impact of an incoming request with $p_i$ prompt tokens and $d_i$ decode tokens on the decode phase of already existing $n$ requests as:
\begin{eqnarray} \label{eqn:2}
    r_{d}^m &=& -\texttt{grad}_2*\sum_{j=1}^{n} \left(p_j^m+d_j^m\right) +p_i + d_i
\end{eqnarray}
With our selection of $\texttt{grad}_1$ as $3.2\times 10^{-4}$ and $\texttt{grad}_2$ as $3.3\times 10^{-5}$, we expect the values $r_d^m$ and $T_{p_i}^m$ to be in the ranges of $[-1, 1]$ and $[-1, 0]$ respectively when there are no requests waiting at the model instance. We  combine \autoref{eqn:1} and  \autoref{eqn:2} to get the final penalty of mixing requests: $r_{\texttt{mixing}}(s_t, s_{t+1}) = \alpha r_{p_i}^m + (1-\alpha)r_{d}^m$ where $m$ is the action taken for the state transition $s_t\to s_{t+1}$. Here, parameter $\alpha \in (0,1)$ balance  priority over the prompt and decode phases. 

\subsection{RL based router}
\label{subsec:RL_router}
We formulate the problem of routing incoming requests to the $m$ model instances as discrete-time Markov Decision Process (MDP) and propose a reinforcement learning-based solution. The discounted  MDP is denoted as a tuple $\mathcal{M} = (\mathcal{S}, \mathcal{A}, P, r, \gamma)$, where $\mathcal{S}$ is the state space, $\mathcal{A}$ is the action space, $P(s' |s, a)$ is the transition dynamics, $r(s, a)$ is the reward function, and $\gamma  \in  [0, 1)$ is the discount factor. The goal is to find the policy $\pi(s|a)$, a distribution of actions over states, which maximizes the discounted return $G_t = \sum_{k=0}^{\infty} \gamma^k r_{t+k+1}$ in expectation. We assume an arrival rate of $\lambda$ for the requests and the ideal estimated time to complete request $i$ as $\hat{T}_i$. Let $o_{jt}$ denote the total number of output tokens produced until time $t$ by request $j$, and $\hat{d}_{jt}$ denote the estimated decode tokens for the $j$-th request. Therefore, we denote the fraction of request $j$ completed at time $t$ by $f_{jt} \coloneqq \frac{o_{jt}}{\hat{d}_{jt}}$. We assume state transition at every $\Delta t$, where $\Delta t$ is the average time to generate a decode token.

\textbf{State Space}: At time $t$, the state of the system, which comprises $m$ model instances and requests waiting in the queue, can be captured by the following: 
1) The number of requests in the queue at time $t$, denoted by $w_{q_t}$;
2) The exact number of prompt tokens, denoted by $p_t \in \mathbb{R}$, and the estimated bucket for the decode tokens, denoted by $d_t \in \{0, \ldots, n_d\}$, corresponding to the next request in the queue. Here, the estimated bucket varies from zero to $n_d$;
3) Matrices, $\mathbf{P}_t \in \mathbb{R}^{mn_p}$ and $\mathbf{D}_t \in \mathbb{R}^{mn_d}$, capturing the prompt and decode distribution of requests at the model instances. 
We represent the prompt (decode) distribution by $n_p$ ($n_d$) buckets. $(\mathbf{P}_t)_{i,j}$ ($(\mathbf{D}_t)_{i,j}$) denotes the number of requests in prompt (decode) phase at the $i$-th model instance that are present in the $j$-th bucket i.e. have $j$ prompt (decode) tokens. We represent the prompt and decode distribution across the model instances as a matrix, which maintains the finite dimensionality of the state space. 4) The capacity available at the model instances at time t is denoted by $\mathbf{C}_t \in \mathbb{R}^{m}$, as a function $g(\texttt{batch size}, \mathbf{P}_t, \mathbf{D}_t)$; and
5) The estimated completion time for the earliest request in model $j$, denoted by $\hat{T}_{c_t}$. 

\textbf{Action Space}: At any given point in time, the agent must decide whether to schedule incoming request $i$ to any of the $m$ model instances or choose to take no action. Therefore, $a \in \{0, \ldots, m\}.$ Here, index $m$ refers to no action being taken by the router.

\textbf{Reward Design}: 
Based on the insights from \autoref{Sec: Observation}, we include the following elements in the reward formulation: a) a negative penalty for requests in the queue, decreasing as requests are processed, to account for the autoregressive nature of requests, b) a positive reward for each completed request, and c) a workload impact estimator-based penalty, which encodes the adverse effect of routing specific requests to a model instance with existing requests, and prevents requests from being queued at each model instance due to a lack of memory. Note that adding the workload impact estimator-based penalty directly to the reward function might introduce bias. Therefore, we propose to augment the prior knowledge using a heuristic-guided formulation \citep{cheng2021heuristic}, and the reward at time t is given by:
\begin{eqnarray} \label{eqn:reward_function}
  r_t &=& 
 - \sum_{j \in \mathcal{J}}\left(\frac{1}{\hat{T}_j}\left( 1- f_{jt} \right)\right) \nonumber \\ 
  & & + \sum_{j=1}^{m} \sum_{i} r_w \times \mathbf{w}_{{mi}_t} \nonumber \\
  & & -(\gamma - \tilde{\gamma}_k) h (\mathbf{s}_t, \mathbf{s}_{t+1})
\end{eqnarray}
where 
\begin{eqnarray}
 h(s_{t}, s_{t+1}) &= & 
 r_{\texttt{mixing}}(\mathbf{s}_t,\mathbf{s}_{t+1}) \nonumber \\
 & & -\max_{l= 1, \ldots, m}{(r_{\texttt{mixing}}(\mathbf{s}_t, \mathbf{s}_{t+1}^l)}) \label{eqn:heuristic}
\end{eqnarray}
Here, $\mathcal{J}$ includes the set of scheduled and unscheduled requests, and $\mathbf{w}_{{mi}_t} \in \{0,1\}$ indicates whether the $i^{th}$ at the $m^{th}$ model completed at time $t$.
$r_w \in \mathbb{Z}^+$ is the positive reward for completing a request. The function $h: \mathcal{\mathbf{S}} \times \mathcal{\mathbf{S}} \to \mathbb{R}$ represents the difference in penalty due to assigning the incoming request to a model other than the one for which the impact is minimum (a function of equations \autoref{eqn:1} and \autoref{eqn:2}). The term "guidance discount" is given by $\tilde{\gamma}_k = \lambda_k \gamma$, where the subscript $k$ denotes the $k$-th episode. Here, $\lambda_k \in [0,1]$ is the mixing coefficient and settles to zero with an increase in episodes \citep{cheng2021heuristic}. The discount factor in the MDP is set to $\tilde{\gamma}_e$ during training.
The function $h()$ returns zero when the request is assigned to the model with the least workload mixing impact. Intuitively, the formulation introduces horizon-based regularization, and its strength diminishes as the mixing coefficient increases, which modifies the original MDP, $\mathcal{M} = (\mathcal{S}, \mathcal{A}, P, r, \gamma)$, to $\tilde{\mathcal{M}} = (\mathcal{S}, \mathcal{A}, P, \tilde{r}, \tilde{\gamma})$. Over the course of training, the agent interacts with the environment, and the effects of the heuristic in the MDP decrease, and the agent eventually optimizes for the original MDP. Guarantees on the boundedness of the reshaped MDP's value function directly translate from \citep{cheng2021heuristic}. 

%% file: sections/6_Experiments.tex
\begin{figure}[t]
    \centering
    \begin{subfigure}{0.33\linewidth}
        \resizebox{\linewidth}{!}{%
   \input{plots/tikzplots/tbb_time_series}
        }
    \caption{Average TBT of requests served}\label{fig:tbb_time_series}
    \end{subfigure}%
     \begin{subfigure}{0.33\linewidth}
        \resizebox{\linewidth}{!}{%
        \input{plots/tikzplots/tbb_distribution}
        }
        \caption{Average TBT distribution}\label{fig:tbt}
    \end{subfigure}
    \begin{subfigure}{0.33\linewidth}
        \resizebox{\linewidth}{!}{%
       \input{plots/tikzplots/waiting_requests}
        }
        \caption{Queue length at model instance}\label{fig:waiting_queue}
    \end{subfigure}
    \caption{\textbf{Experimental Results.} We simulate the arrival of 2000 requests, each with distinct characteristics, at an arrival rate of 20 requests per second and average results over 20 episodes Round-Robin performs better initially in terms of average TBT, but the value increases over time as more requests with different characteristics accumulate.  Workload Guided RL minimizes the variance in average number of waiting requests and TBT values, with fewer requests in the waiting queue compared to other methods. 
  }
    \label{fig:exp_results}
\end{figure}
\section{Experiments}
\label{sec:experiments}


We conducted an extensive evaluation of the proposed framework to evaluate the following:\\
\begin{enumerate}
\item What is the performance improvement of the intelligent router compared to different heuristics on the dataset presented in Section \ref{Sec: Observation}?
\item How does the performance improvement change with an increase in the number of model instances available for serving inference requests and for different hardware and LLM combinations?
\item How does the intelligent router perform and adapt in the presence of different optimizations available at the model-instance level?
\item How does the performance of an intelligent router vary with different datasets and when some information, such as prompt content, is missing?
\end{enumerate}
\textbf{Evaluation metrics}: For all the experiments, we report the end-to-end latency, Time-To-First-Token (TTFT), which is the time taken for the user to see the initial response, and Time-Between-Tokens (TBT), which is the average token streaming latency \citep{patel2023splitwise}. Additionally, we report the throughput achieved by different approaches. We included three variants of the RL formulation, including the baseline RL formulation with a reward function consisting of only the first and second terms from Equation \ref{eqn:reward_function}, workload-augmented RL which simply adds the penalty from $r_{\texttt{mixing}}$ to the baseline RL (workload knowledge augmented), and workload-guided RL that uses the heuristic-guided formulation \ref{eqn:reward_function}. 

\textbf{Setup}
We route requests between four instances of LLama-2-7b-hf model \citep{touvron2023llama} on a cluster of four V100 GPUs using vLLM \citep{vllm} with its default First-Come-First-Served (FCFS) scheduler for iteration-level scheduling. We assume an average request arrival rate of $\lambda=20/s$, with requests uniformly sampled at random from the dataset in Section \ref{Sec: Observation}. The routing of requests to model instances is asynchronous, and we take actions every 0.02 seconds, which is the minimum decode batch execution time. 

{For baseline RL, we set $\gamma - \tilde{\gamma}_e = 0$ in the reward function from \autoref{eqn:reward_function}. For workload-aware RL, we directly augment the penalty for mixing requests to the reward function. Therefore, we set $\gamma - \tilde{\gamma}_e = 1$. For all experiments, we give equal weight to the impact on the prompt and decode phase. Therefore, $\alpha$ for equations \autoref{eqn:1} and \autoref{eqn:2} is set to 0.5. For workload-guided RL, we use the guidance mechanism from \autoref{eqn:reward_function}. We set $\lambda_k = e^{-\beta_d k}$ (exponential decay over each episode) with $\beta_d=0.5$, and the guided discount factor for training $\gamma$ as $\hat{\gamma}=(1-e^{-\beta_d k})\gamma$.} Additional details on the model training are added to \autoref{sec: appendix}.
\subsection{Performance evaluation of Intelligent router}
Here we compare the performance improvements with respect to various heuristics.

\textbf{End-to-end latency}: 
We evaluate RL based  approaches  over 20 episodes, each comprising 2,000 requests with distinct characteristics. As shown in Figure \ref{fig:overall_improvement}, our methods outperformed Round-Robin in terms of end-to-end latency for servicing all requests. Baseline RL surpassed Round Robin by an average of 7.53 seconds (4.35\%). Incorporating the workload-aware penalty into the reward function enhanced this advantage to 13.50 seconds (7.79\%), and utilizing the penalty as heuristic guidance for the RL agent improved the advantage to 19.18 seconds (11.43\%).  This is intuitive as heuristics should only be employed as a warm start and should be reduced as the agent collects more information about the environment.

Classical heuristics such as Join Shortest Queue, Maximum Capacity Usage, and Min-Min Algorithm \citep{6621389} only marginally outperformed Round Robin by 0.46\%, 2.60\%, and 1.50\%, respectively, in terms of end-to-end latency. These results are intuitive as classical heuristics do not translate well for LLM workloads due to their unique nature. Due to this, we only provide further results in comparison to Round Robin. We provide further details on these algorithms in the appendix \autoref{app:additional_baselines}.
\\
\textbf{Improvements in TTFT and TBT}: 
RL-based approaches outperformed the Round-robin router in terms of average TTFT (Figure \ref{fig:ttft_time_series}), with significant improvement as the number of accumulated requests increased over time.
Baseline RL halved the TTFT for late-arriving requests by finding a better assignment than Round-robin. 
 Workload-aware penalty further enhanced these decisions, but not optimally, as it diluted the urgency to complete requests promptly and introduced constant bias. 
 Workload-guided RL performed the best by selecting more optimal model instances and mitigating spikes in TBT of existing requests (Figure \ref{fig:tbb_time_series}). Although Round-robin performed better initially in terms of average TBT, the value increased over time as more requests with different characteristics accumulated. Workload awareness effectively reduced the number of outliers and the variance of the distribution (Figure \ref{fig:tbt}).

\textbf{Queuing at Router and Model Instance}: 
 \autoref{fig:waiting_queue} illustrates the average length of the waiting queue at the model instances. While Baseline RL exhibited a shorter average waiting time of 0.59 seconds at the router, the requests got preempted at the model instances and accumulated substantial delays. This approach was suboptimal since postponing the routing decision could have resulted in a better model instance getting assigned and resulted in faster processing of the request. In contrast, Workload-aware RL, with an average router wait time of 4.41 seconds, addressed this issue by incorporating a penalty based on the workload. Workload Guided RL further refined this strategy by utilizing the penalty as a guidance mechanism, resulting in an average router wait time of 2.05 seconds and improved overall performance.

\begin{table}
    \adjustbox{width=0.5\textwidth}{%
    \centering
    \begin{tabular}{cccc}
    \hline
         Routing Algorithm & Prefill Chunking & Avg. E2E Latency (s) & Improvement  \\
         \hline
         Round Robin & No & 248.41 & - \\
         Baseline RL & No & 240.58 & 3.15\% \\
         Workload Aware RL & No & 231.66 & 6.74\% \\
         Workload Guided RL & No & 221.80 & 10.71\% \\
         Round Robin & Yes & 247.30 & 0.45\% \\
         Baseline RL & Yes & 240.68 & 3.11\% \\
         Workload Aware RL & Yes & 231.12 & 6.96\% \\
         Workload Guided RL & Yes & 220.93 & 11.06\% \\
         \hline
    \end{tabular}
    }
    \caption{{Intelligent router was able to generalize the approach across different model and hardware combinations, outperforms heuristics, and shows additional improvements even with chunked prefills.}}
    \label{tab:a100_experiments}
\end{table}

\begin{figure}[t]
    \centering
    \begin{subfigure}{0.33\linewidth}
        \resizebox{\linewidth}{!}{%
        \includegraphics[]{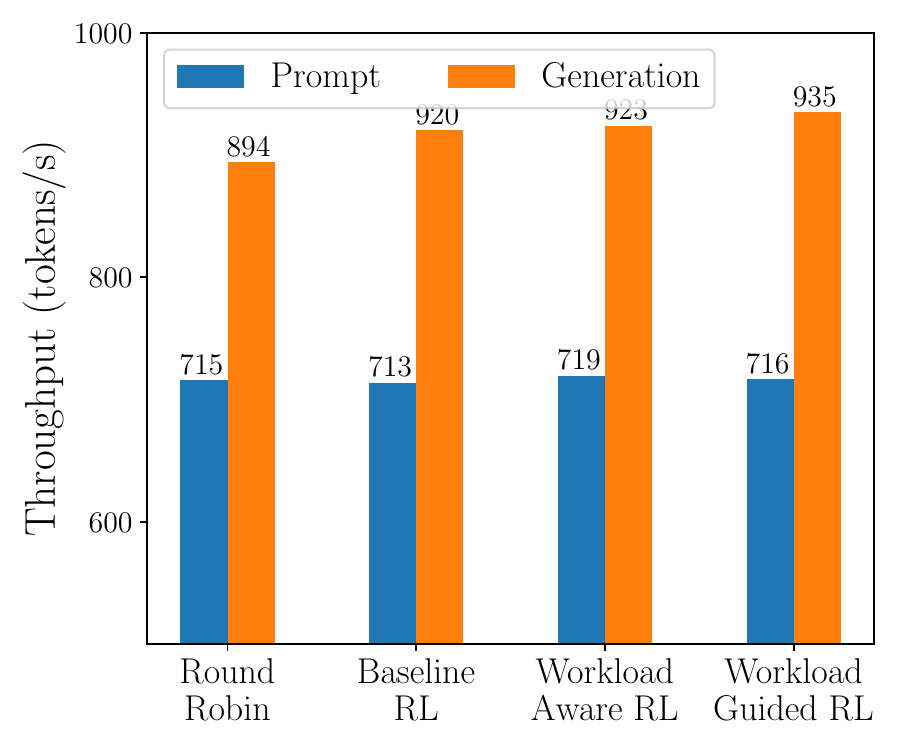}
        }
    \caption{Throughput on A100s with Llama 3.1}\label{fig:throughput}        
    \end{subfigure}
    \begin{subfigure}{0.33\linewidth}
        \resizebox{\linewidth}{!}{%
   \includegraphics[]{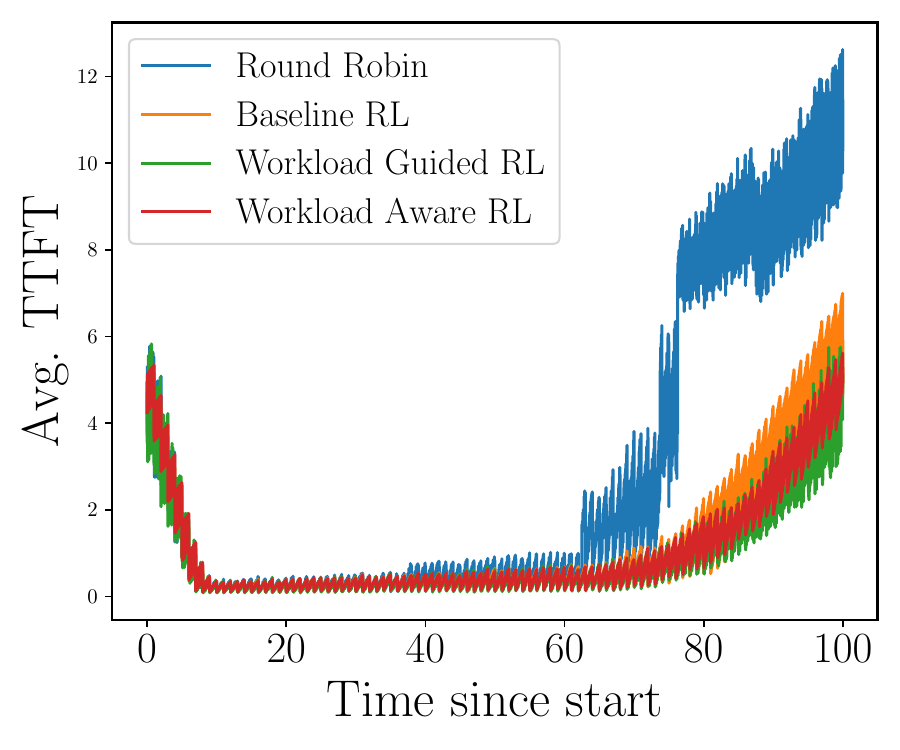}
        }
    \caption{Average TTFT with chunking}\label{fig:ttft_chunked}
    \end{subfigure}%
     \begin{subfigure}{0.33\linewidth}
        \resizebox{\linewidth}{!}{%
   \includegraphics[]{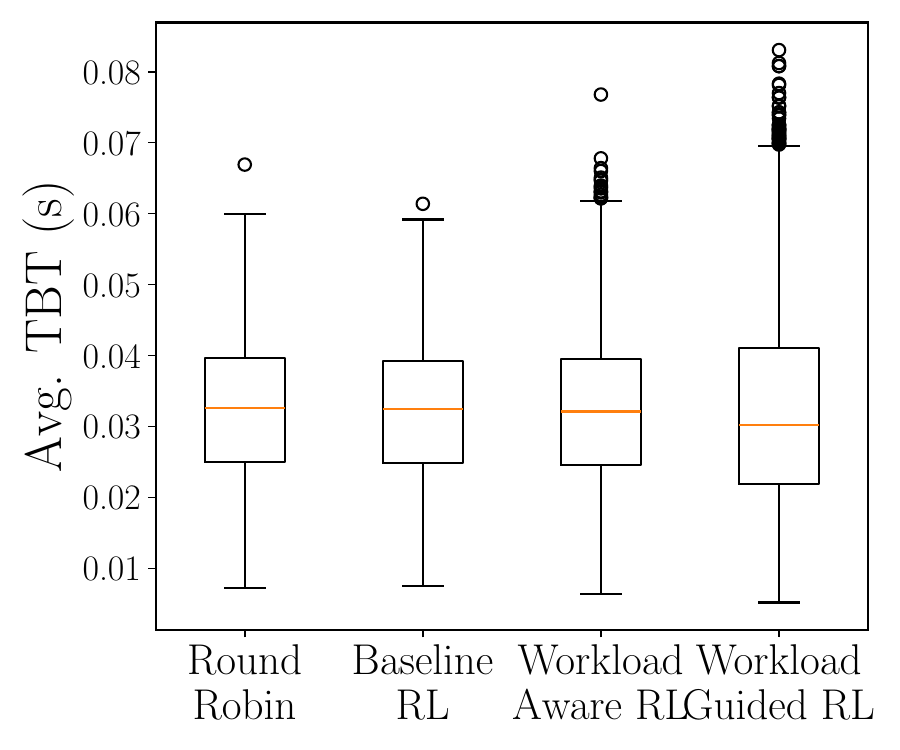}
        }
    \caption{TBT distribution with chunking}\label{fig:TBT_chunked}        
    \end{subfigure}
    \caption{{Model and hardware generalizability: Experiments on A100s with Llama 3.1 8B shows that intelligent router maintains prompt and generation throughput similar to Round Robin. Intelligent router still outperform Round Robin in the presence of optimizations such as chunked prefills.} 
    }
    \label{fig:exp_results_a100}
\end{figure}
\subsection{Different LLM and Hardware combination}
We conducted experiments on different LLM and hardware combinations, specifically testing on A100 with Llama-3.1-8B.
Due to better processing capabilities, we increased the arrival rate to 80 RPS, benchmarked the gradients again for the hardware/model combination, and retrained our agent with the same remaining hyperparameters. 
With more requests coming in, the router had many more decisions to make. 
Even then, our strategies were able to outperform Round Robin by similar margins (10.81\%), as shown by the first four rows of Table \ref{tab:a100_experiments}.
We observe in \autoref{fig:throughput} that our methods maintain prompt and generation throughput similar to Round Robin. Round Robin exhibits similar throughput over a longer period of time, indicating that it generates more tokens to service the same number of requests, highlighting the impact of request preemption.
Additional experiments were conducted to validate the scalability of the proposed approach, and the results are presented in \autoref{exp:scale}. The intelligent router outperformed Round Robin by 11.62\% when evaluated on a setting with eight LLM instances.
\subsection{Performance in the presence of SOTA Optimizations}
Next, we will evaluate the performance of the intelligent router in the presence of chunked prefill tokens \citep{agrawal2023sarathi}. The aim is to assess the performance improvements achieved by the intelligent router in the context of optimizations at the instance-level scheduler.

For Round-Robin, chunked prefill tokens only improves performance by 0.45\%, which could be due to experimental noise. Chunking is not primarily intended to improve E2E latency but rather to enhance user experience by reducing TBT/decode throughput at the expense of TTFT. However, we observe that our method is able to adapt well to this new setting and maintain its lead over Round Robin. \autoref{fig:ttft_chunked} shows that the intelligent router still improves TTFT with chunking, despite the fact that chunking is supposed to harm TTFT. \autoref{fig:TBT_chunked} shows that TBT has much less variance now, and the average TBT across methods is the same. 
The performance gains are intuitive, as the intelligent router prevents preemptions of requests and selects the best suitable LLM instance for each request based on the request characteristics and other requests currently being served by each instance.
Additional experiments that validate performance improvements on a different dataset, which is the real production trace from Cloud provider X, have been added to \autoref{exp:realtrace}.

%% file: plots/tikzplots/tbb_distribution.tex
\begin{tikzpicture}

\definecolor{darkgray176}{RGB}{176,176,176}
\definecolor{darkorange25512714}{RGB}{255,127,14}

\begin{axis}[
tick align=outside,
tick pos=left,
x grid style={darkgray176},
xmin=0.5, xmax=3.5,
xtick style={color=black},
xtick={1,2,3},
xticklabels={Baseline\\RL,Workload\\Aware RL,Workload\\Guided RL},
y grid style={darkgray176},
ylabel={Avg. TBT (s)},
ymin=-0.0239586365422564, ymax=0.529790052971201,
ytick style={color=black},
xticklabel style={align=center,font=\Large},
ylabel style={font=\LARGE},
yticklabel style={font=\Large},
]
\addplot [black]
table {%
0.85 0.0407286669864937
1.15 0.0407286669864937
1.15 0.082800487908282
0.85 0.082800487908282
0.85 0.0407286669864937
};
\addplot [black]
table {%
1 0.0407286669864937
1 0.00135651940391177
};
\addplot [black]
table {%
1 0.082800487908282
1 0.14531890205715
};
\addplot [black]
table {%
0.925 0.00135651940391177
1.075 0.00135651940391177
};
\addplot [black]
table {%
0.925 0.14531890205715
1.075 0.14531890205715
};
\addplot [black, mark=o, mark size=3, mark options={solid,fill opacity=0}, only marks]
table {%
1 0.323725501696269
1 0.153163740190409
1 0.504619657993317
1 0.188177696296147
1 0.148269581794739
1 0.288360646792821
1 0.154800999164581
1 0.169582631852892
1 0.150679119297715
1 0.235846245730365
1 0.171844278063093
1 0.170143153932359
1 0.168394786970956
1 0.195580676198006
1 0.391155600547791
1 0.163979813456535
1 0.174341613596136
1 0.174594865125768
1 0.162343115103049
1 0.201539218425751
1 0.155369566546546
1 0.159807613917759
1 0.435972189903259
1 0.164982914924622
1 0.253758972341364
1 0.184776939451694
1 0.151847626481737
1 0.166599545478821
1 0.146065606389727
1 0.15750223940069
1 0.195332399634428
1 0.258061895003686
1 0.206575877526227
1 0.208897829055786
1 0.344532273032448
1 0.381097123736427
1 0.211893172825084
1 0.211892766111037
1 0.150471605238367
1 0.459498473576137
1 0.249885984829494
1 0.151982998847961
1 0.196107195269677
1 0.212156559739794
1 0.158855593204498
1 0.190044075250626
1 0.14678430557251
1 0.150998288934881
1 0.187520839549877
1 0.14660237232844
1 0.202327023381772
1 0.169549485047658
1 0.159416417280833
};
\addplot [black]
table {%
1.85 0.0354705061792766
2.15 0.0354705061792766
2.15 0.0650177073274922
1.85 0.0650177073274922
1.85 0.0354705061792766
};
\addplot [black]
table {%
2 0.0354705061792766
2 0.00121175843562806
};
\addplot [black]
table {%
2 0.0650177073274922
2 0.109271734952927
};
\addplot [black]
table {%
1.925 0.00121175843562806
2.075 0.00121175843562806
};
\addplot [black]
table {%
1.925 0.109271734952927
2.075 0.109271734952927
};
\addplot [black, mark=o, mark size=3, mark options={solid,fill opacity=0}, only marks]
table {%
2 0.118486881256104
2 0.114593359140249
2 0.11098237991333
2 0.127648135026296
2 0.176854511102041
2 0.127177969884064
2 0.114183268457089
2 0.118288964033127
2 0.278444349765778
2 0.113973736763
2 0.122089564800262
2 0.345491528511047
2 0.163970159159766
2 0.137196122622881
2 0.226937099739357
2 0.112113173191364
2 0.119030032839094
2 0.18798197640313
2 0.109753535718334
2 0.110060040156047
2 0.11268695195516
2 0.119473274548848
2 0.133862235329368
2 0.159919171106248
2 0.11327380993787
2 0.113273543470046
2 0.110282623342105
2 0.123105881644077
2 0.212313890457153
2 0.109413683414459
2 0.145991333893367
2 0.125868928432465
2 0.1197440094418
2 0.124597330888112
2 0.11771940148395
2 0.154508666558699
2 0.146717982632773
};
\addplot [black]
table {%
2.85 0.0330830334960269
3.15 0.0330830334960269
3.15 0.0629121428994346
2.85 0.0629121428994346
2.85 0.0330830334960269
};
\addplot [black]
table {%
3 0.0330830334960269
3 0.00182719308821881
};
\addplot [black]
table {%
3 0.0629121428994346
3 0.107031329037392
};
\addplot [black]
table {%
2.925 0.00182719308821881
3.075 0.00182719308821881
};
\addplot [black]
table {%
2.925 0.107031329037392
3.075 0.107031329037392
};
\addplot [black, mark=o, mark size=3, mark options={solid,fill opacity=0}, only marks]
table {%
3 0.113829793930054
3 0.227748493353526
3 0.118498446577686
3 0.359208017587662
3 0.134245212872823
3 0.141406229564122
3 0.184493456568037
3 0.146934783017194
3 0.109987046037401
3 0.150702693245628
3 0.121790117687649
3 0.234086990356445
3 0.108088771502177
3 0.143993276816148
3 0.115789771080017
3 0.132849920363653
3 0.108836236752962
3 0.125862998120925
};
\addplot [darkorange25512714]
table {%
0.85 0.0602912844357565
1.15 0.0602912844357565
};
\addplot [darkorange25512714]
table {%
1.85 0.0493966890621267
2.15 0.0493966890621267
};
\addplot [darkorange25512714]
table {%
2.85 0.0475890879941763
3.15 0.0475890879941763
};
\end{axis}

\end{tikzpicture}

%% file: plots/tikzplots/waiting_requests.tex
\begin{tikzpicture}

\definecolor{darkgray176}{RGB}{176,176,176}
\definecolor{steelblue31119180}{RGB}{31,119,180}

\begin{axis}[
tick align=outside,
tick pos=left,
x grid style={darkgray176},
xmin=-0.54, xmax=2.54,
xtick style={color=black},
xtick={0,1,2},
xticklabels={Baseline\\RL,Workload\\Aware RL,Workload\\Guided RL},
y grid style={darkgray176},
ylabel={Avg. waiting requests},
ymin=0, ymax=1.2075,
ytick style={color=black},
xticklabel style={align=center, font=\Large},
ylabel style={font=\LARGE},
yticklabel style={font=\Large},
]
\draw[draw=none,fill=steelblue31119180] (axis cs:-0.4,0) rectangle (axis cs:0.4,1.15);
\draw[draw=none,fill=steelblue31119180] (axis cs:0.6,0) rectangle (axis cs:1.4,0.3);
\draw[draw=none,fill=steelblue31119180] (axis cs:1.6,0) rectangle (axis cs:2.4,0);
\end{axis}

\end{tikzpicture}

%% file: sections/7_Conclusion.tex
\section{Limitations and Conclusion}\label{sec:conclusions}

We propose a heuristic-guided Reinforcement Learning (RL) based router for efficiently scheduling requests across homogeneous LLM instances. Our approach introduces and models the novel notion of performance impact resulting from serving workloads with distinct characteristics concurrently. By incorporating prior knowledge on mixing workloads with distinct characteristics and their related adverse effects into the router, we are able to improve overall end-to-end latency over current approaches. Our formulation is sufficiently generalized to improve request-level metrics such as Time-To-First-Token (TTFT) and Time-Between-Two-Tokens (TBT), and can be extended to include additional requirements such as serving throughput.

Our extensive experimental evaluations demonstrate the superior performance of our proposed approach over other baselines, different model-hardware combinations, and with respect to different datasets. Although the RL based approach imposes additional overhead on the LLM inference serving compared to heuristics, our framework enables the identification of optimal load balancing strategies for a given model-hardware combination and for a given LLM instance level scheduler. As such, we hope that it will serve as a standard for future benchmarking of inference schedulers for the community.

\newpage

%% file: sections/Appendix.tex
\clearpage

\appendix

\section{Appendix / supplemental material}
\label{sec: appendix}
\subsection{Batching and Routing Algorithms}\label{app:heuristic}
All the batching algorithms are non-preemptive in nature, meaning that once the processing of a request has started, it is prioritized over requests which have not. Next, we discuss different batching and routing algorithms defined in \autoref{Sec: Observation}.

\subsubsection{Batching: Bin Packing Algorithm}
When a new request's processing can be started, we select the largest request that can fit into the memory available. Ties are broken by FCFS.

\subsubsection{Batching: Least Work Left}
Among the requests available, we select the request with the smallest number of decode tokens.

\subsubsection{Batching: FCFS}
The request which arrives first is processed first.

\subsubsection{Routing: Dedicated Small-Large}
For the two LLM model instances, we dedicate one instance for servicing only the heavy-decode requests while the other model instance services only the light-decode requests.

\subsubsection{Routing: Round Robin}
Each of the two model is user alternatively by the router to send requests to.

\subsubsection{Routing: Decode Balancer}
We assume that the total number of output tokens is known beforehand for the request and we balance the sum of decode tokens on both the model instances.

\subsection{Additional Baselines}\label{app:additional_baselines}
We implemented three baselines other than Round Robin and the Light-weight Heuristic:
\subsubsection{Join Shortest Queue}
Each arriving request is routed to the model with the least number of prompt and decode tokens yet to be processed.
\subsubsection{Maximum Capacity Usage}
Request at the front of the queue is routed to the model with the maximum capacity available, given that it can process this particular request, at intervals of one second.
\subsubsection{Min-Min Algorithm}
We implemented the classical Min-min algorithm, using the number of prompt tokens and the upper bund of the predicted decode token buckets to calculate the time for finishing each job. Since we have homogenous model instances, this strategy becomes similar to shortest job first.

\subsection{Overhead of the Router}
For each decision, the router has to perform two additional steps in our approach: (i) inference from DistillBERT for output length bucket and (ii) inference from the neural network being used. The approach can parallelize these modules when the number of requests in the queue is large (and the request being routed has already been processed by the length predictor). (i) takes us 0.01 (on GPU) and 0.8 (on CPU) seconds per batch of size 64 and (ii) takes $<10^6$ operations (within miliseconds) to process.

\subsection{Details of Dataset}\label{app:dataset_details}
From each dataset, we take the subset of prompts that have a maximum prompt length of 1000 tokens.
\subsubsection{Prompts}
For each task, the prompt is created in the following manner:
\paragraph{Sentiment Analysis (IMDb dataset)} For each review in the dataset, we randomly select one of the following sentences and add it to the review:
\begin{enumerate}
    \item "Based on this review, judge if the user liked this movie or not?"
    \item "Please identify if the review is positive or negative?"
    \item "Based on this review, should we recommend this movie to other users with similar tastes?"
\end{enumerate}
We add these tasks either at the beginning or at the end of the prompts, again randomly.

\paragraph{QnA (Eli5 Reddit subset)} We pick the question in the title as it is and provide it as the prompt to the LLM.

\paragraph{Entity Recognition (WNUT dataset)} We add the suffix "Can you identify the <entity> mentioned in the above sentence?" where <entity> is selected ranomdly from "person", "place", and "object".

\paragraph{In context QnA (SQuAD dataset)} We add the question as well as the four options of the answer to the prompt and ask the LLM to select the correct option and provide reasoning with it as well. 

\paragraph{Translation (Books dataset)} We provide the text and add the phrase "Please translate this text into <language>" either at the start or at the end of the text. <language> is selected from the ones provided in the dataset itself.
\subsubsection{Task Hints}
For providing a hint of the task to the model, we add the phrase "This is a <task> task" at the end of each prompt before providing it to the classifier.

\subsection{Prompt-Decode Distribution}
\begin{figure}[t]
    \centering
    \includegraphics[width=0.5\linewidth]{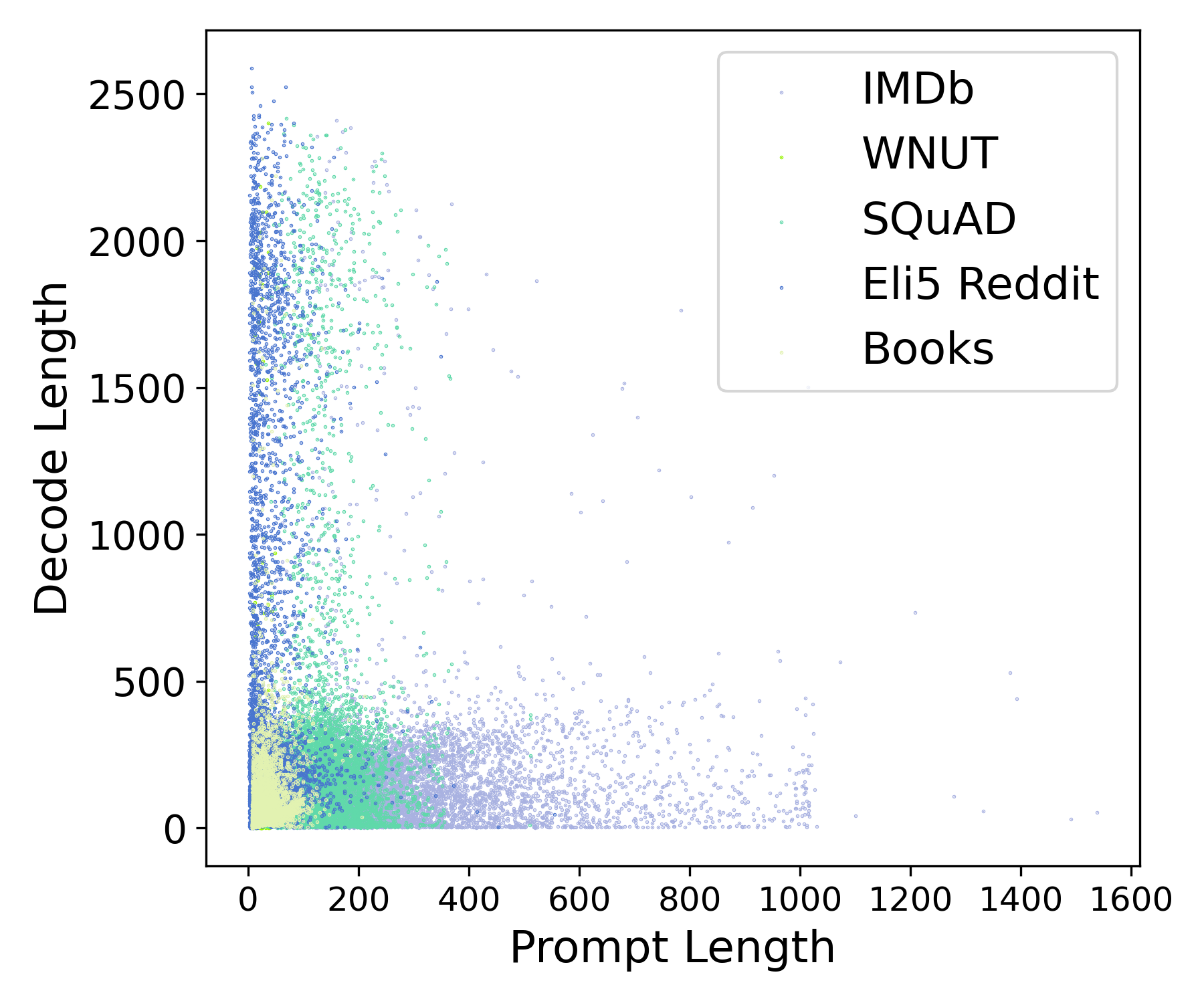}
    \caption{Prompt decode distribution for our dataset with responses generated from Llama 2 7B model.}
    \label{fig:prompt_decode_distribution}
\end{figure}

\begin{figure}[t]
    \centering
    \begin{subfigure}{0.5\linewidth}
        \resizebox{\linewidth}{!}{%
    \includegraphics[]{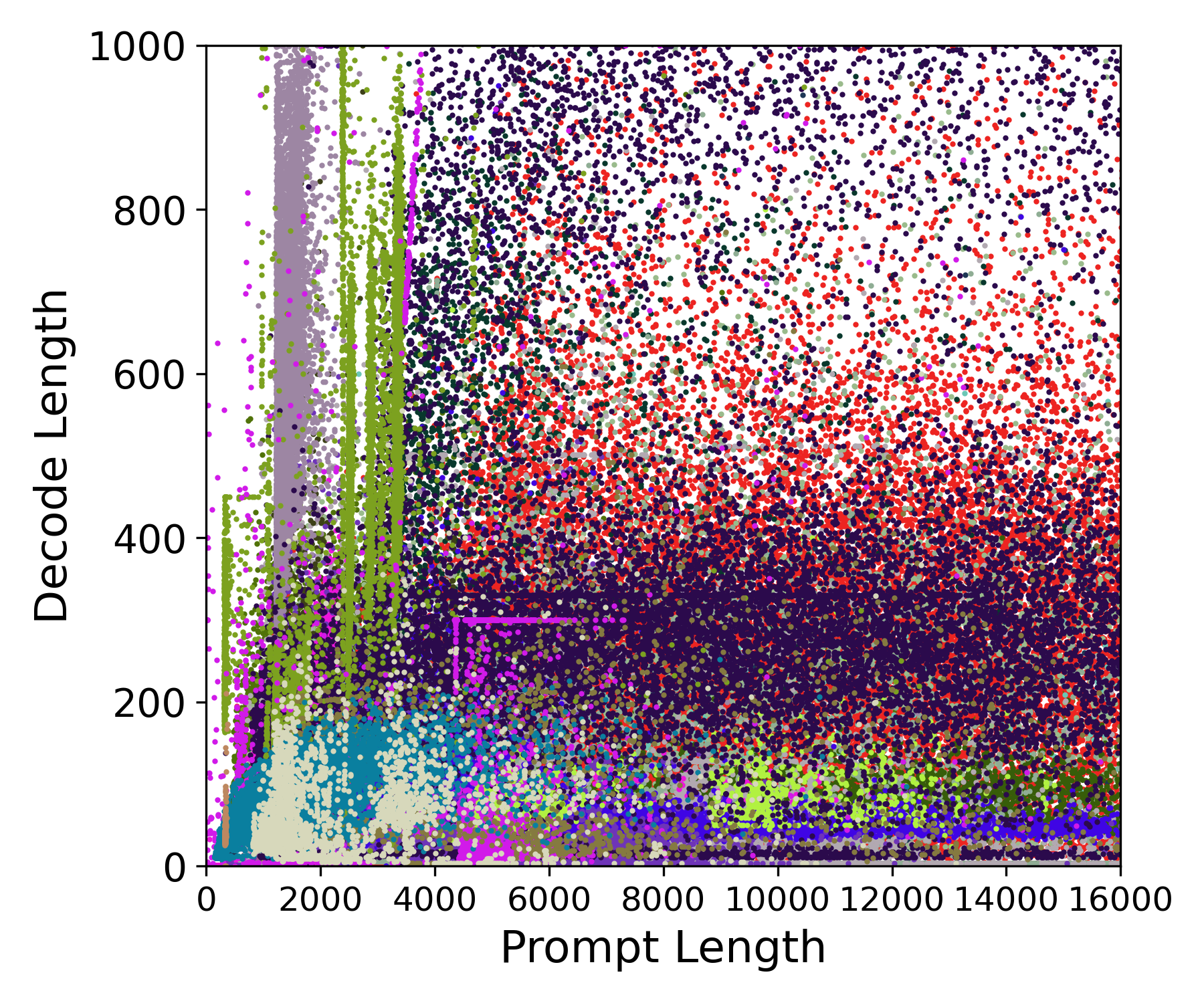}
        }
    \caption{Distribution of entire trace}\label{fig:entire_trace_distribution}
    \end{subfigure}%
    
    \begin{subfigure}{0.5\linewidth}
        \resizebox{\linewidth}{!}{%
    \includegraphics[]{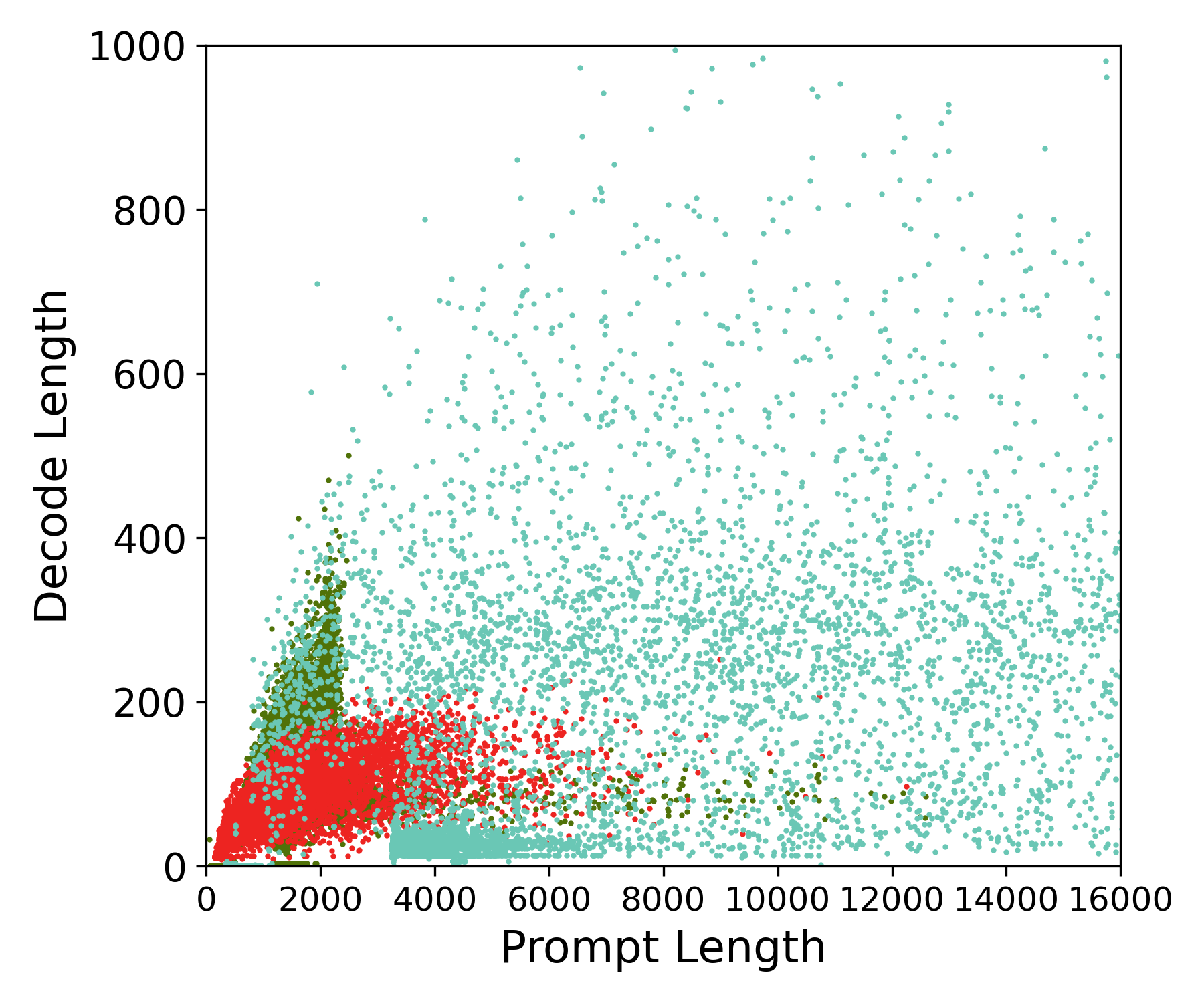}
        }
    \caption{Distribution of certain applications}\label{fig:entire_trace_distribution1}
    \end{subfigure}%
    \caption{Prompt-decode distribution of the production traffic from Cloud provider X.}
    \label{fig:trace_input_output_distribution}
\end{figure}

~\autoref{fig:prompt_decode_distribution} shows the distribution of prompts and decode tokens across the different datasets we mixed. We can clearly see the different distributions each dataset has. Prompts from Eli-5 Reddit subset are shorter in length and have longer responses than the rest of the dataset, while the IMDb distribution on the other hand has longer prompt lengths and shorter responses. Such a varied distriubtion contributes to the low accuracy of the current SOTA model by~\citet{jin2023s3}.

\subsection{Training details of the output length predictor} 
We had a total of 31329 samples in our mixed dataset, from which we had an 80:20 train-test split.  We had a train time accuracy of 81\% after performing 6 epochs of fine-tuning with the entire training set.

\subsection{Task Predictability} \label{subsec:task_predictability}
We predict the task of a prompt sampled from our dataset described in~\autoref{Sec: Observation} using DistillBERT, the same methodology we use to predict their output length bucket as discussed in~\autoref{sec:output_length_predictor}. We observe an accuracy of  93.79\%.

This allows us to proceed safely with the assumption that we can provide task type as part of the prompt to the output length predictor.

\subsection{Licenses}\label{app:licenses}
\begin{enumerate}
    \item WNUT Dataset: CC-by-4.0
    \item SQuAD dataset: CC-by-SA-4.0
    \item vLLM: Apache-2.0
\end{enumerate}

\subsection{Details of RL training} \label{app:rl_details}
For our experiments, we use 4 LLM model instances to route among. This results in our state space having 27 dimensions (6 for each model instance and 4 for the request queue at the router). In order to bound our state space, we round the estimated capacity available at each model instance and the estimated completion time for the earliest request to two decimal places. We also upper bound the waiting queue length that we provide to the DQN to $4\times (\texttt{max batch size}) = 4 \times 128 = 512$. We provide the DQN with 3 buckets: 0-256, 256-2048, $\geq$ 2048. 

\subsubsection{Q-Learning}
Q-Learning yields poor performance for our task due to the size of the state space. If we upper bound the total number of requests that can be present at a model instance to 150 (even though there can be infinitely many) and the prompt and decode length to 4096 (maximum content window of LLama-2 7B), each model instance can be in $150 \times 150 \times 100 \times (4096\times 150 \times \texttt{grad}_2 \times 100) = 3.0405\times 10^{9}$ different states. This would result in a total of $(3.0405\times 10^9)^4 \times 512 \times 4096 \times 3 \approx 5\times 10^{44}$. Even though we will never visit most of these states, the possible states that be visited are large enough to make Q-Learning infeasible.

\subsubsection{Training Rewards}
\begin{figure}
    \centering
    \includegraphics[width=0.3\linewidth]{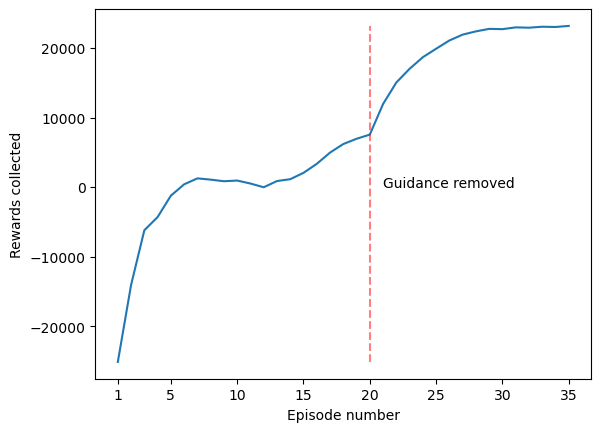}
    \caption{Training reward for workload guided RL}
    \label{fig:training_workload_guided}
\end{figure}
\autoref{fig:training_workload_guided} Shows the rewards collected during training of the RL model. We see that the guidance heuristic helps the agent converge. After episode 20, we no longer use explore with random actions and exploit this knowledge.

\subsubsection{Double-DQN}
We take a double DQN approach for our RL agent. We set the request completion reward to 60 and train our DQN with a batch size of 512. We use a neural network with layer sizes $(27, 64), (64, 64), (64, 5)$ and ReLU activation function for layers 1 and 2.

\begin{figure}
    \centering
    \begin{subfigure}{0.32\linewidth}
    \resizebox{\linewidth}{!}{
        \includegraphics[]{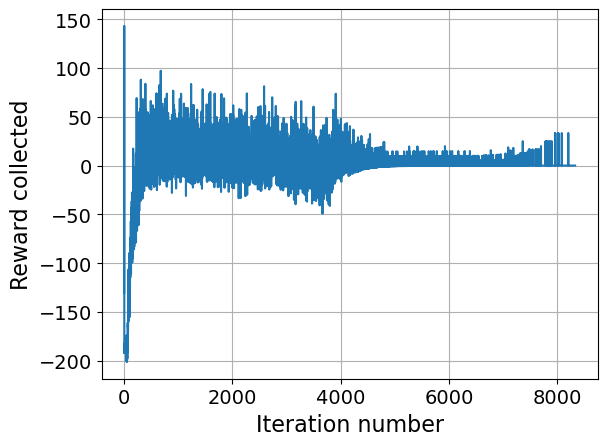}
    }
    \caption{Baseline RL}
    \end{subfigure}
    \begin{subfigure}{0.32\linewidth}
    \resizebox{\linewidth}{!}{
        \includegraphics[]{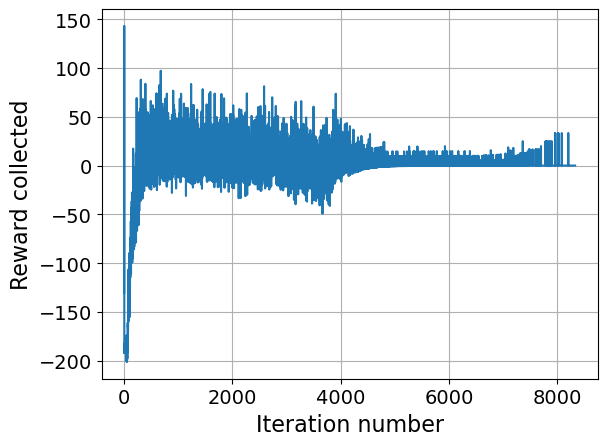}
    }
    \caption{Workload Aware RL}
    \end{subfigure}
    \begin{subfigure}{0.32\linewidth}
    \resizebox{\linewidth}{!}{
        \includegraphics[]{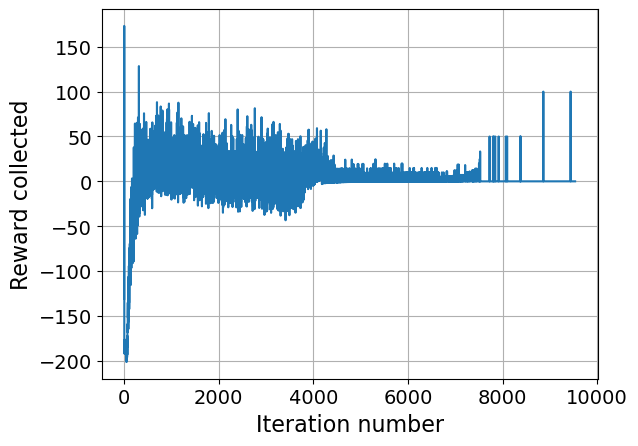}
    }
    \caption{Workload Guided RL}
    \end{subfigure}
    \caption{Rewards collected during testing for each strategy }
    \label{fig:rewards}
\end{figure}

~\autoref{fig:rewards} shows the  rewards collected by each strategy during testing. Requests stop arriving at iteration number 4000, after which, we see the rewards tend to positive values due to the high request completion reward.

\subsection{Overhead of the router}
For each decision, the router has to perform two additional steps in our approach: (i) inference from DistillBERT for output length bucket and (ii) inference from the neural network being used. The approach can parallelize these modules when the number of requests in the queue is large (and the request being routed has already been processed by the length predictor). (i) takes us 0.01 (on GPU) and 0.8 (on CPU) seconds per batch of size 64 and (ii) takes $<10^6$ operations (within milliseconds) to process.  
\subsection{Additional experiments to validate the scalability of proposed framework}  \label{exp:scale}
To further test the scalability of our approach, we tested our methods with eight model instances. We increased the number of processed requests to 4,000 and the request arrival rate to 40/s to remain consistent with previous experiments. To scale our approach, we needed to increase the parameters in our neural network. Our methods outperformed the Round-Robin approach in this setup as well. On average, Baseline RL, Workload Aware RL, and Workload Guided RL outperformed Round Robin by 5.84\%, 6.64\%, and 11.62\%, respectively.
\subsection{Experiments on Real Production Trace from Cloud Provider X} \label{exp:realtrace}
Next, we validate our approach using one hour production trace from Cloud provider X. We use 4000 requests for our experiments, with average prompt length of 5526.64 tokens and average decode length of 112.69 tokens. We do our experiments at 80 requests per second, again using Llama-3.1-8B model. We enable chunking for this experiment with maximum number of batched tokens set to 1024. Round robin takes 1005.31 seconds on average (across 20 random iterations). We see that the advantages of our algorithms are less pronounced when the prompt length becomes much longer than the decode length, with advantages of baseline RL, workload aware RL and workload guided RL reducing to 2.28\% (982.38 seconds), 4.39\% (961.17 seconds) and 7.84\% (926.49 seconds) respectively. This can also be attributed to the lesser number of preemptions happening as the decode length has gotten shorter.

To reduce the overhead of output length prediction, we assume the unavailability of prompt content and only assume the availability of prompt token count. Therefore, for the bucket prediction module, we train a Random Forest which takes the prompt length of the request along with the application name associated with the request. Using the same bucket sizes as before, this module is able to achieve 79\% accuracy (while 68.44\% of the requests were in bucket 0) due to the predictable nature of production traffic. \autoref{fig:trace_input_output_distribution} shows the prompt and decode distribution from the production trace. The prompt and decode distribution of applications from the production trace show distinct trend as shown in \autoref{fig:entire_trace_distribution1} which makes the decode length predictable with prompt length and application type.
\subsection{Additional Proofs}

Reshaping the MDP ($\mathcal{M}$) with heuristic guided RL  preserves the value bounds and linearity of the original MDP: 1) If
$h(s)$  $\in  [0, \frac{1}{1-\gamma}]$, then value function corresponding to the policy, $\pi$, $\tilde{V}^{\pi} (s) \in [0, \frac{1}{1-\gamma}] $ for all $\pi$ and $s \in \mathcal{S}$. 2) If $\mathcal{M}$ is a linear MDP with
feature vector $\phi(s, a)$ (i.e. $r(s, a)$ and $\mathbb{E}_{s'|s,a}\left[g(s')\right]$ for any $g$ that can be linearly parameterized in $\phi(s,a)$ \citep{cheng2021heuristic}.